\documentclass[pdflatex,sn-mathphys-num]{sn-jnl}% Math and Physical 

\usepackage{amsmath}
\usepackage{adjustbox}
\usepackage{caption}
\usepackage{subcaption}

\usepackage{undertilde}
\usepackage{accents}

\usepackage{tikz}
\usepackage{graphicx}%
\usepackage{multirow}%
\usepackage{amssymb,amsfonts}%
\usepackage{amsthm}%
\usepackage{mathrsfs}%
\usepackage[title]{appendix}%
\usepackage{xcolor}%
\usepackage{textcomp}%
\usepackage{manyfoot}%
\usepackage{booktabs}%
\usepackage{algorithm}%
\usepackage{algorithmicx}%
\usepackage{algpseudocode}%
\usepackage{listings}%
\usepackage{enumitem}
\usepackage[colorinlistoftodos]{todonotes}

\usepackage{hyperref}
\newcommand{\downloadlink}[1]{%
  \begin{center}
    \Large\bfseries
    \href{#1}{Click here to download source files}
  \end{center}
}

\newcommand{\black}[1]{\textcolor{black}{#1}}

\begin{document}

\title[Article Title]{A plastic correction algorithm for full-field elasto-plastic finite element simulations : critical assessment of predictive capabilities and improvement by machine learning}

%\title{A plastic correction algorithm for full-field elasto-plastic finite element simulations : critical assessment of predictive capabilities and improvement by machine learning

%\vspace{30}
%Abhishek Palchoudhary, 
%Simone Peter,
%Vincent Maurel, 
%Cristian Ovalle,
%Pierre Kerfriden
%}


\author*[1]{\fnm{Abhishek} \sur{Palchoudhary}}\email{abhishek.palchoudhary@minesparis.psl.eu}

\author[1,2]{\fnm{Simone} \sur{Peter}}
\equalcont{These authors contributed equally to this work.}

\author[1]{\fnm{Vincent} \sur{Maurel}}
\equalcont{These authors contributed equally to this work.}

\author[1]{\fnm{Cristian} \sur{Ovalle}}
\equalcont{These authors contributed equally to this work.}

\author[1]{\fnm{Pierre} \sur{Kerfriden}}\email{pierre.kerfriden@minesparis.psl.eu}
\equalcont{These authors contributed equally to this work.}

\affil*[1]{\orgdiv{Centre for Material Sciences (CMAT), CNRS UMR 7633, BP 87}, \orgname{Mines Paris, PSL University}, \orgaddress{\city{Evry}, \postcode{91003}, \country{France}}}

\affil[2]{\orgname{Technical University of Munich}, \orgaddress{\street{Arcisstr. 21}, \city{Munich}, \postcode{80333}, \country{Germany}}}

\abstract{\black{This paper introduces a new local plastic correction algorithm that is aimed at accelerating elasto-plastic finite element (FE) simulations for structural problems exhibiting localised plasticity (around e.g. notches, geometrical defects). The proposed method belongs to the category of generalised multi-axial Neuber-type methods, which process the results of an elastic prediction point-wise in order to calculate an approximation of the full elasto-plastic solution. The proposed algorithm relies on a rule of local proportionality, which, in the context of J2 plasticity, allows us to express the plastic plastic correction problem in terms of the amplitude of the full mechanical tensors only. This lightweight correction problem can be solved for numerically using a fully implicit time integrator that shares similarities with  the radial return algorithm.
The numerical  capabilities of the proposed algorithm are demonstrated for a notched structure and a specimen containing a distribution of spherical pores, subjected to monotonic and cyclic loading. 
As a second point of innovation, we show that the proposed local plastic correction algorithm can be further accelerated by employing a simple meta-modelling strategy, with virtually no added errors. At last, we develop and investigate the merits of a deep-learning-based corrective layer designed to the approximation error of the plastic corrector. A convolutional architecture is used to analyse the neighbourhoods of material points and outputs a scalar correction to the point-wise Neuber-type predictions. This optional brick of the proposed plastic correction methodology relies on the availability of a set of full elasto-plastic finite element solutions to be used as training data-set.}
}

\maketitle

\color{black}
\section*{List of Symbols}
\begin{description}[labelwidth=3cm, leftmargin=4cm, style=nextline]
    \vspace{1cm}
    \item[\normalfont\textbf{Material behavior constants}]
    \item[$\lambda, \mu$] Lamé coefficients
    \item[\(\sigma_y\)] Yield stress
    \item[\(C,D\)] Kinematic hardening parameters
    \item[\(Q,b\)] Isotropic hardening parameters
    \vspace{1cm}
    
    \item[\normalfont\textbf{Quantities from the elasto-static computation}]
    \item[\(\bar{\sigma}_{\mathrm{VM}}^{\#}\)] von Mises stress coming from the elasto-static computation at a fixed loading
    \item[\(\utilde{\sigma}^{\#}\),
    \(\utilde{\varepsilon}^{\#}\)] Stress and strain tensors coming from the elasto-static computation
    \item[\(\utilde{{\sigma}}_d^{\#}\), \(\utilde{{\varepsilon}}_d^{\#}\)] Deviatoric stress and strain tensors coming from the elasto-static computation
    \item[\(\utilde{\bar{\sigma}}_d^{\#}\), \(\utilde{\bar{\varepsilon}}_d^{\#}\)] Deviatoric stress and strain tensors coming from the elasto-static computation calculated at a fixed loading
    \item[\(\utilde{{\sigma}}_{d,o}^{\#}\), \(\utilde{{\varepsilon}}_{d,o}^{\#}\)] Deviatoric stress and strain tensors coming from the elasto-static computation at the last peak of cyclic loading
    \vspace{1cm}
    
    \item[\normalfont\textbf{Quantities from a reference  elasto-plastic computation}]
    
    \item[\(\utilde{\sigma}\), \(\utilde{\varepsilon}\)] Reference stress and strain tensors
    \item[\(\utilde{\sigma}_d\), \(\utilde{\varepsilon}_d\)] Reference deviatoric stress and strain tensors
    \item[\(\utilde{\varepsilon}^p\)] Reference plastic strain tensor
    \item[\(\utilde{\varepsilon}^p_d\)] Reference deviatoric plastic strain tensor
    \item[\(f_y\)] Reference yield surface function
    \item[\(\mathcal{J}\)] Reference von Mises stress
    \item[\(\utilde{X}\)] Reference nonlinear kinematic hardening tensor
    \item[\(R\)] Reference nonlinear isotropic hardening function
    \item[\(p\)] Reference cumulative plastic strain
    \vspace{1cm}
    
    \item[\normalfont\textbf{Quantities from an elasto-plastic computation by the plastic corrector}]
    
    \item[\(\hat{\utilde{\sigma}}_d\), \(\hat{\utilde{\varepsilon}}_d\)] Deviatoric stress and strain tensors computed by the plastic corrector
    \item[\(\hat{\utilde{\sigma}}_{d,o}\), \(\hat{\utilde{\varepsilon}}_{d,o}\)]  Deviatoric stress and strain tensors at the last peak of cyclic loading computed by the plastic corrector
    \item[\(\utilde{\hat{\varepsilon}}_d\)] Deviatoric stress and strain tensors computed by the plastic corrector
    \item[\(\utilde{\hat{\varepsilon}}^p\)] Plastic strain tensor computed by the plastic corrector
    \item[\(\utilde{\hat{\varepsilon}}^p_d\)] Deviatoric plastic strain tensor computed by the plastic corrector
    \item[\(\hat{f}_y\)] Yield surface function computed by the plastic corrector
    \item[\(\mathcal{\hat{J}}\)] von Mises stress computed by the plastic corrector
    \item[\(\utilde{\hat{X}}\)] Nonlinear kinematic hardening tensor computed by the plastic corrector
    \item[\(\hat{p}\)] Cumulative plastic strain computed by the plastic corrector
    \vspace{1cm}
    
    \item[\normalfont\textbf{Scalar variables for proportional evolution rule}]

    \item[\(s\)] Ratio of the approximated deviatoric stress tensor to the deviatoric stress tensor from the elasto-static computation
    \item[\(e\)] Ratio of the approximated deviatoric strain tensor to the deviatoric strain tensor from the elasto-static computation
    \item[\(e^p\)] Ratio of the approximated deviatoric plastic strain tensor to the deviatoric strain tensor from the elasto-static computation
    \item[\(x\)] Ratio of the approximated kinematic hardening tensor to the deviatoric strain tensor from the elasto-static computation
    \item[\(s_o\)] Ratio of the approximated deviatoric stress tensor to the deviatoric stress tensor from the elasto-static computation at the last peak of cyclic loading
    \item[\(e_o\)] Ratio of the approximated deviatoric strain tensor to the deviatoric strain tensor from the elasto-static computation at the last peak of cyclic loading
    \item[\(e^p_o\)] Ratio of the approximated deviatoric plastic strain tensor to the deviatoric strain tensor from the elasto-static computation at the last peak of cyclic loading
    \item[\(f\)] Loading function
    \item[\(f_o\)] Loading function at the last peak of cyclic loading
    \vspace{1cm}
    
    \item[\normalfont\textbf{Other miscellaneous notation}]
    
    \item[\(\hat{\hat{\utilde\sigma}}_d\)] Projected deviatoric stress tensor using the local proportionality rule
    
    \item[\(\underbar{u}\)] Displacement vector
    \item[$\mathbb{I}$] Identity tensor
    
    \item[\(\Delta p\)] Cumulative plastic strain range in the 20\textsuperscript{th} cycle
    \item[\(\phi\)] Intrinsic dissipation in a loading cycle
    \vspace{1cm}

    % Add more symbols here
\end{description}
\color{black}

\section{Introduction}
\medskip
\color{black}
The computer simulation of industrial components is often based on plasticity analysis around critical areas with high stress concentrations such as notches or defects like pores
% PK: unnecessarily specific
%, for the calibration of associated \textcolor{black}{failure} criteria 
\cite{Kugel1961, Morel2002, Hooreweder2011, Krzyzak2014, Driss2019,Dezecot2017,Paul2014,Branco2018}.
These analyses usually require finite element simulations. However, plasticity simulations are computationally expensive due to the number of elements required to accurately represent the geometry of stress concentrators, and due to the number of time increments needed to integrate non-linear material laws over time. In this context, we propose a new methodology, which belongs to the class of plastic correction approaches, to rapidly obtain an approximation of the full-field elasto-plastic response of structures subjected to proportional loading sequences,  from a single 3D elasto-static finite element solution.

\color{black}
Several types of plastic correction methodologies have been developed in the past to post-process the elasto-plastic response of structures from elastic finite element solutions. One family of methods uses homogenization theory, which consists in viewing the plastic zone as an inclusion in an elastic matrix and deriving local constraints to simulate the evolution of plastic quantities at the notch tip under load, starting from an elastic finite element solution \cite{Herbland2009, Chouman2014, Levieil2019}.
\black{Another family of methods relies on the use of Neuber-type rules. These rules are heuristics that relate the stresses and strains in an elasto-plastic body to those in a geometrically similar elastic body undergoing similar loading conditions.} While originally developed for a uni-axial stress state at a notch subjected to monotonic loading \cite{Neuber1961} an then cyclic loading \cite{Topper1969,Chaudonneret1985}, the Neuber rule (and other Neuber-type methods like the Equivalent Strain Energy Density approach \cite{Molski1981}) have been generalized to multi-axial loading states at notch tips. 
\black{One sub-family of Neuber-type methods for multi-axiality is based upon independently employing Neuber-type rules for every scalar component of the stress and strain tensors \cite{Buczynski2003, Ince2013}.} The other sub-family of Neuber-type approaches reduces the complexity of the previous approach by employing variations of proportional evolution rules for the stress tensor, strain tensor, or for a combination of stress and strain \cite{Moftakhar1995, Demorat2002, Ye2008, McDonald2011}. \black{As far as we are aware, these existing pieces of work concentrate specifically on the development of Neuber-type plastic correction methods for the prediction of multiaxial stress and strain states at notch tips.} One specific piece of work by Desmorat et al. \cite{Demorat2002} suggests using locally proportional evolutions together with Neuber-type rules to predict the evolution of elasto-plastic fields. Yet we could not find any detailed analysis of the accuracy of the suggested methodology away from notches and free boundaries, nor could we find the derivations of general-purpose algorithms to integrate the resulting plastic correction equations numerically, i.e. under arbitrary (proportional) load histories.

\black{
The first point of innovation proposed in this paper is the development of a multiaxial Neuber-type plastic correction method that may be used to produce approximations of elasto-plastic quantities at every quadrature point of a finite element mesh. As such, the developed methodology will not rely on particular strain state assumptions regarding the recovery of elasto-plastic solutions at traction-free boundaries. Our proposal is to use a local proportionality rule for the deviatoric stress and strain tensors, which will therefore be linearly related to their counterparts as calculated using an elasto-plasticity analysis. Remarkably, in the context of J2 plasticity (we use a Chaboche model with kinematic and isotropic hardening), the local proportionality rule results in scalar constitutive equations for the deviatoric strain and stress amplitudes, without further assumption on plastic flow, which was already identified in \cite{Demorat2002}. Complemented by a scalar Neuber-type rule, the constitutive equations may be solved locally, the load stemming directly from that heuristic rule. We use the standard change of peaks method \cite{Chaudonneret1985} to account for load cycles with non-zero mean stresses.  We will analyse the accuracy of full 3D elasto-plastic solutions computed using the Neuber-type approximation, i.e. at and away from notches and free boundaries. This full-field aspect is particularly relevant to fields whereby fracture criteria are based on full-field elasto-plastic solutions, for example non-local fatigue models that use elasto-plastic fields around critical points \cite{Taylor2005}, or weakest-link models that compute probabilities of failure via a weighted average of mechanical stresses over the entire computational domain \cite{Beremin87,Zok2017}.
}

\black{The second point of novelty is a machine-learning-based acceleration of the proposed plastic corrector, which is particularly useful for finite element models exhibiting a large number of degrees of freedom and long time analyses. This is because the local time integration of the elasto-plastic equations, using the plastic correction methodology mentioned previously, remains computationally expensive. We show that under the previously stated rules (J2 plasticity, proportionality of deviatoric strain and stress tensors, scalar Neuber rule), and for a given load history, any output of the plastic correction algorithm exhibits a (nontrivial) scalar dependency to the von Mises stress of the elastic finite element simulation. Therefore, we suggest a meta-modelling strategy in the form of a Gaussian process regression \cite{rasmussen2006}, that will be trained to produce the elasto-plastic quantity of interest given a von Mises stress \black{stemming from the elasto-static simulation} as input. A dataset is generated by populating the real positive axis and computing the corresponding outputs using the plastic corrector. Few (30 to 150) such datapoints are necessary for a one-dimensional regression. We show that training the Gaussian process regression on such a small dataset is sufficient to act as a virtually cost-free surrogate for all remaining local plastic corrections, without sacrificing accuracy. As a consequence, the cost of acquiring approximated elasto-plastic quantities for specimens with single or multiple stress concentrators reduces to the cost of the elastic finite element simulation, as the number of degrees of freedom of the finite element model increases.
}

\black{As a third and more exploratory part of the development, we propose a methodology of plastic correction using neural networks, which we coin Neural plastic corrector (NPC). The Neural Network developed here will be used to correct the output of the proposed Neuber-type methodology in order to better reproduce the mechanical fields delivered by a full elasto-plastic finite element analysis. While the Neuber-type method described thus far never requires elasto-plastic finite element simulations to be performed, the NPC is based on training with examples (supervised learning), and is therefore based on the availability of reference elasto-plastic solutions that will be used as a dataset. The Neuber corrector being local by nature, we aim to correct it by using information about the local topology of material point neighbourhoods, which will be analysed in an end-to-end fashion by a Convolutional Neural Network (CNN). More precisely, we will first voxelise the neighbourhood of any point of interest. We will then project the result of the previously described Neuber-type methodology to be improved onto the voxelised neighbourhood. Finally, we will provide this gridded data as input to a Neural network and train it to predict a correction to the elasto-plastic quantity of interest obtained approximately using the Neuber-type plastic corrector. Whilst the previously introduced Neuber-type method may be applied to general structures exhibiting stress concentrations, the Neural plastic corrector will be dedicated to the analysis of specimens with randomly placed pores, such as those that may be of interest when simulating the failure of porous alloys obtained by casting, welding or additive manufacturing \cite{Bercelli2021, Merot2024, Palchoudhary2024}. In this context, the necessity to analyse specimens, or batches of specimens, containing large quantities of defects may justify the deployment of an AI that learns from a small to medium quantity of \black{reference} elasto-plastic simulations in order to predict the \black{elasto-plastic} output of subsequent simulations. We do not anticipate a strong generalisation ability of the NPC, consistently with observations and analyses made in our previous work \cite{Krokos2022, Krokos2024}, albeit in a  different mechanical context. However, we will show that the output of the Neuber-type methodology may be improved upon by the proposed AI, using a reasonable amount of data for training, and that using the output of our Neuber-type methodology as full-field input to the CNN is indeed beneficial as compared to using inputs of the NPC with a lesser mechanical content.
}

This paper is divided into four sections: the first section introduces the proposed Neuber-type plastic corrector scheme.
The second section is devoted to results and error analysis of the plastic correction algorithm \textcolor{black}{for two different problems : A structure with a notch undergoing monotonic loading, and a specimen made of a porous alloy with a spherical pore population undergoing monotonic and cyclic loading.
The third section introduces the meta-modelling strategy proposed to accelerate the plastic correction algorithm, and the fourth section presents the Neural Plastic Correction strategy. The three elements introduced in the paper are separately validated in the section where they are respectively presented. The paper is concluded by a discussion and perspectives of future studies.}

\section{Full field Neuber-type plastic corrector}
\subsection{Linear elasticity problem}\label{sec:LE}

Neuber-type methods approximate plasticity by locally processing the stress and strain fields stemming from a single elasto-static finite element analysis. We set up a linear elastic problem in an isotropic material, with stiffness tensor $\mathcal{C}$ and \textcolor{black}{Lamé coefficients $\lambda, \mu$}, over a time interval  $[0, \, T]$, whereby proportional loading conditions are assumed \textcolor{black}{at the global level, i.e. that of the structure}. \textcolor{black}{The time-dependent displacement, stress and strain tensors for the elasto-static simulation are denoted by $\underbar{u}^{\#}(t)$, ${\utilde\sigma}^{\#}(t)$ and $\utilde{\varepsilon}^{\#}(t)$, respectively. The identity tensor is denoted by $\mathbb{I}$. The boundary $\partial \Omega$ of computational domain $\Omega$ is additively split into a Dirichlet part $\partial \Omega_u$ and a Neumann part $\partial \Omega_t$. The equations of linearised elasticity are introduced as follows:}\color{black}
\begin{equation}
    \textrm{div}\, \utilde{\sigma}^{\#}(t) + f(t) \, {\underline{\bar{\xi}}} = 0
\end{equation}
\begin{equation}
    {\utilde\sigma}^{\#} (t) = \mathcal{C}:\utilde{\varepsilon}^{\#}(t) = \lambda \textrm{Tr}(\utilde{\varepsilon}^{\#}(t)) \mathbb{I} + 2\mu\utilde{\varepsilon}^{\#}(t)
\end{equation}
\begin{equation}
    \utilde{\varepsilon}^{\#}(t) = \frac{1}{2} ( \nabla \underbar{u}^{\#}(t) + \nabla \underbar{u}^{\#}(t)^{T})
\end{equation}
\begin{equation}
   \underbar{u}^{\#}(t) = f(t) \, \bar{\underbar{u}}_a \quad \quad \text{over} \quad \partial \Omega_u
\end{equation}
\begin{equation}
   \utilde{\sigma}^{\#}(t) \cdot \underbar{n} = f(t)  \, \bar{\underbar{t}}_a \quad \quad \text{over} \quad \partial \Omega_t
\end{equation}
In the previous set of equations, $f : [0, \, T] \mapsto \mathbb{R}$ is an arbitrary function of time that may be set to simulate, for example, proportionally monotonic and cyclic loading. $\underline{\bar{\xi}}$ is a vector-valued field of volume forces,  $\bar{\underbar{u}}_a$ is a vector-valued field of applied displacements, and  $\bar{\underbar{t}}_a$ is a vector-valued field of applied traction loads. The previous elastic time-dependent problem is proportional, in the sense that all prescribed loading conditions, be it Neumann, Dirichlet conditions or volume sources, are introduced as fixed vector-valued fields multiplied by a function of time $f$. 

Therefore, $\underbar{u}^{\#}(t)$, ${\utilde\sigma}^{\#}(t)$ and $\utilde{\varepsilon}^{\#}(t)$ may be obtained by solving the equation of elasto-statics for $f=1$, and processed in the following way to recover the history of elastic solutions over $[0, \, T]$:
\begin{equation}
    \underbar{u}^{\#}(t) = f(t) \, \bar{\underbar{u}}^{\#} 
\end{equation}
\begin{equation}
    \utilde\sigma^{\#}(t) = f(t) \, \utilde{\bar{\sigma}}^{\#}
\end{equation}
\begin{equation}
    \utilde{\varepsilon}^{\#}(t) = f(t)  \, \utilde{\bar{\varepsilon}}^{\#}
\end{equation}
where $\bar{\underbar{u}}^{\#}$, $\utilde{\bar{\sigma}}^{\#}$ and $\utilde{\bar{\varepsilon}}^{\#}$ are, respectively, the displacement field, the stress tensor and the strain tensor fields obtained by setting $f=1$. (note: a bar symbol $(\bar{\bullet})$ is used to denote quantities that are obtained by solving the equations of elasto-statics with $f=1$).

If the elastic deviatoric stress and strain tensors obtained for $f=1$ are denoted by $\utilde{\bar{\sigma}}_d^{\#}$ and $\utilde{\bar{\varepsilon}}_d^{\#}$ respectively, then the elastic deviatoric tensors at any time $t$ (denoted by $\utilde{\sigma}^{\#}_d(t)$ and $\utilde{\varepsilon}^{\#}_d(t)$) are obtained by the following scaling:
\begin{equation}\label{eq10}
\utilde{{\sigma}}_d^{\#}(t) = f(t) \, \textcolor{black}{\utilde{\bar{\sigma}}_d^{\#}}
\end{equation}
\begin{equation}\label{eq11}
    \utilde{{\varepsilon}}_d^{\#}(t) = f(t) \, \textcolor{black}{\utilde{\bar{\varepsilon}}_d^{\#}}
\end{equation}
\color{black}
\subsection{\black{Von Mises plasticity constitutive model}}\label{sec:plas}

\black{A von Mises plasticity model} with a non-linear kinematic hardening and a non-linear isotropic hardening is chosen for the description of the evolution of the yield surface \cite{Chaboche1989}. The deviator of the stress tensor $\utilde{\sigma}$ is henceforth denoted by $\utilde{\sigma}_d$, the total strain by $\utilde{\varepsilon}$ and the plastic strain tensor by $\utilde{{\varepsilon}}^p$. Elasticity is given by
\color{black}
\begin{equation}\label{eq:plas:stressfull}
    \utilde{\sigma} = \mathcal{C}:(\utilde{\varepsilon} - \utilde{\varepsilon}^p)
\end{equation}
The von Mises stress is given as:
\begin{equation}\label{eq:plas:VMStressfull}
    \mathcal{J}(\utilde{\sigma}_d - \utilde{X}) = \sqrt{\frac{3}{2} (\utilde{\sigma}_d - \utilde{X}):(\utilde{\sigma}_d - \utilde{X})}
\end{equation}
where $\utilde{X}$ is the non-linear kinematic hardening tensor. The expression for the plastic strain rate $\dot{\utilde{\varepsilon}}^p$ is given as:
\begin{equation}\label{eq:full_plasticstrainrate}
    \dot{\utilde{\varepsilon}}^p = \dot{p} \left(  \frac{3}{2} \frac{\utilde{\sigma}_d - \utilde{X}}{\mathcal{J}(\utilde{\sigma}_d - \utilde{X})} \right)
\end{equation}
The evolution of $\utilde{X}$ is given by the following expression:
\begin{equation}\label{eq:plas:kinematicfull}
    \dot{\utilde{X}} = \frac{2}{3}C\dot{\utilde{\varepsilon}}^p - D \utilde{X}\dot{p}
\end{equation}
where $\dot{p}$ is the time evolution of the cumulative plastic strain $p$, and $C$ and $D$ are kinematic hardening material parameters. The isotropic hardening $R(p)$ is  given by the following expression:
\begin{equation}\label{eq:isotropic}
    R(p)=Q\big(1-\exp{(-b p)}\big)
\end{equation}
where $Q$ and $b$ are isotropic hardening material parameters. The deviator of the plastic strain tensor is used to define the cumulative plastic strain:
\begin{equation}
 \dot{p} = \sqrt{\frac{2}{3}\dot{\utilde{\varepsilon}}_d^p:\dot{\utilde{\varepsilon}}_d^p}
\end{equation}

The evolution of the yield surface $f_y(\utilde{\sigma}_d;\utilde{X},p)$ is required to satisfy the following two constraints:
\begin{equation}\label{eq:plas:yieldfull}
    f_y(\utilde{\sigma}_d;\utilde{X},p) = \mathcal{J}(\utilde{\sigma}_d - \utilde{X}) - \sigma_y - R(p) \leq 0
\end{equation}
\begin{equation}\label{eq:plas:yieldfull2}
    f_y \dot{p} = 0
\end{equation}
\color{black}
\subsection{Modified Neuber rule}\label{sec:ModNeuber}
\medskip

\paragraph*{Neuber rule for deviatoric stress and strain tensors}\mbox{}\\
\color{black}
The Neuber-type rule proposed in this paper operates on the deviatoric parts of the stress and strain tensors \cite{Buczynski2003}:
\begin{equation}\label{eq:neuberfull_monotone}
 \hat{\utilde{\sigma}}_d: \hat{\utilde{\varepsilon}}_d
= \utilde{{\sigma}}_d^{\#}: \utilde{{\varepsilon}}_d^{\#}
\end{equation}
where $\hat{\utilde{\sigma}}_d$, $\hat{\utilde{\varepsilon}}_d$ stand for the approximated deviatoric stress and strain tensors, respectively (note: a hat symbol $(\hat{\bullet})$ is used to denote quantities that are approximated by the plastic corrector). 

Tensors $\utilde{{\sigma}}_d^{\#}$, $\utilde{{\varepsilon}}_d^{\#}$, which are obtained by solving elasto-statics instead of elasto-plasticity, are time-dependent, but the $(t)$ notation has been dropped for conciseness. 

Cyclic loading is handled with the classical change of origin at every peak, as proposed by Chaudonneret \cite{Chaudonneret1985}:
\begin{equation}\label{eq:neuberfull}
( \hat{\utilde{\sigma}}_d - \hat{\utilde{\sigma}}_{d,o} ):( \hat{\utilde{\varepsilon}}_d - \hat{\utilde{\varepsilon}}_{d,o} ) 
= ( \utilde{{\sigma}}_d^{\#} - \utilde{{\sigma}}_{d,o}^{\#} ):( \utilde{{\varepsilon}}_d^{\#} - \utilde{{\varepsilon}}_{d,o}^{\#} )
\end{equation}
In the previous equation, quantities $\hat{\utilde{\sigma}}_{d,o}$, $\hat{\utilde{\varepsilon}}_{d,o}$ stand for the approximated deviatoric stress and strain tensors at the last peak of loading. The tensors $\utilde{{\sigma}}_{d,o}^{\#}$, $\utilde{{\sigma}}_{d,o}^{\#}$ stand for the deviatoric stress and strain tensors coming from the elastic finite element solution evaluated at the last peak.
\color{black}

\paragraph*{Proportional evolution rule for deviatoric stress and strain tensors}\mbox{}\\
It is important to note that stresses and strains in a structure depend on the geometry, the material behaviour and the boundary conditions. \textcolor{black}{When there is a change in the ratio of any two components of the stress tensor at a given material point, these local stresses become non-proportional, by definition. If there is plastic flow anywhere in the structure, local non-proportionality may arise even in the case where the external loading is proportional.}

\black{The Neuber rule developed so far and the constitutive relations operate on symmetric second order tensors.} We assume local proportionality \cite{Demorat2002}, that is, while the local evolution of the actual deviatoric stress and strain tensors at a point may be arbitrarily complex, we postulate that the evolution is well approximated by assuming that it remains in the direction of the deviatoric stress and strain tensors obtained from elasto-statics, i.e. there is no shift in the direction in which plasticity develops. 

Therefore, the deviatoric stress tensor may be written as a scaling of the stress tensor stemming from elasto-statics, which reads as
\begin{equation}\label{eq:sigprop}
\hat{\utilde{\sigma}}_d(t) = s(t)\textcolor{black}{\utilde{\bar{\sigma}}_d^{\#}}
\end{equation}
and similarly for the deviatoric strain tensor,
\begin{equation}\label{eq:epsprop}
\hat{\utilde{\varepsilon}}_d(t) = e(t)\textcolor{black}{\utilde{\bar{\varepsilon}}_d^{\#}}
\end{equation}
\black{where the scalar variables $s$ and $e$ are introduced as scaling factors to be determined.}

\paragraph*{Neuber rule in terms of proportionality ratios}\mbox{}\\
\textcolor{black}{Taking the local proportionality rules into account, the Neuber rule reduces to a constraint on scalar variables $s$ and $e$. By substituting equations \eqref{eq10} and \eqref{eq11}, and \eqref{eq:sigprop} and \eqref{eq:epsprop} into equation \eqref{eq:neuberfull_monotone}, the monotonic Neuber rule, in the case of monotonic loading functions, becomes:}
\begin{equation}\label{eq:eq100}
\textcolor{black}{
 s\,e \, {\utilde{\bar{\sigma}}}_d^{\#}: {\utilde{\bar{\varepsilon}}}_d^{\#}
= f^2 \, \utilde{{\bar{\sigma}}}_d^{\#}: \utilde{{\bar{\varepsilon}}}_d^{\#}
}
\end{equation}
which yields
\begin{equation}\label{eq:eq101}
 \textcolor{black}{s\,e = f^{2}}
\end{equation}
\textcolor{black}{where we remind the reader that $s$, $e$ and $f$ are time-dependent variables (the $(t)$ notation has been dropped for conciseness).}

\textcolor{black}{For cyclic loading, we introduce  $s_o$ and $e_o$, which are the values of the scalar variables $s$ and $e$ at the last peak. Substituting equations \eqref{eq10} and \eqref{eq11}, and \eqref{eq:sigprop} and \eqref{eq:epsprop} into equation \eqref{eq:neuberfull}:}
\begin{equation}\label{eq12}
    \textcolor{black}{
    ( s{\utilde{\bar\sigma}}_d^{\#} - s_o{\utilde{\bar\sigma}}_d^{\#} ):( e{\utilde{\bar\varepsilon}}_d^{\#} - e_o{\utilde{\bar\varepsilon}}_d^{\#} ) 
= ( f\utilde{\bar{\sigma}}_d^{\#} - f_o\utilde{\bar{\sigma}}_d^{\#} ):( f\utilde{\bar{\varepsilon}}_d^{\#} - f_o\utilde{\bar{\varepsilon}}_d^{\#} )
}
\end{equation}
\begin{equation}\label{eq13}
\textcolor{black}{
{\utilde{\bar\sigma}}_d^{\#}:{\utilde{\bar\varepsilon}}_d^{\#}( s - s_o )( e - e_o ) 
= \utilde{\bar{\sigma}}_d^{\#}:\utilde{\bar{\varepsilon}}_d^{\#}( f - f_o)( f - f_o )
}
\end{equation}

\begin{equation}\label{eq:neuberreduced}
( s - s_o )( e - e_o ) = ( f - f_o)^2
\end{equation}

\begin{figure*}
  \begin{center}
    \def\svgwidth{.95\textwidth}
\begingroup%
  \makeatletter%
  \providecommand\color[2][]{%
    \errmessage{(Inkscape) Color is used for the text in Inkscape, but the package 'color.sty' is not loaded}%
    \renewcommand\color[2][]{}%
  }%
  \providecommand\rotatebox[2]{#2}%
  \newcommand*\fsize{\dimexpr\f@size pt\relax}%
  \newcommand*\lineheight[1]{\fontsize{\fsize}{#1\fsize}\selectfont}%
  \ifx\svgwidth\undefined%
    \setlength{\unitlength}{725.7253418bp}%
    \ifx\svgscale\undefined%
      \relax%
    \else%
      \setlength{\unitlength}{\unitlength * \real{\svgscale}}%
    \fi%
  \else%
    \setlength{\unitlength}{\svgwidth}%
  \fi%
  \global\let\svgwidth\undefined%
  \global\let\svgscale\undefined%
  \makeatother%
  \begin{picture}(1,0.55614042)%
    \lineheight{1}%
    \setlength\tabcolsep{0pt}%
    \put(0,0){\includegraphics[width=\unitlength,page=1]{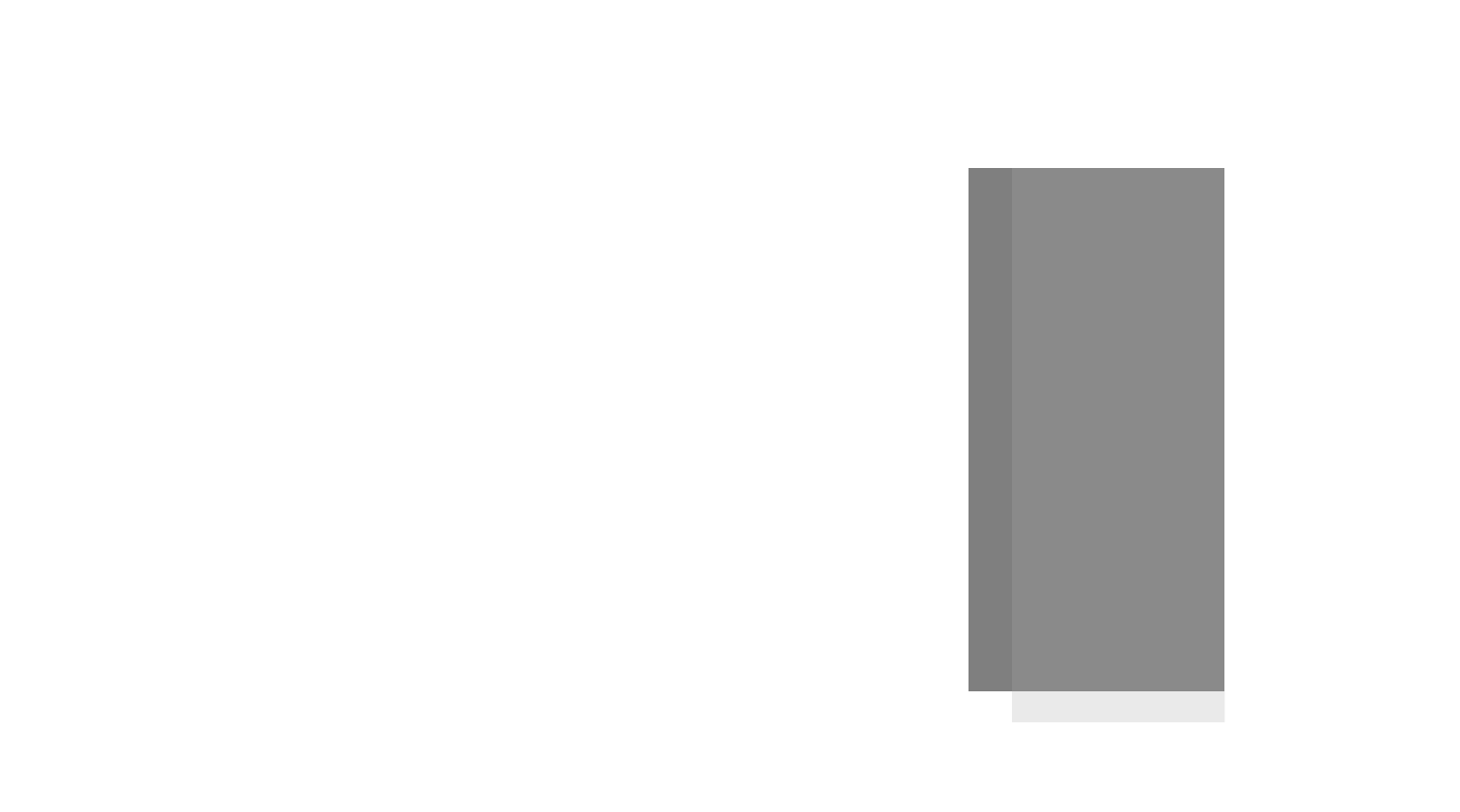}}%
    \put(0.00232773,0.52413371){\color[rgb]{0,0,0}\makebox(0,0)[lt]{\lineheight{0}\smash{\begin{tabular}[t]{l}$s$\end{tabular}}}}%
    \put(0.56618391,0.52470988){\color[rgb]{0,0,0}\makebox(0,0)[lt]{\lineheight{0}\smash{\begin{tabular}[t]{l}$s$\end{tabular}}}}%
    \put(0.3959563,0.21307646){\color[rgb]{0,0,0}\makebox(0,0)[lt]{\lineheight{0}\smash{\begin{tabular}[t]{l}$e$\end{tabular}}}}%
    \put(0.96801711,0.2084746){\color[rgb]{0,0,0}\makebox(0,0)[lt]{\lineheight{0}\smash{\begin{tabular}[t]{l}$e$\end{tabular}}}}%
    \put(0,0){\includegraphics[width=\unitlength,page=2]{Figures/Neuber_proportionality_constants_evolution.pdf}}%
    \put(0,0){\includegraphics[width=\unitlength,page=3]{Figures/Neuber_proportionality_constants_evolution.pdf}}%
    \put(0,0){\includegraphics[width=\unitlength,page=4]{Figures/Neuber_proportionality_constants_evolution.pdf}}%
    \put(0,0){\includegraphics[width=\unitlength,page=5]{Figures/Neuber_proportionality_constants_evolution.pdf}}%
    \put(0.00035254,0.42523963){\color[rgb]{0,0,0}\makebox(0,0)[lt]{\lineheight{0}\smash{\begin{tabular}[t]{l}$s^1$\end{tabular}}}}%
    \put(0.85210846,0.07956816){\color[rgb]{0,0,0}\makebox(0,0)[lt]{\lineheight{0}\smash{\begin{tabular}[t]{l}$s^2$\end{tabular}}}}%
    \put(0.85309683,0.03983976){\color[rgb]{0,0,0}\makebox(0,0)[lt]{\lineheight{0}\smash{\begin{tabular}[t]{l}$s'$\end{tabular}}}}%
    \put(0.46007814,0.40702724){\color[rgb]{0,0,0}\makebox(0,0)[lt]{\lineheight{0}\smash{\begin{tabular}[t]{l}$e'$\end{tabular}}}}%
    \put(0.16489831,0.20973301){\color[rgb]{0,0,0}\makebox(0,0)[lt]{\lineheight{0}\smash{\begin{tabular}[t]{l}$e^1$\end{tabular}}}}%
    \put(0.65678664,0.45595555){\color[rgb]{0,0,0}\makebox(0,0)[lt]{\lineheight{0}\smash{\begin{tabular}[t]{l}$e^2$\end{tabular}}}}%
    \put(0.8376136,0.45636815){\color[rgb]{0,0,0}\makebox(0,0)[lt]{\lineheight{0}\smash{\begin{tabular}[t]{l}$s_o^2$, $e_o^2$\end{tabular}}}}%
    \put(0.01421361,0.21069139){\color[rgb]{0,0,0}\makebox(0,0)[lt]{\lineheight{0}\smash{\begin{tabular}[t]{l}$0$\end{tabular}}}}%
    \put(0.57927899,0.20866297){\color[rgb]{0,0,0}\makebox(0,0)[lt]{\lineheight{0}\smash{\begin{tabular}[t]{l}$0$\end{tabular}}}}%
    \put(-0.00083936,0.17974453){\color[rgb]{0,0,0}\makebox(0,0)[lt]{\lineheight{0}\smash{\begin{tabular}[t]{l}$s_o^1$, $e_o^1$\end{tabular}}}}%
  \end{picture}%
\endgroup%
\end{center}
\caption{An illustration of the plastic correction algorithm during the (a) first branch of loading (b) second branch of loading, with $s_o$ and $e_o$ updated to their respective values at the last peak }
\label{Fig:IllustrationNeuber}
\end{figure*}

\textcolor{black}{The scalar constraint between (deviatoric) stress and strain amplitudes is illustrated in Figure \ref{Fig:IllustrationNeuber} in the context of cycling loading. During the first branch of loading (denoted by the superscript $1$), the quantities $s_o$ and $e_o$ take on the initial values of $s_o^1$ and $e_o^1$, i.e. $0$, and $s$ and $e$ evolve as $s^1$ and $e^1$. When the peak of loading is reached, and a load reversal is made for a second branch of loading (denoted by the superscript $2$), the values of $s_o$ and $e_o$ are updated to the previous peak values, shown in the figure as $s_o^2$ and $e_o^2$. Next, $s$ and $e$ evolve as $s^2$ and $e^2$ according to the updated origin. This process is repeated, with the quantities $s_o$ and $e_o$ successively taking on the values of $s$ and $e$ at every peak of loading following all the load reversals defined in the function $f(t)$.} Equation \eqref{eq:neuberreduced} will be complemented by the elasto-plastic constitutive equations to yield a constitutive update algorithm.

\color{black}
\subsection{Elasto-plastic constitutive equations under proportional tensor evolutions}\label{sec:plasreduced}
\medskip
\color{black}

\textcolor{black}{The aim of this section is to reduce the tensorial equations of the constitutive law (presented in section \ref{sec:plas}) to scalar equations, using the rules of proportionality stated in equations \eqref{eq:sigprop} and \eqref{eq:epsprop}, leading to a set of reduced constitutive equations in a set of scalar variables including $s$ and $e$. The reader is reminded that a hat symbol $(\hat{\textbf{.}})$ is used to denote quantities approximated by the plastic corrector.}

\paragraph{Stress-strain relation}\mbox{}\\
\color{black}
The approximated stress tensor is given by:
\begin{equation}
   \hat{\utilde\sigma} = 2\mu(\hat{\utilde\varepsilon} - \hat{\utilde\varepsilon}^p) + \lambda{\textrm{Tr}(\hat{\utilde\varepsilon} - \hat{\utilde\varepsilon}^p)}\mathbb{I}
\end{equation}
The approximated stress tensor $\hat{\utilde\sigma}$ can be split into its approximated deviatoric $\hat{\utilde\sigma}_d$ and hydrostatic $\hat{\utilde\sigma}_h$ parts:
\begin{equation}
    \hat{\utilde\sigma} = \hat{\utilde\sigma}_d + \hat{\utilde\sigma}_h
\end{equation}
where the approximated deviatoric stress tensor $\hat{\utilde\sigma}_d$, under isochoric plastic flow, becomes:
\begin{equation}
\label{eq:dev01}
    \hat{\utilde\sigma}_d = 2\mu({\hat{\utilde\varepsilon}}_d - \hat{\utilde\varepsilon}_d^p)
\end{equation}
The deviatoric strain is approximated by assuming proportional evolution to \textcolor{black}{the deviatoric strain coming from the elasto-static simulation} (refer to Section \ref{sec:ModNeuber}). \black{By virtue of equations \eqref{eq:full_plasticstrainrate} and \eqref{eq:dev01} the plastic deviatoric strain evolves proportionally to the deviatoric strain coming from the elasto-static simulation, $\utilde{\bar{\varepsilon}}_d^{\#}$. A scalar variable $e^p$ can be introduced to encode this:}
\begin{equation}\label{eq:scalarfn_e_p}
\hat{\utilde{\varepsilon}}_d^p(t) = e^p(t)\textcolor{black}{\utilde{\bar{\varepsilon}}_d^{\#}}
\end{equation}
The approximated stress and strain tensors in equation \eqref{eq:dev01} are replaced with the scalar variables $s$, $e$ and $e^p$ and the corresponding quantities from elasto-statics \textcolor{black}{${\utilde{\bar\sigma}}_d^{\#}$ and $\utilde{\bar\varepsilon}_d^{\#}$}:
\begin{equation}
\label{eq:dev02}
\textcolor{black}{
    s{\utilde{\bar\sigma}}_d^{\#} = 2\mu(e{\utilde{\bar\varepsilon}}_d^{\#} - e^p{\utilde{\bar\varepsilon}}_d^{\#})
    }
\end{equation}
A projection of equation \eqref{eq:dev02} in the direction of ${\bar{\utilde{\varepsilon}}}_d^{\#}$ is carried out:
\begin{equation}
\label{eq:dev03}
\textcolor{black}{
s{\utilde{\bar\sigma}}_d^{\#}:{\utilde{\bar\varepsilon}}_d^{\#} = 2\mu(e{\utilde{\bar\varepsilon}}_d^{\#}:{\utilde{\bar\varepsilon}}_d^{\#} - e^p{\utilde{\bar\varepsilon}}_d^{\#}:{\utilde{\bar\varepsilon}}_d^{\#})
}
\end{equation}
\textcolor{black}{As $\utilde{\bar{\sigma}}_d^{\#} : \utilde{\bar{\varepsilon}}_d^{\#} = 2\mu \utilde{\bar{\varepsilon}}_d^{\#} : \utilde{\bar{\varepsilon}}_d^{\#}$ (owing to the properties of isotropic linear elasticity)}, equation \eqref{eq:dev03} reduces to:
\begin{equation}\label{eq:monotonic_stresstrain_scalar}
s = e - e^p
\end{equation}

This equation for $s$ can be extended to the cyclic variant ($s - s_o$) by writing:
\color{black}
\begin{equation}
\label{eq:algo_delta_o_e}
    s - s_o = (e - e_o) - (e^p - e^p_o)
\end{equation}
\black{Here, alongside the updating strategy of $s_o$ and $e_o$ previously described in section \ref{sec:ModNeuber}, $e_o^p$ is also updated with the value of $e^p$ at the last peak, every time a load reversal takes place.}

\paragraph{Von Mises Stress}\mbox{}\\
\black{By virtue of equations \eqref{eq:plas:kinematicfull} and \eqref{eq:scalarfn_e_p}, the approximated kinematic hardening tensor evolves proportionally to the deviatoric strain stemming from the elasto-static computation. A scalar variable $x$ can be introduced to encode this:}
\begin{equation}\label{eq:Xx}
\utilde{X}(t) = x(t)\textcolor{black}{{\utilde{\bar\varepsilon}}_d^{\#}}
\end{equation}
Then, equation \eqref{eq:plas:VMStressfull}, with equations \eqref{eq:sigprop} and \eqref{eq:Xx}, now reads:
\textcolor{black}{
\begin{equation}
    \hat{\mathcal{J}}(s,x) = \sqrt{\frac{3}{2}({s^2\utilde{\bar{\sigma}}}_d^{\#}:{\utilde{\bar{\sigma}}}_d^{\#} - 2{sx\utilde{\bar{\sigma}}}_d^{\#}:\utilde{\bar{\varepsilon}}_d^{\#} + x^2\utilde{\bar{\varepsilon}}_d^{\#}:\utilde{\bar{\varepsilon}}_d^{\#})}
\end{equation}
}

\color{black}
From $\bar{\sigma}_{VM}^{\#} = \sqrt{\frac{3}{2}\utilde{\bar{\sigma}}_d^{\#}:\utilde{\bar{\sigma}}_d^{\#}}$ and $\utilde{\bar{\sigma}}_d^{\#} = 2\mu \utilde{\bar{\varepsilon}}_d^{\#}$ we retrieve $\utilde{\bar{\sigma}}_d^{\#}:\utilde{\bar{\sigma}}_d^{\#} = \frac{2}{3} (\bar{\sigma}_{VM}^{\#})^2 $, $\utilde{\bar{\sigma}}_d^{\#} : \utilde{\bar{\varepsilon}}_d^{\#} = \frac{1}{3\mu}({\bar{\sigma}_{\text{VM}}^{\#}})^2$ and $\utilde{\bar{\varepsilon}}_d^{\#} : \utilde{\bar{\varepsilon}}_d^{\#} = \frac{1}{6\mu^2}({\bar{\sigma}_{\text{VM}}^{\#}})^2$.

This finally leads to:
\begin{equation}\label{eq:approx_J2vm_scalar}
    \hat{\mathcal{J}}(s,x) = \left| s - \frac{x}{2 \mu}\right|{\bar{\sigma}_{\text{VM}}^{\#}} 
\end{equation}

\color{black}
\paragraph{Kinematic hardening}\mbox{}\\
\black{We now need to derive the scalar evolution equation for $\dot{x}$ as a function of $x$ and $e^p$.} From equation \eqref{eq:plas:kinematicfull}, with equations \eqref{eq:scalarfn_e_p} and \eqref{eq:Xx}, we have
\begin{equation}
\textcolor{black}{
 {\dot{x}}{\utilde{\bar\varepsilon}}_d^{\#} = \frac{2}{3}C\dot{e}^p{\utilde{\bar{\varepsilon}}}_d^{\#} - D \dot{\hat{p}} x  {\utilde{\bar{\varepsilon}}}_d^{\#}
 }
\end{equation}
\textcolor{black}{and therefore, after multiplying the right and left hand-side by ${\utilde{\bar{\varepsilon}}}_d^{\#}$ and dividing both sides by ${\utilde{\bar{\varepsilon}}}_d^{\#}:{\utilde{\bar{\varepsilon}}}_d^{\#}$, we found that
\begin{equation}
 \dot{x} = \frac{2}{3}C\dot{e}^p - Dx\dot{\hat{p}}
\end{equation}
}
\color{black}
\paragraph{Flow rule}\mbox{}\\

We now reformulate the flow rule in terms of scalar variables $s$, $x$ and $e^p$. From equation \eqref{eq:full_plasticstrainrate}, with equations \eqref{eq:sigprop}, \eqref{eq:scalarfn_e_p} and \eqref{eq:Xx}, we have that:
\begin{equation}\label{eq:scalar_plasticstrainrate_dev01}
    \dot{e}^p \utilde{\bar{\varepsilon}}_d^{\#} = \dot{\hat{p}} \left(  \frac{3}{2} \frac{s\utilde{\bar{\sigma}}_d^{\#} - x\utilde{\bar{\varepsilon}}_d^{\#}}{\mathcal{\hat{J}}(s,x)} \right)
\end{equation}
A projection of this equation \eqref{eq:scalar_plasticstrainrate_dev01} in the direction of ${\bar{\utilde{\varepsilon}}}_d^{\#}$ is carried out:
\begin{equation}\label{eq:scalar_plasticstrainrate_dev02}
    \dot{e}^p \utilde{\bar{\varepsilon}}_d^{\#} : {\bar{\utilde{\varepsilon}}}_d^{\#}  = \dot{\hat{p}} \left(  \frac{3}{2} \frac{s\utilde{\bar{\sigma}}_d^{\#} : {\bar{\utilde{\varepsilon}}}_d^{\#} - x\utilde{\bar{\varepsilon}}_d^{\#} : {\bar{\utilde{\varepsilon}}}_d^{\#}}{\mathcal{\hat{J}}(s,x)} \right)
\end{equation}
Substituting the contracted products for $\utilde{\bar{\sigma}}_d^{\#} : \utilde{\bar{\varepsilon}}_d^{\#}$ and $\utilde{\bar{\varepsilon}}_d^{\#} : \utilde{\bar{\varepsilon}}_d^{\#}$, we get:
\begin{equation}\label{eq:scalarflowrule}
    \dot{e}^p = \dot{\hat{p}}\left(\dfrac{3}{2}\dfrac{2\mu s - x}{\mathcal{\hat{J}}(s,x)}\right)
\end{equation}

\color{black}
\paragraph{Cumulative plastic strain}\mbox{}\\
The deviator of the plastic strain tensor is used for the computation of \textcolor{black}{the approximated cumulative plastic strain rate $\dot{\hat{p}}$}. The reformulation from the tensorial equation to the scalar equation uses the scalar variable $e^p$, and reads as follows
\textcolor{black}{
\begin{equation}
 \dot{\hat{p}} = \sqrt{\frac{2}{3}{(\dot{e}^p})^2{\bar{\utilde{\varepsilon}}}_d^{\#}:{\bar{\utilde{\varepsilon}}}_d^{\#}}
\end{equation}
}
\color{black}
or in a simplified form:
\begin{equation}\label{eq:approx_pdot_scalar}
 \dot{\hat{p}} = \frac{1}{3\mu} |\dot{e}^{p}| {\bar{\sigma}_{\text{VM}}^{\#}}
\end{equation}

\paragraph{Solution algorithm}\mbox{}\\
The Neuber rule and constitutive equations derived under the local proportionality rule are summarised in table \ref{tab:tensortoscalarplascorr}. We provide these relationships in terms of mechanical tensors, but also in terms of reduced scalar variables $s$, $e$, $e^p$, $x$ and $\hat{p}$. \black{It is clear that the reduced scalar variables depend solely on the elasto-static solution through the von Mises stress ($\bar{\sigma}_{\text{VM}}^{\#}$). This dependency to a scalar quantity from the elasto-static simulation is key to the success of the meta-modelling approach developed later on in the paper.} 

\black{The system of plastic correction equations needs to be solved for each quadrature point of the finite element mesh. For arbitrary (proportional) load functions $f(t)$, this is done by computing the value of $e^p(t)$ at a series of time steps, using the fully implicit time stepping scheme described in the appendix \hyperref[appendixA]{A}.}

\black{
We can finally reconstruct the approximated elasto-plastic tensor variables $(\hat{\utilde{\sigma}}(t),\hat{\utilde{\varepsilon}}(t),\hat{\utilde{\varepsilon}}^p(t),\hat{p}(t),\hat{\utilde{X}}(t))$ from the scalar quantities computed using the plastic corrector, i.e. $(s(t),e(t),e^p(t),\hat{p}(t),x(t))$. For example, to compute a component of the stress tensor, we can first reconstruct $\hat{\utilde{\sigma}}(t) = s(t) \hat{\utilde{\sigma}}_d + f(t) \text{Tr}(\bar{\utilde{\sigma}}^{\#})\mathbb{I}$ and then report the time evolution of desired component of this approximated stress tensor.
}

\begin{table}[!htbp]\color{black}
\caption{\black{Plastic correction equations written in terms of tensor variables ($\hat{\protect\utilde{\sigma}}_d$,$\hat{\protect\utilde{\varepsilon}}_d$,$\hat{\protect\utilde{\varepsilon}}^p$,$\hat{p}$,$\hat{\protect\utilde{X}}$)
to the left, and in terms of scalar variables ($s$,$e$,$e^p$,$\hat{p}$,$x$) to the right. $f$ is the global loading function, which is time dependent, and ${\bar{\sigma}_{\text{VM}}^{\#}}$ denotes the von Mises stress stemming from the elastic finite element simulation performed with $f=1$. }}
\label{tab:tensortoscalarplascorr}
\centering
\begin{tabular}[t]{llll}
\hline\noalign{\smallskip}
 Type & Tensorial variables & Scalar variables \\
\noalign{\smallskip}\hline\noalign{\smallskip}
 Neuber rule & $\hat{\utilde{\sigma}}_d: \hat{\utilde{\varepsilon}}_d
= \utilde{{\sigma}}_d^{\#}: \utilde{{\varepsilon}}_d^{\#}$ & $se = f^{2}$ \\
 \noalign{\smallskip}\hline\noalign{\smallskip}
 Elasticity & $\hat{\utilde\sigma} = 2\mu(\hat{\utilde\varepsilon} - \hat{\utilde\varepsilon}^p) + \lambda{\textrm{Tr}(\hat{\utilde\varepsilon} - \hat{\utilde\varepsilon}^p)}\mathbb{I}$ & $s = (e - e^p)$ \\
 \noalign{\smallskip}\hline\noalign{\smallskip}
Yield function & ${f_y}\textrm(\utilde{\hat{\sigma}};\utilde{\hat{X}},\hat{p}) = {\mathcal{J}} \left( {\utilde{\hat{\sigma}}} - {\utilde{\hat{X}}} \right) - \sigma_y - {R}(\hat{p})$ & ${\hat{f_y}}(s;x,\hat{p}) = \mathcal{\hat{J}}(s,x) - {\sigma}_{y} - {R}(\hat{p})$ \\
\noalign{\smallskip}\hline\noalign{\smallskip}
\textcolor{black}{Evolution of yield function} & ${{f_y}}(\utilde{\sigma}_d;\utilde{X},\hat{p}) \dot{\hat{p}} = 0 \quad \textrm{and} \quad {{f_y}}(\utilde{\sigma}_d;\utilde{X},\hat{p}){\leq 0}$ & ${\hat{f_y}}(s;x,\hat{p}) \dot{\hat{p}} = 0 \quad \textrm{and} \quad {\hat{f_y}}(s;x,\hat{p}){\leq 0}$ \\
\noalign{\smallskip}\hline\noalign{\smallskip}
von Mises stress & $\mathcal{{J}}(\utilde{\hat{\sigma}}_d - \utilde{\hat{X}}) = \sqrt{\frac{3}{2} (\utilde{\hat{\sigma}}_d - \utilde{\hat{X}}):(\utilde{\hat{\sigma}}_d - \utilde{\hat{X}})}$ & $\mathcal{\hat{J}}(s,x) = \left| s - \frac{x}{2 \mu}\right|{\bar{\sigma}_{\text{VM}}^{\#}} $ \\
\noalign{\smallskip}\hline\noalign{\smallskip}
 Isotropic hardening & ${R}(\hat{p})=Q\big(1-\exp{(-b \hat{p})}\big)$ & ${R}(\hat{p})=Q\big(1-\exp{(-b \hat{p})}\big)$ \\
\noalign{\smallskip}\hline\noalign{\smallskip}
 Kinematic hardening & $\dot{\hat{\utilde{X}}} = \frac{2}{3}C\dot{\hat{\utilde{\varepsilon}}}^p - D\utilde{\hat{X}}\dot{\hat{p}}$ & $\dot{x} = \frac{2}{3}C\dot{e}^p - Dx\dot{\hat{p}}$ \\
\noalign{\smallskip}\hline\noalign{\smallskip}
Flow rule & $\dot{\hat{\utilde{\varepsilon}}}^p = \dot{\hat{p}} \left(  \dfrac{3}{2} \dfrac{\utilde{\hat{\sigma}}_d - \hat{\utilde{X}}\;}{\;\mathcal{{J}}(\utilde{\hat{\sigma}}_d - \utilde{\hat{X}})} \right)$ & $\dot{e}^p = \dot{\hat{p}}\left(\dfrac{3}{2}\dfrac{2\mu s - x}{\mathcal{\hat{J}}(s,x)}\right)$\\
\noalign{\smallskip}\hline\noalign{\smallskip}
Cumulative plastic strain & $\dot{\hat{p}} = \sqrt{\frac{2}{3}\dot{\hat{\utilde{\varepsilon}}}^p_d:\dot{\hat{\utilde{\varepsilon}}}^p_d}$ & $\dot{\hat{p}} = \frac{1}{3\mu} |\dot{e}^{p}| {\bar{\sigma}_{\text{VM}}^{\#}}$ \\
\noalign{\smallskip}\hline\noalign{\smallskip}
\end{tabular}
\end{table}

\clearpage
\section{Numerical investigations and results}
\color{black}
The aim of this section is to evaluate the accuracy of the plastic corrector with respect to reference computations, which consist of full elasto-plastic FE simulations using the Z-Set suite \cite{Besson1998}. Two test cases, a notched structure and a specimen with spherical pores, will be presented with detailed full-field and point-wise comparisons of variables like the cumulative plastic strain. A third test case, that of a specimen with pores of realistic morphology, meshed based on images from X-ray tomography, is presented in Appendix \hyperref[appendixB]{B}.
The meshes for the test cases were created with GMSH \cite{gmsh2020}, and elasto-static simulations needed for the plastic corrector methodology were performed with FEniCS \cite{fenics2023}.

The parameters used for the reference elasto-plastic equations (equations \eqref{eq:plas:stressfull}-\eqref{eq:plas:yieldfull2}) and the plastic corrector equations (detailed in table \ref{tab:tensortoscalarplascorr}) chosen for this investigation, for test cases 1 and 2, are summarized in table \ref{tab:PlasticityModelParameters}.
\color{black}
\begin{table}[htbp]
\caption{\black{Parameters of the reference elasto-plastic equations detailed in equations \eqref{eq:plas:stressfull}-\eqref{eq:plas:yieldfull2} and the plastic corrector equations detailed in table \ref{tab:tensortoscalarplascorr}, used for test cases 1 and 2}}
\label{tab:PlasticityModelParameters}
\centering
\begin{tabular}[t]{llllllll}
\hline\noalign{\smallskip}
E & $\sigma_y$ & b &  Q  & C & D  \\
MPa & MPa      &   & MPa & MPa &    \\
\noalign{\smallskip}\hline\noalign{\smallskip}
 200000 & 100 & 10 & 100 & 40000 & 400\\
\noalign{\smallskip}\hline
\end{tabular}
\end{table}

\color{black}

\subsection{Test cases}

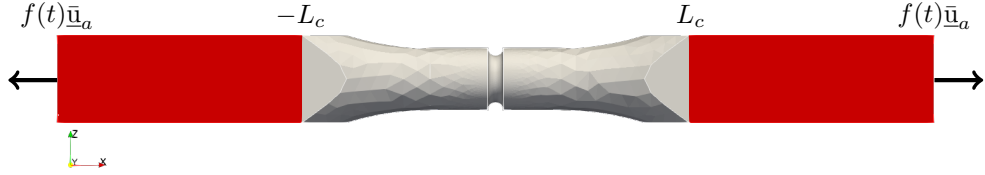
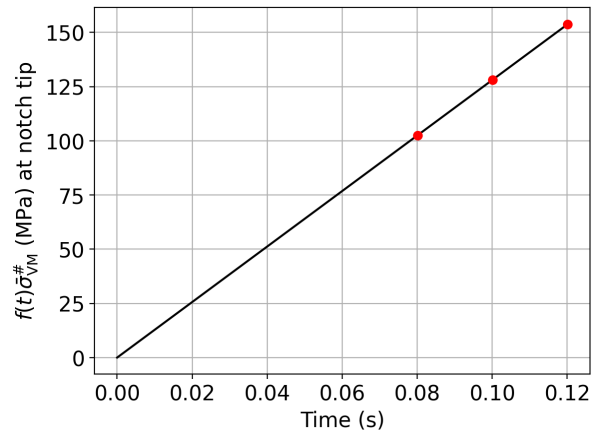
\begin{figure}[htbp]
    \centering
    \begin{subfigure}[b]{0.98\textwidth}
        \centering
        \begin{tikzpicture}
            \node[anchor=south west,inner sep=0] (image) at (0,0) {\includegraphics[width=\textwidth, trim=0 150 0 0, clip]{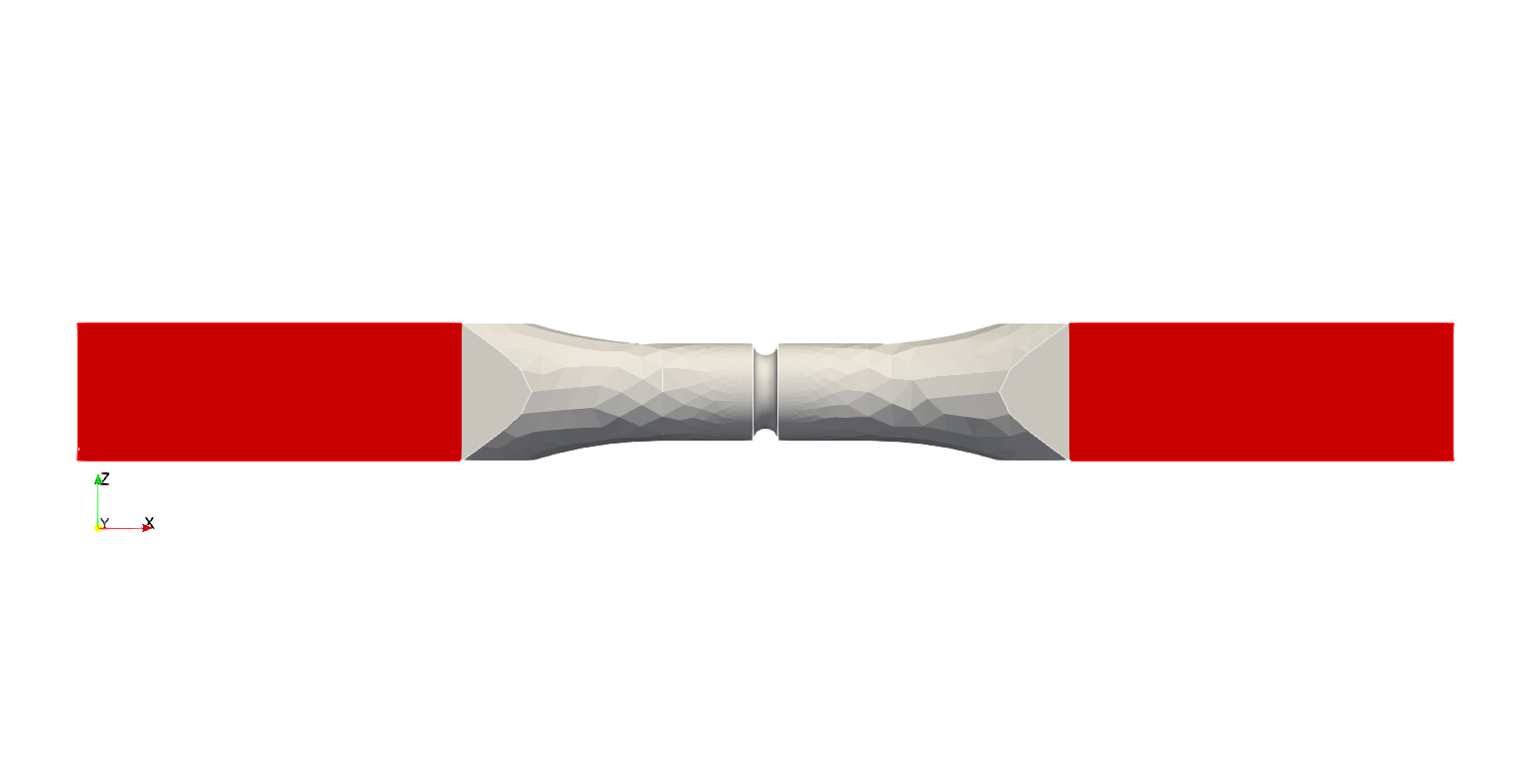}};
            \begin{scope}[x={(image.south east)},y={(image.north west)}]
                % Draw arrows
                \draw[black, ultra thick, <-] (0.0, 0.38) -- (0.05, 0.38); % Left arrow
                \node at (0.05, 0.54) {$f(t)\bar{\underbar{u}}_a$}; % Left arrow text
                \draw[black, ultra thick, ->] (0.95, 0.38) -- (1.0, 0.38); % Right arrow
                \node at (0.95, 0.54) {$f(t)\bar{\underbar{u}}_a$}; % Right arrow text
                \node at (0.30, 0.54) {$-L_c$}; % BC text
            \node at (0.70, 0.54) {$L_c$}; % BC text
            \end{scope}
        \end{tikzpicture}
        \caption{Notched structure, with displacement $\bar{u}_a$ at $f=1$ scaled by $f(t)$ on highlighted red regions, in opposite directions}
    \end{subfigure}    
    \hfill
    \begin{subfigure}[b]{0.6\textwidth}
        \centering
        \includegraphics[width=\textwidth]{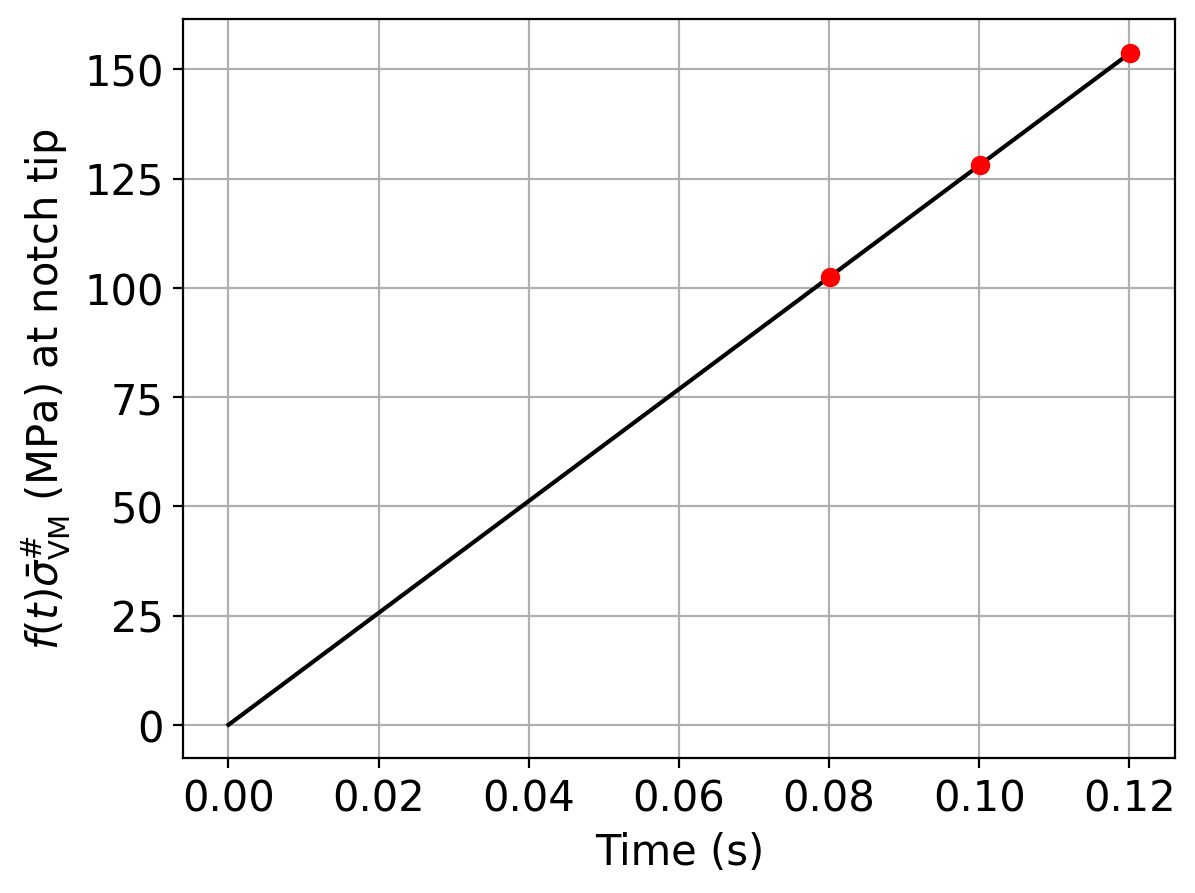}
        \caption{\black{At the notch tip, the von Mises stress from the elastic computation $\bar{\sigma}_{\text{VM}}^{\#}$ at $f=1$ is scaled by $f(t)$. $f(t)\bar{\sigma}_{\text{VM}}^{\#}$ goes up to 155\% of the yield stress at the peak of loading. The red points indicate the specific time-steps at which the cumulative plastic strain results from the full-field plastic corrector will be compared against the reference cumulative plastic strain.}}
    \end{subfigure}
    \caption{\textcolor{black}{Test case 1: Geometry, boundary conditions and applied loading}}
    \label{fig:notchMeshBCandLoad}
\end{figure}

\color{black}
\paragraph{Notched structure}\mbox{}\\
Test case 1 is a notched structure, illustrated in Figure \ref{fig:notchMeshBCandLoad}(a). For the plastic corrector, an elasto-static FEA computation is computed with prescribed displacements $\bar{\underbar{u}}_a = [u_x,0,0]$ for $x\geq L_c$ and $\bar{\underbar{u}}_a = [-u_x,0,0]$ for $x\leq -L_c$ such that $\bar{\sigma}_{\text{VM}}^{\#}$ at the notch area is at the yield stress of the material (this computation corresponds to $f=1$). The function $f(t)$ is then chosen to monotonically increase such that $f(t)\bar{\sigma}_{\text{VM}}^{\#}$ reaches 155\% of the yield stress of the material at the notch tip. This monotonic load function is shown in Figure \ref{fig:notchMeshBCandLoad}(b). The full-field $\bar{\sigma}_{\text{VM}}^{\#}$ in the structure is used for the plastic corrector approximations. The reference elasto-plastic computation is performed in such a way as to achieve the same loading level as in the plastic corrector computation.

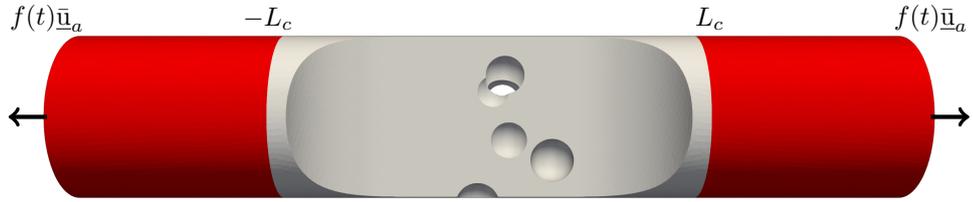
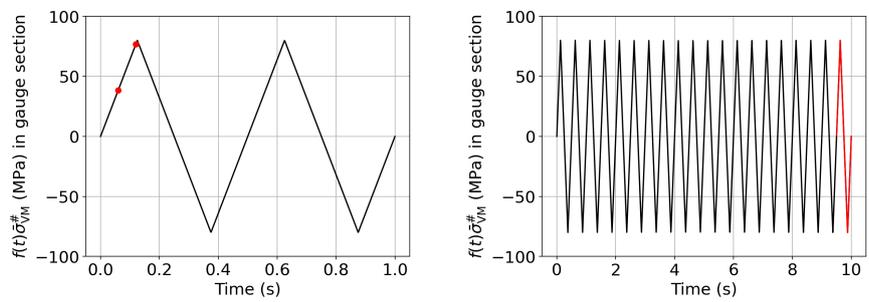
\begin{figure}
        \centering
        \begin{subfigure}[b]{0.98\textwidth}
            \begin{tikzpicture}
            \node[anchor=south west,inner sep=0] (image) at (0,0) {\adjincludegraphics[width=12.63cm,Clip={0.0\width} {0.3\height} {0.0\width} {0.3\height}]{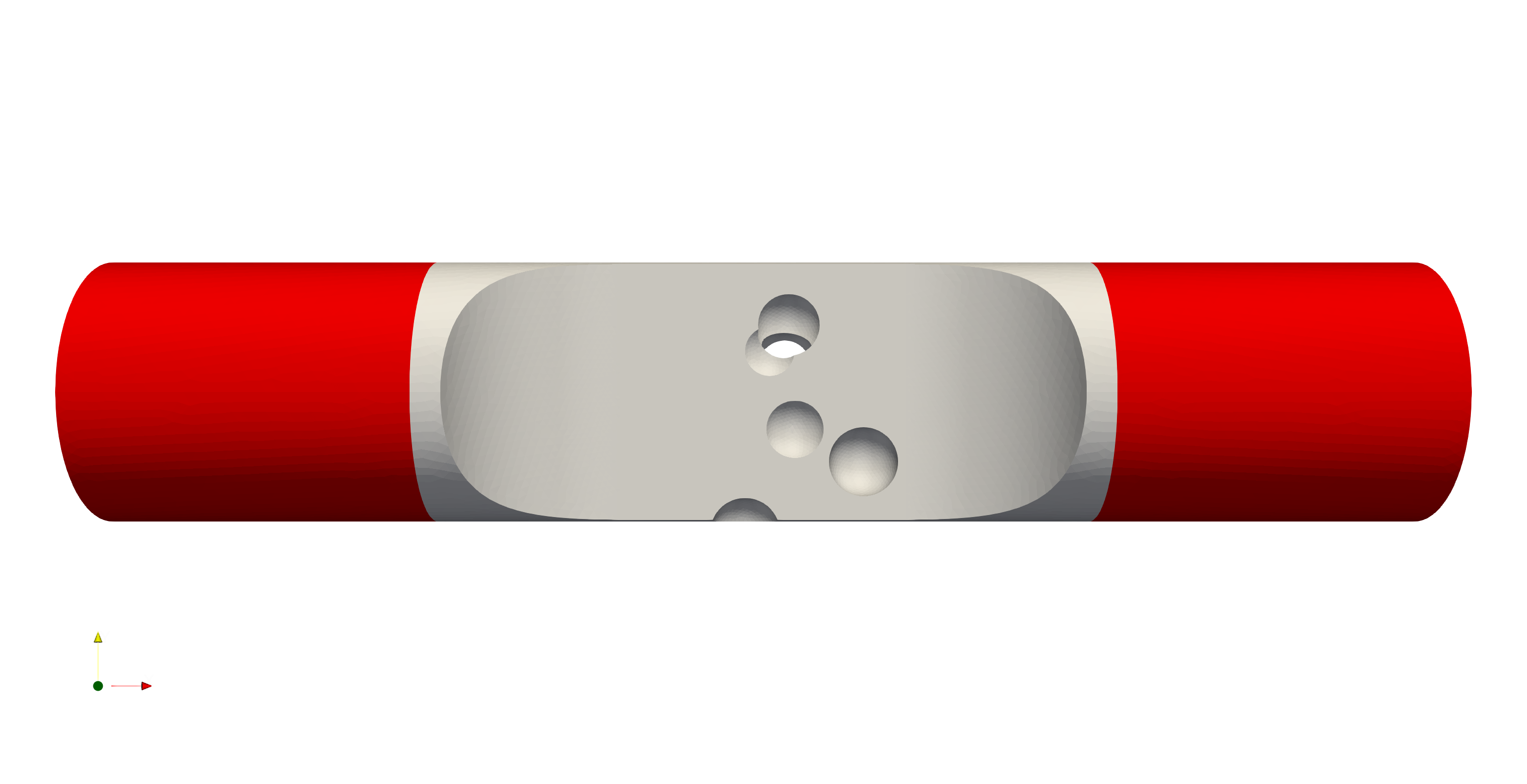}};
            \begin{scope}[x={(image.south east)},y={(image.north west)}]
                % Draw arrows
                \draw[black, ultra thick, <-] (0.0, 0.5) -- (0.04, 0.5); % Left arrow
                \node at (0.04, 1.) {$f(t)\bar{\underbar{u}}_a$}; % Left arrow text
                \draw[black, ultra thick, ->] (0.96, 0.5) -- (1.0, 0.5); % Right arrow
                \node at (0.96, 1.) {$f(t)\bar{\underbar{u}}_a$}; % Right arrow text
                \node at (0.27, 1.) {$-L_c$}; % BC text
                \node at (0.73, 1.) {$L_c$}; % BC text
            \end{scope}
        \end{tikzpicture}
            \caption{\black{Specimen with spherical pores in gauge section, with displacement $\bar{u}_a$ at $f=1$ scaled by $f(t)$ on highlighted red regions, in opposite directions}}
        \end{subfigure}    
        \centering
        \begin{subfigure}[t]{0.45\textwidth}
            \adjincludegraphics[width=5.41cm,Clip={0.0\width} {0.0\height} {0.0\width} {0.0\height}]{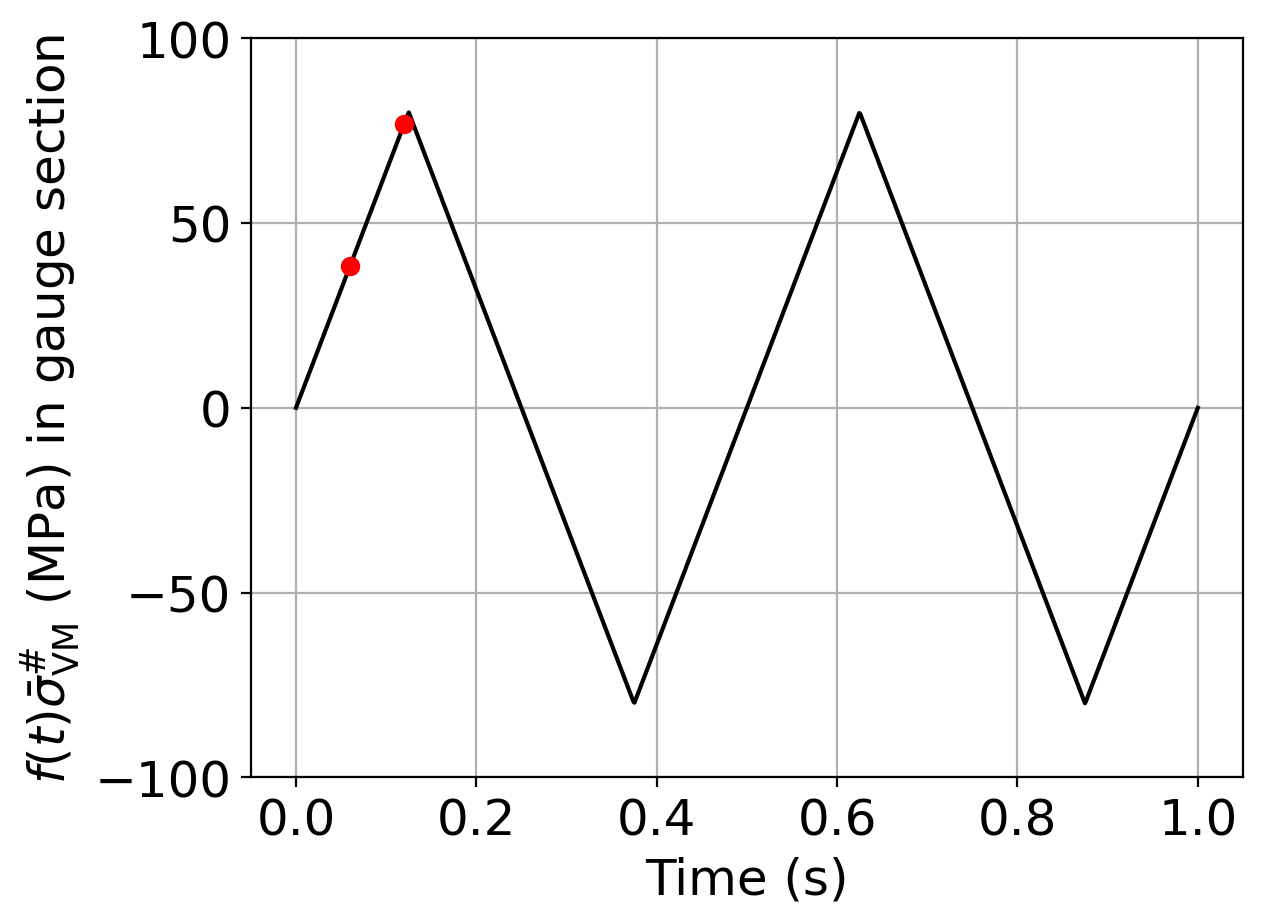}
            \caption{\black{Load function 1: In the gauge section away from pores, the von Mises stress from the elastic computation $\bar{\sigma}_{\text{VM}}^{\#}$ at $f=1$ is scaled by $f(t)$. The plastic corrector and reference cumulative plastic strain will be compared at the specific time-steps indicated in red}}
        \end{subfigure}
        \begin{subfigure}[t]{0.45\textwidth}
            \adjincludegraphics[width=5.41cm,Clip={0.0\width} {0.0\height} {0.0\width} {0.0\height}]{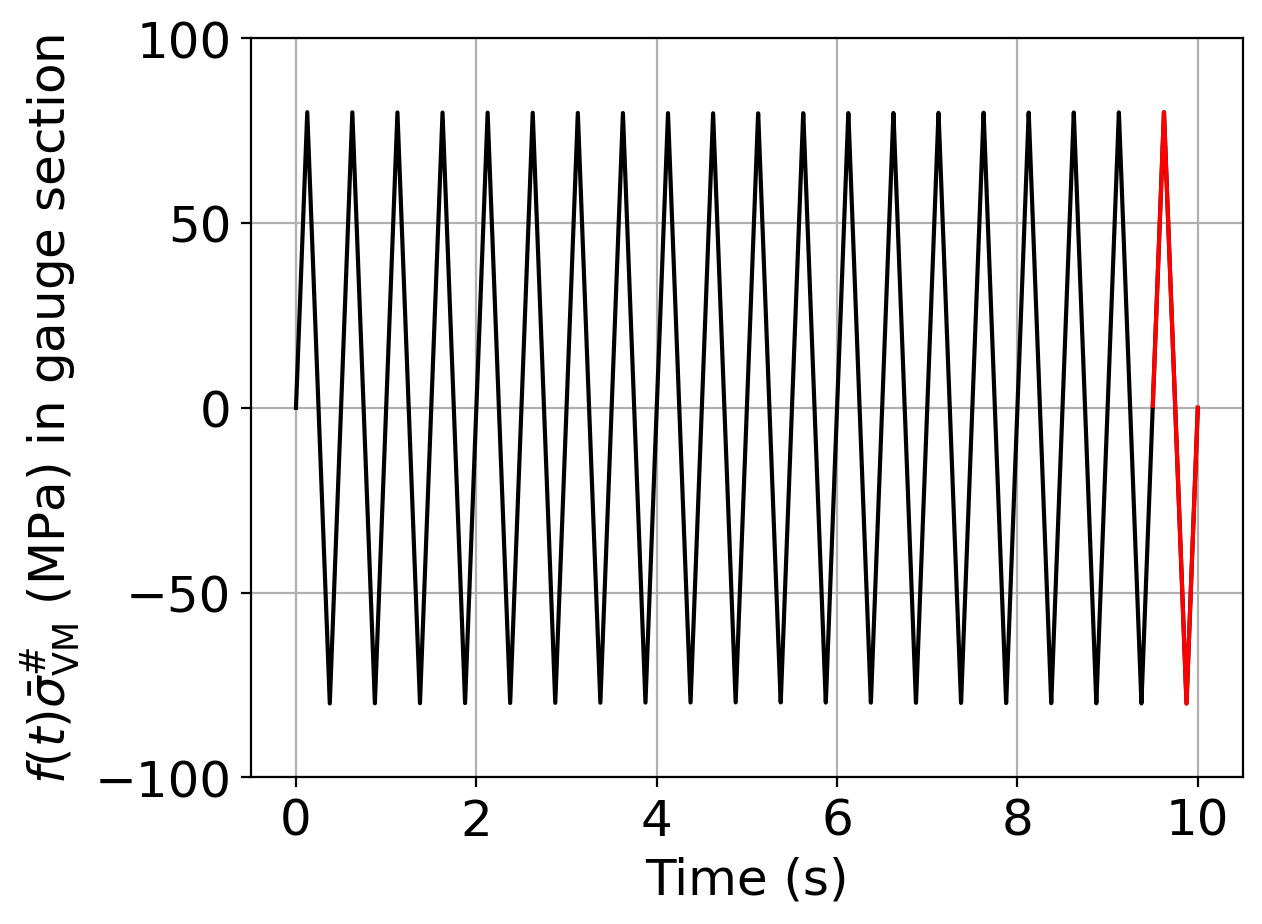}
            \caption{\black{Load function 2: In the gauge section away from pores, the von Mises stress from the elastic computation $\bar{\sigma}_{\text{VM}}^{\#}$ at $f=1$ is scaled by $f(t)$. The plastic corrector and the reference cumulative plastic strain range over the 20\textrm{th} cycle (red), will be compared}}
        \end{subfigure}
        \caption{\textcolor{black}{Test case 2: Geometry, boundary conditions and applied loading}}
        \label{Fig:SpheresPorousMeshBCandLoad}
\end{figure}

\paragraph{Specimen with spherical pores}\mbox{}\\
\color{black}
Test case number 2 is a specimen with a spatial distribution of spherical pores in the gauge section shown in Figure \ref{Fig:SpheresPorousMeshBCandLoad}. The spherical pores are of a fixed size, and their centers were generated using random sampling. For the plastic corrector, an elasto-static FEA computation is computed with prescribed displacements $\bar{\underbar{u}}_a = [u_x,0,0]$ for $x\geq L_c$ and $\bar{\underbar{u}}_a = [-u_x,0,0]$ for $x\leq -L_c$ such that $\bar{\sigma}_{\text{VM}}^{\#}$ in the gauge section, away from pores, is at the yield stress of the material (this computation corresponds to $f=1$). The function $f(t)$ is evaluated twice, generating loading sequences of 2 cycles and 20 cycles. At the peak of these cyclic loads, $f(t) \bar{\sigma}_{\text{VM}}^{\#}$ is chosen to reach 80\% of the yield stress of the material in the gauge section away from pores. These cyclic load functions are shown in Figure \ref{Fig:SpheresPorousMeshBCandLoad}(b-c). Due to the presence of pores in the specimen, the local loading exceeds the yield stress in several regions. The full-field $\bar{\sigma}_{\text{VM}}^{\#}$ in the specimen is used for the plastic corrector approximations. For the reference elasto-plastic computation, the same type of loading as in the plastic corrector is applied.

\color{black}

\clearpage
\subsection{Accuracy of the plastic corrector predictions}

\begin{figure}[htbp]
    \centering
    \begin{subfigure}[b]{0.32\textwidth}
        \centering
        \includegraphics[width=\textwidth, trim=350 150 350 50, clip]{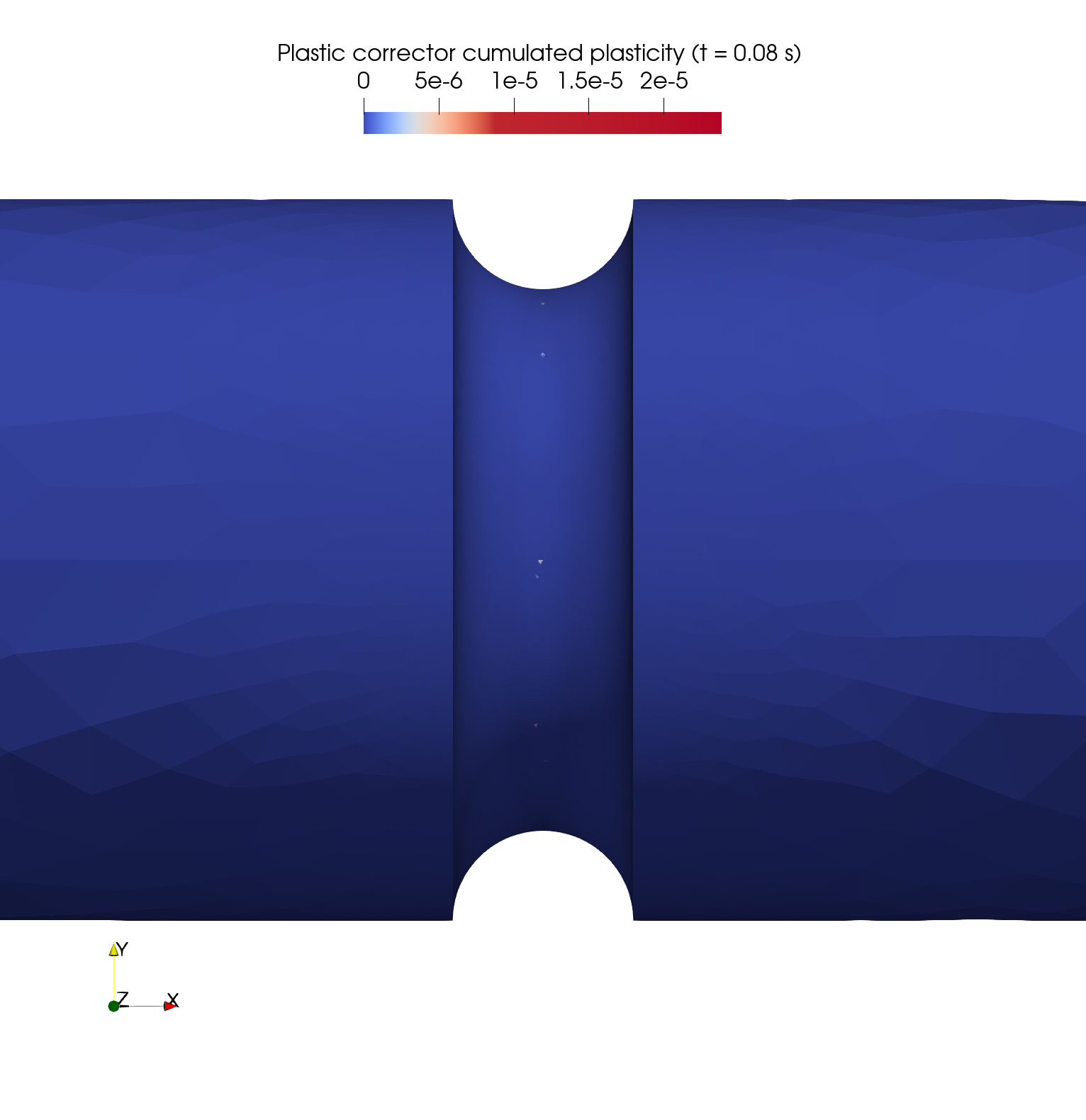}
        \caption{Plastic corrector, t=0.08 s, notch load 102\%$\sigma_y$}
    \end{subfigure}
    \hfill
    \begin{subfigure}[b]{0.32\textwidth}
        \centering
        \includegraphics[width=\textwidth, trim=350 150 350 50, clip]{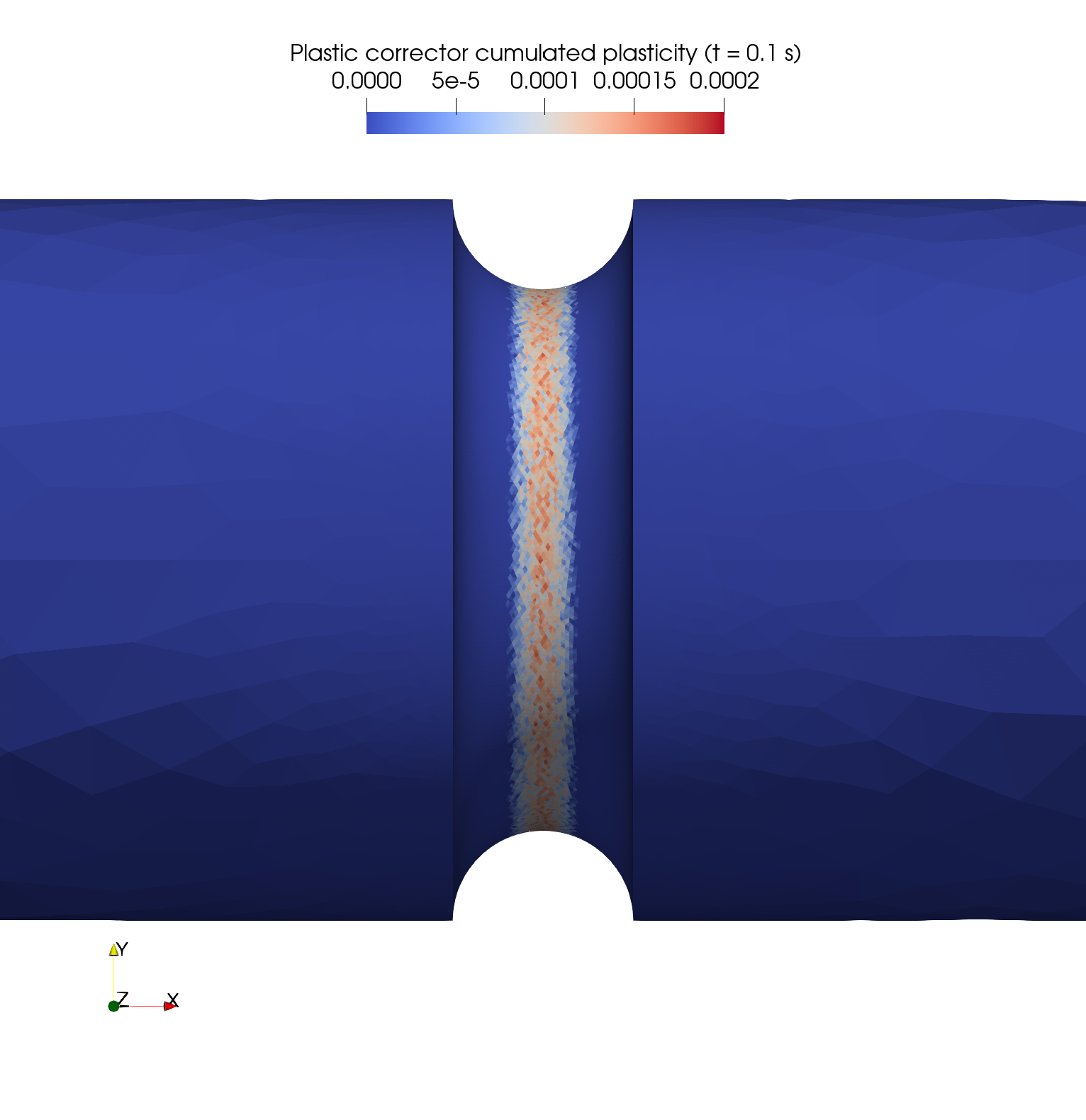}
        \caption{Plastic corrector, t=0.10 s, notch load 127\%$\sigma_y$}
    \end{subfigure}
    \hfill
    \begin{subfigure}[b]{0.32\textwidth}
        \centering
        \includegraphics[width=\textwidth, trim=350 150 350 50, clip]{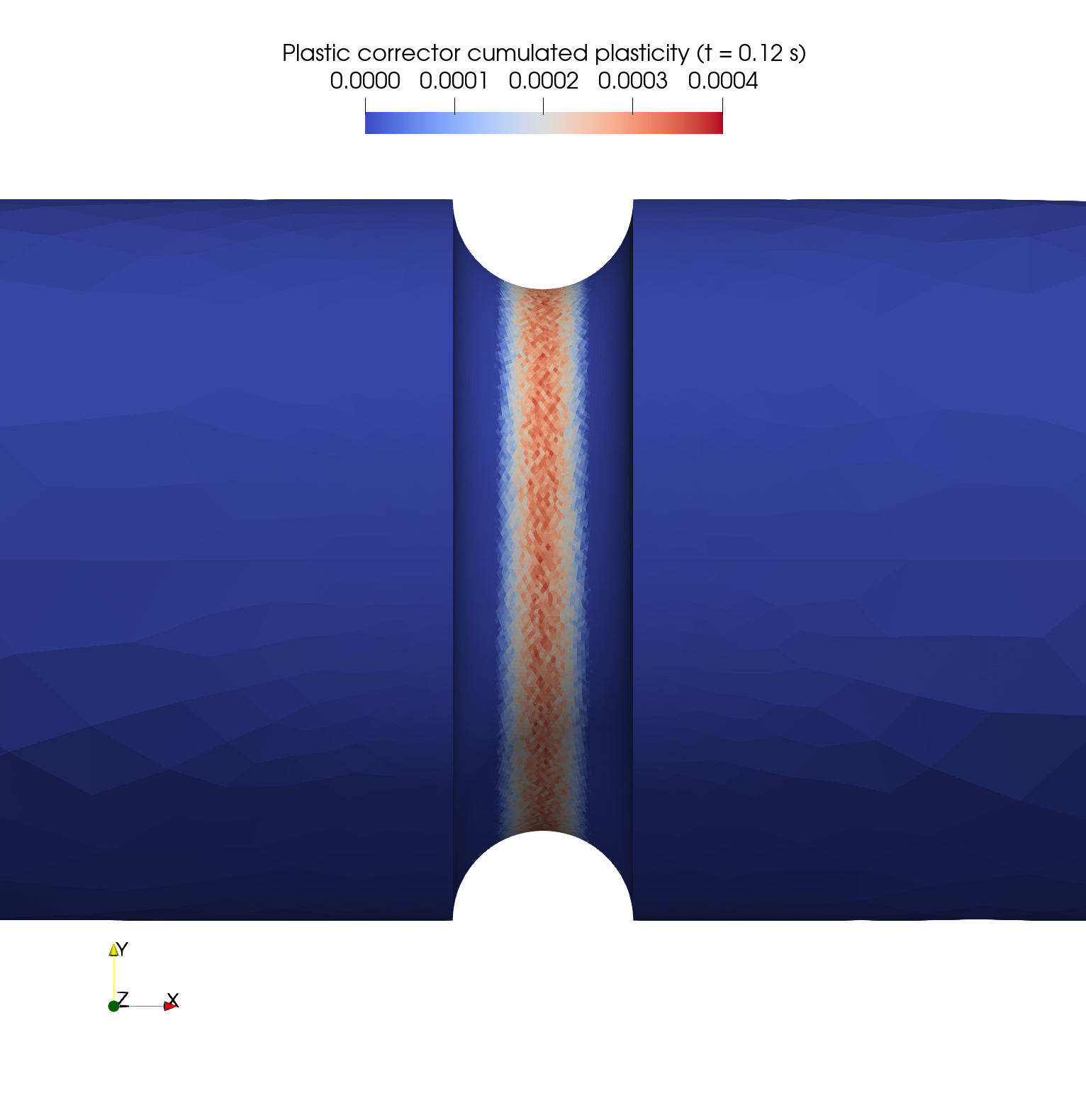}
        \caption{Plastic corrector, t=0.12 s, notch load 152\%$\sigma_y$}
    \end{subfigure}
    
    \vspace{0.5cm} % Add some vertical space between the rows

    \begin{subfigure}[b]{0.32\textwidth}
        \centering
        \includegraphics[width=\textwidth, trim=350 150 350 50, clip]{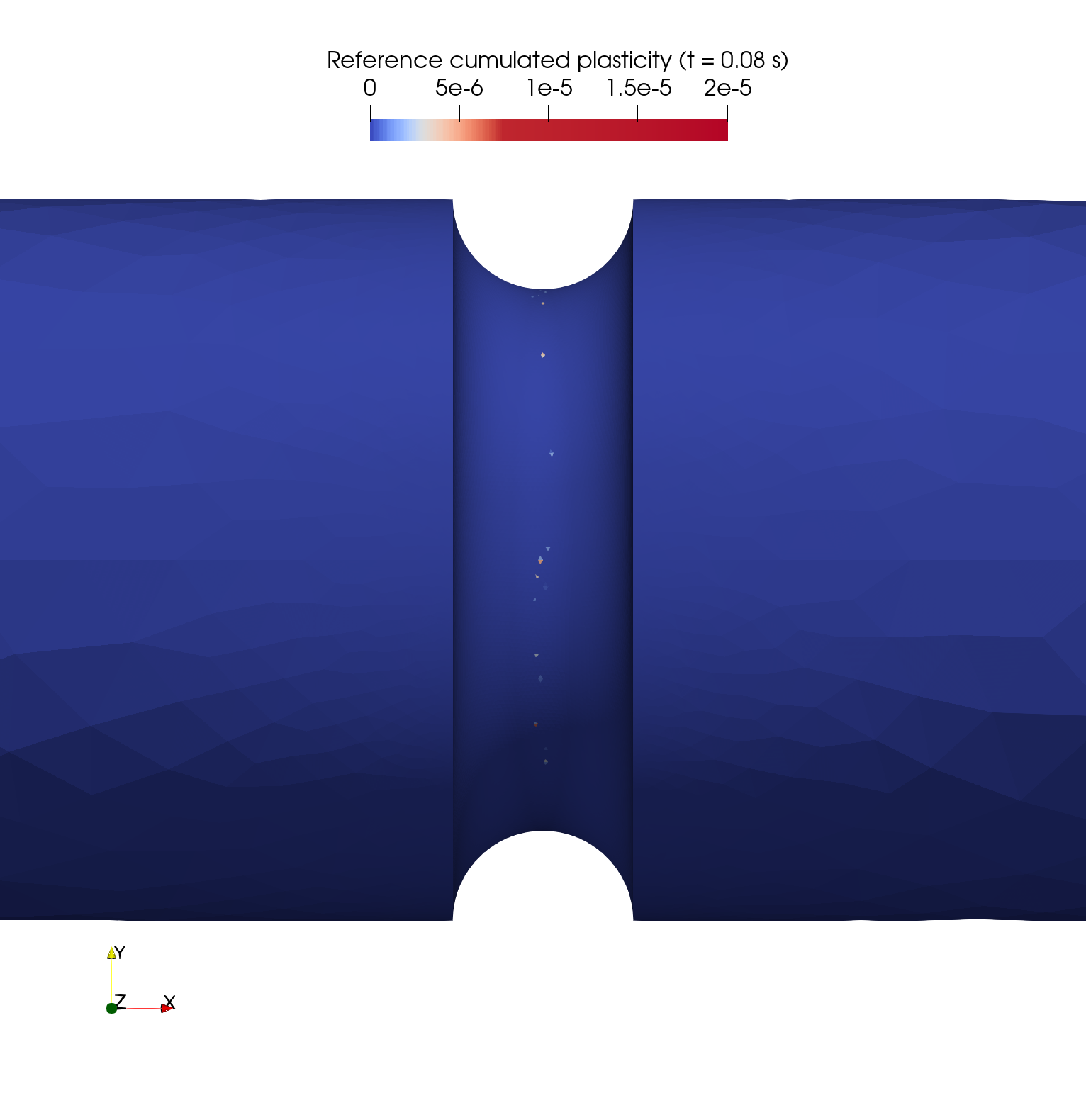}
        \caption{Reference, t=0.08 s, notch load 102\%$\sigma_y$}
    \end{subfigure}
    \hfill
    \begin{subfigure}[b]{0.32\textwidth}
        \centering
        \includegraphics[width=\textwidth, trim=350 150 350 50, clip]{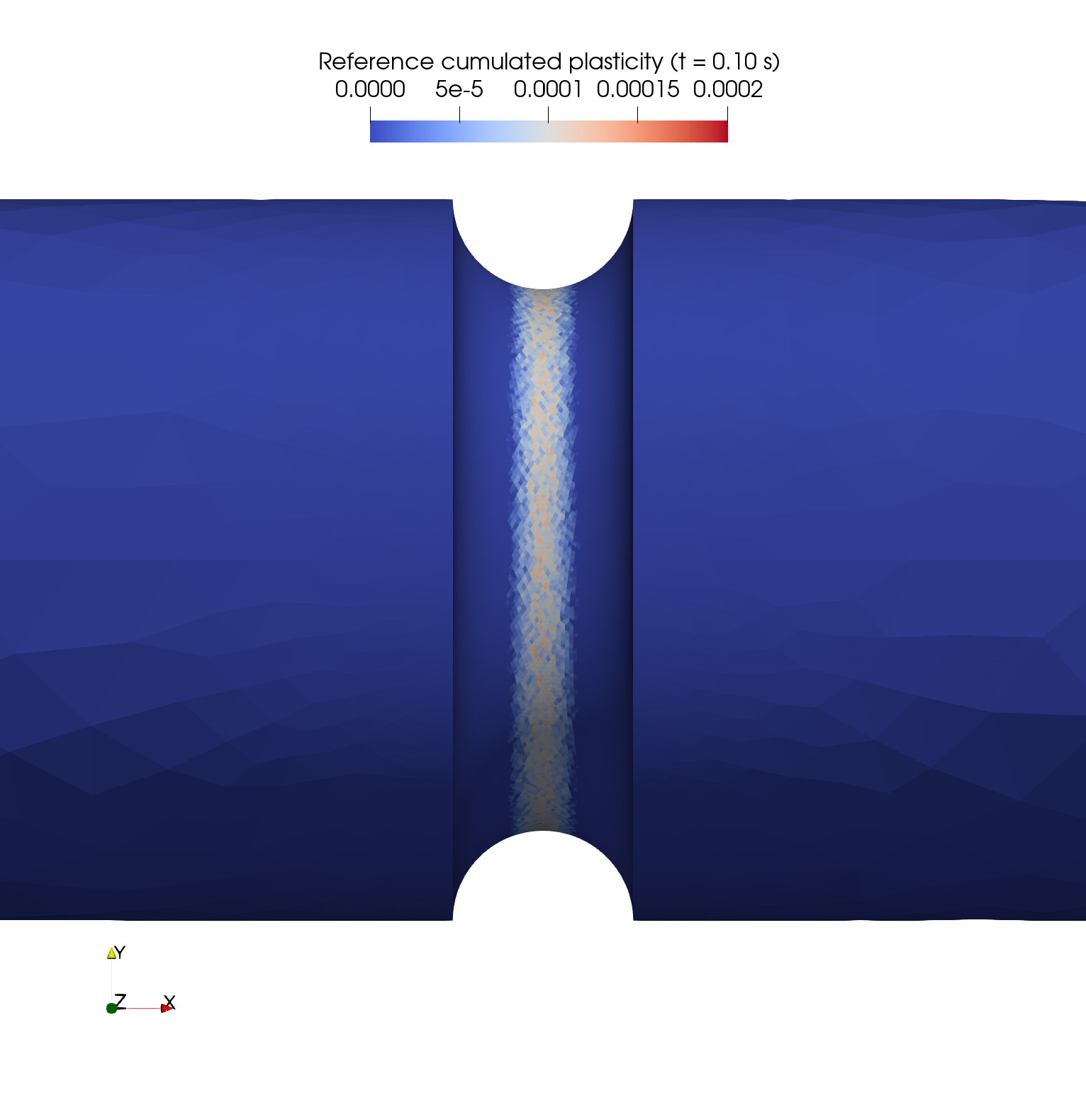}
        \caption{Reference, t=0.10 s, notch load 127\%$\sigma_y$}
    \end{subfigure}
    \hfill
    \begin{subfigure}[b]{0.32\textwidth}
        \centering
        \includegraphics[width=\textwidth, trim=350 150 350 50, clip]{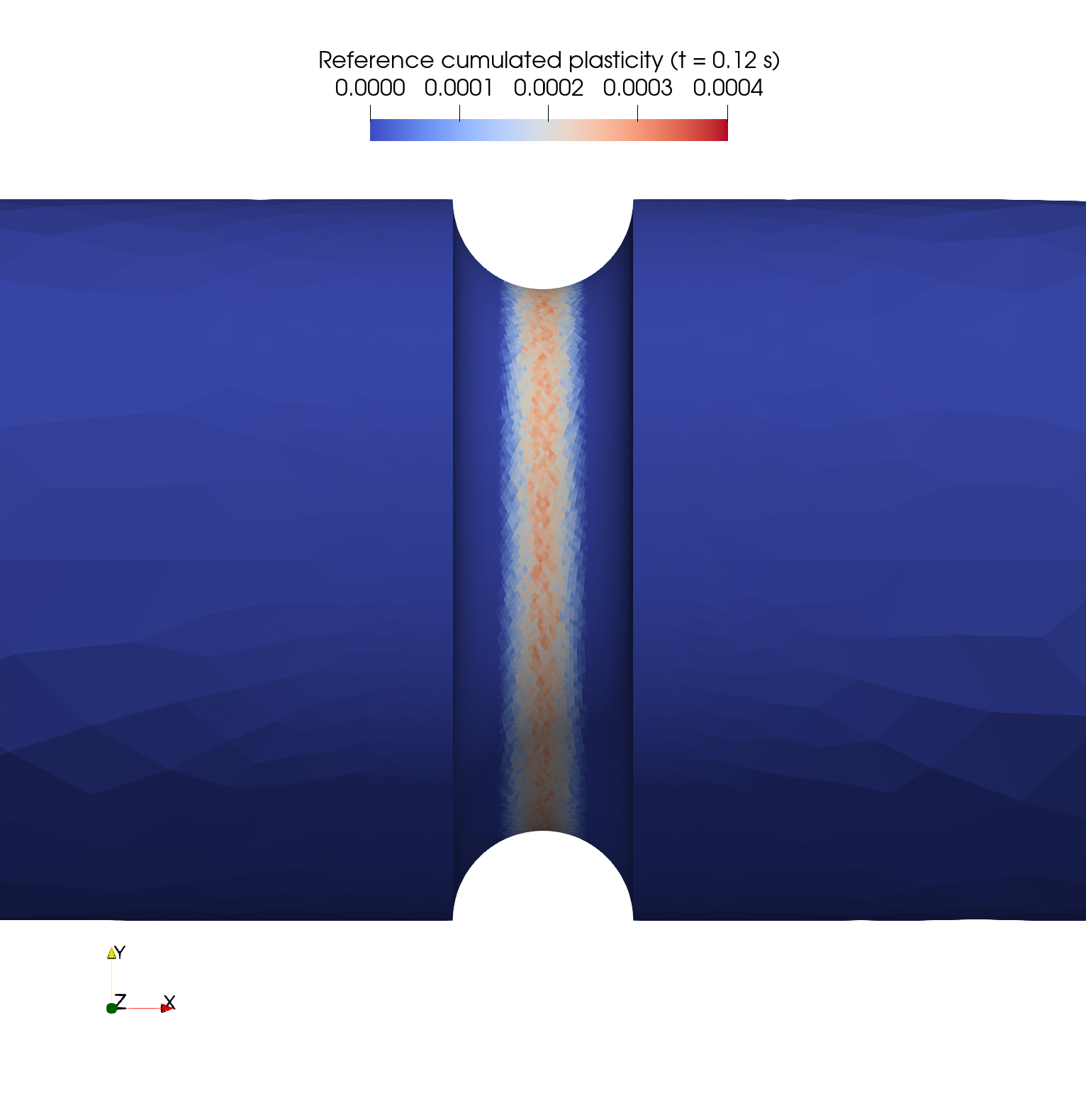}
        \caption{Reference, t=0.12 s, notch load 152\%$\sigma_y$}
    \end{subfigure}
    
    \caption{\textcolor{black}{Test case 1 (notched geometry): Time-evolution of cumulative plastic strain.}}
    \label{fig:notch_p_cumul_time steps1}
\end{figure}

\begin{figure}[htbp]
    \centering
        \begin{subfigure}[b]{0.49\textwidth}
            \includegraphics[width=\textwidth]{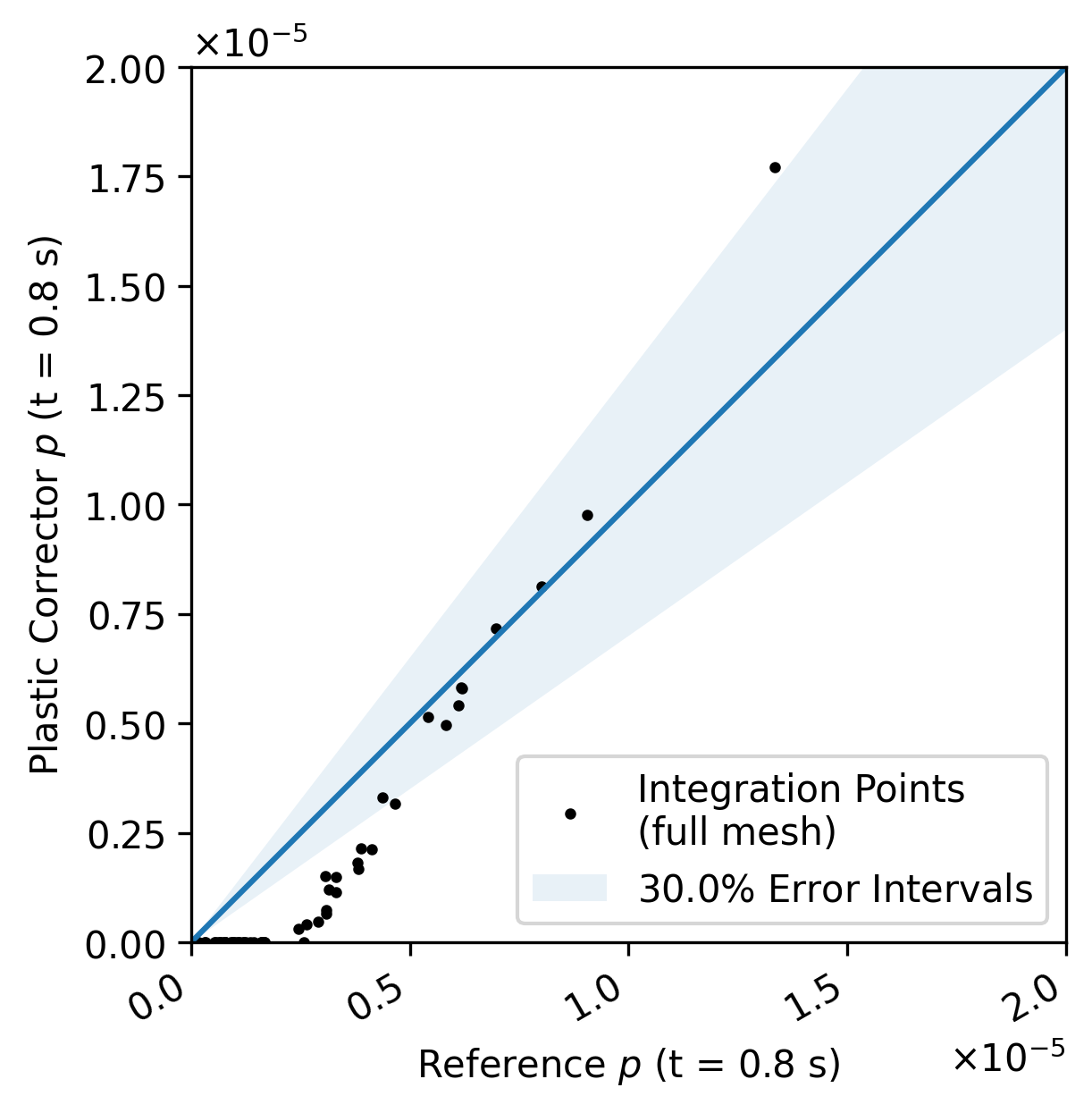}
            \caption{Notch load 102\%$\sigma_y$ (t=0.08 s)}
        \end{subfigure}           
        \begin{subfigure}[b]{0.49\textwidth}
            \includegraphics[width=\textwidth]{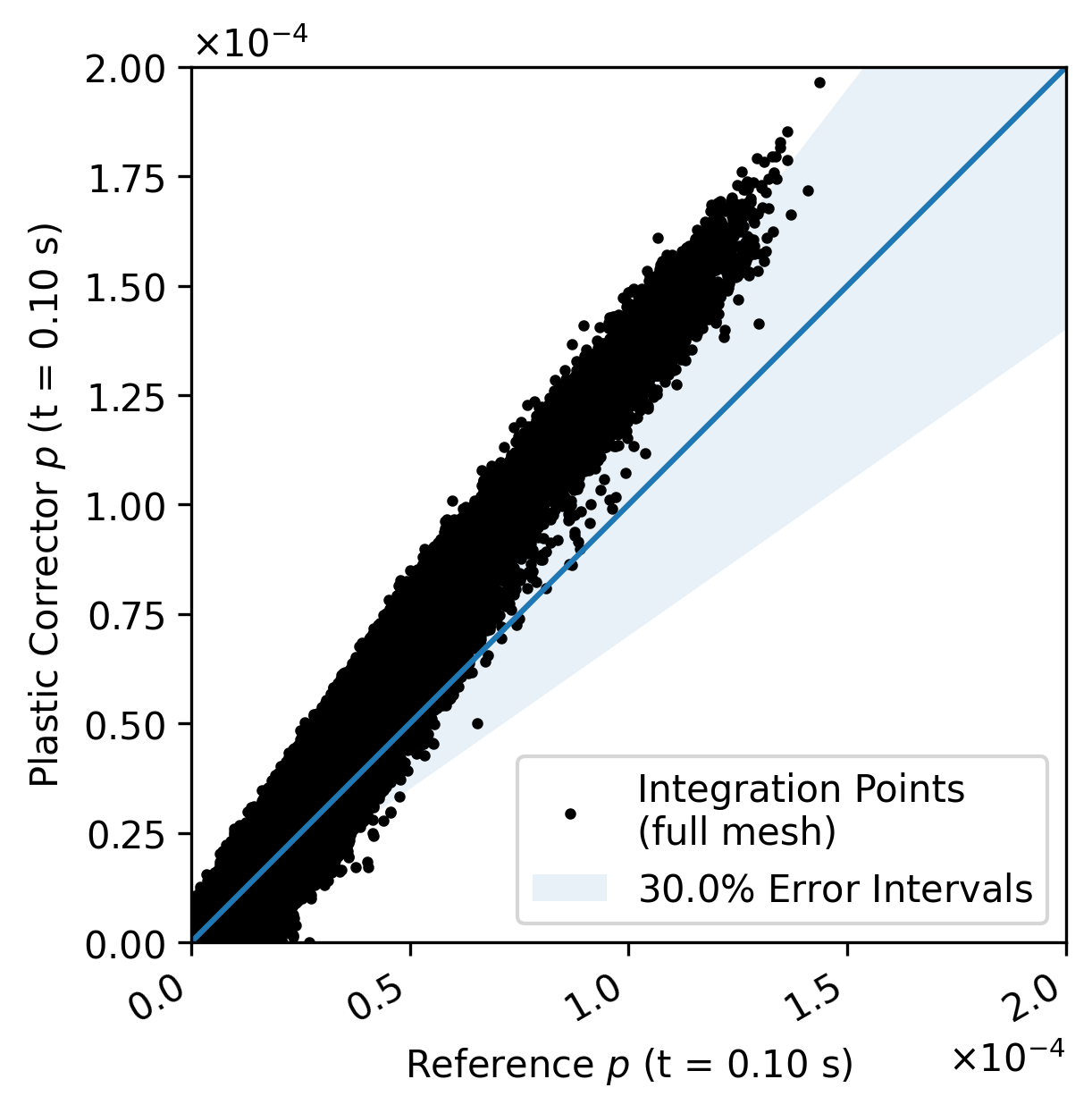}
            \caption{Notch load 127\%$\sigma_y$ (t=0.10 s)}
        \end{subfigure}
        \begin{subfigure}[b]{0.49\textwidth}
            \includegraphics[width=\textwidth]{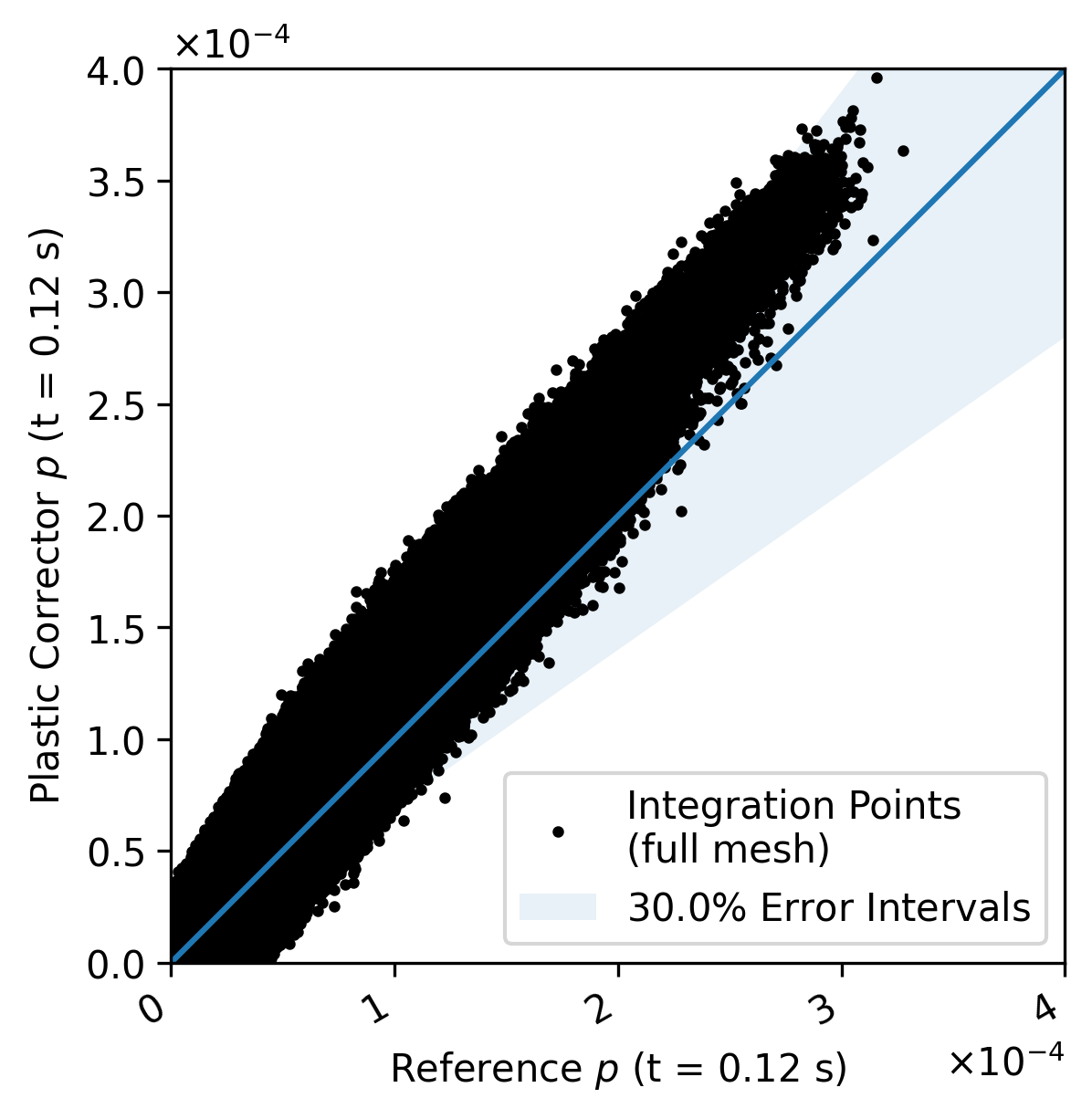}
            \caption{Notch load 152\%$\sigma_y$ (t=0.12 s)}
        \end{subfigure}
        \caption{\textcolor{black}{Test case 1 (notched geometry): Scatter plots of cumulative plastic strain in all integration points, at three time-steps during monotonic loading}}
        \label{fig:notch_p_scatter_comparison_timeframes}
\end{figure}

\black{
The accuracy of some quantities of interest computed using the plastic corrector will now be examined for the previously presented test cases. For test case 1, the evolution of the cumulative plastic strain, at a point of the notch tip and for the full 3D geometry will be shown, for monotonic loading. For test case 2, the evolution of the cumulative plastic strain will be shown in the 3D geometry encompassing the pores, for both monotonic and cyclic loading. The accuracy of the deviatoric stresses and strains will also be evaluated for a few points around the pores. Finally, the accuracy of the cumulative plastic strain range (denoted by $\Delta p = p_{\text{max}}^{\text{cycle}} - p_{\text{min}}^{\text{cycle}}$) over the $20^\textrm{th}$ cycle will be evaluated. The $20^\textrm{th}$ cycle is chosen here because in subsequent cycles, the stress strain responses do not change significantly from one cycle to the next, i.e. the cycling response is stabilised. Quantities of interest extracted from such stabilised cycles are notably suitable for the computation of fatigue criteria \cite{schijve2009}.
}
\paragraph*{Time-evolution of cumulative plastic strain}\mbox{}\\
\color{black}
The field of cumulative plastic strain approximated by the plastic corrector, at three time steps of the loading sequence, is shown for test case 1 in Figure \ref{fig:notch_p_cumul_time steps1} and compared with reference computations at the same levels of applied loading. The time steps correspond to the notch von Mises stress (of the elastic FEA computation) reaching 102\%, 127\% and 152\% of the elastic yield limit. The plastic corrector predicts overall higher cumulative plastic strain in the notch region, as compared to the reference elasto-plastic solution. Scatter plots of the cumulative plastic strain in all integration points in the mesh are shown in Figure \ref{fig:notch_p_scatter_comparison_timeframes}.  The approximated solution of the plastic corrector algorithm aligns closely with the reference solution. \black{However, there exists some scatter, which is expected due to the cumulative errors linked to the rules of the plastic correction algorithm. The reference computation has points that undergo plastic accommodation, which is not captured in the Neuber-type computation, therefore leading to over-estimation by the plastic correction algorithm. This is a known phenomenon and has been reported before \cite{Chouman2014}.
Secondly, the Neuber approach also under-estimates the cumulative plastic strain at some points as it is localised and does not take into account redistribution of stresses that occur in the reference computation \cite{Molski1981}.}

\begin{figure}[htbp]
    \centering
        \begin{subfigure}[b]{0.48\textwidth}
            \includegraphics[width=\textwidth, trim=0 0 0 0, clip]{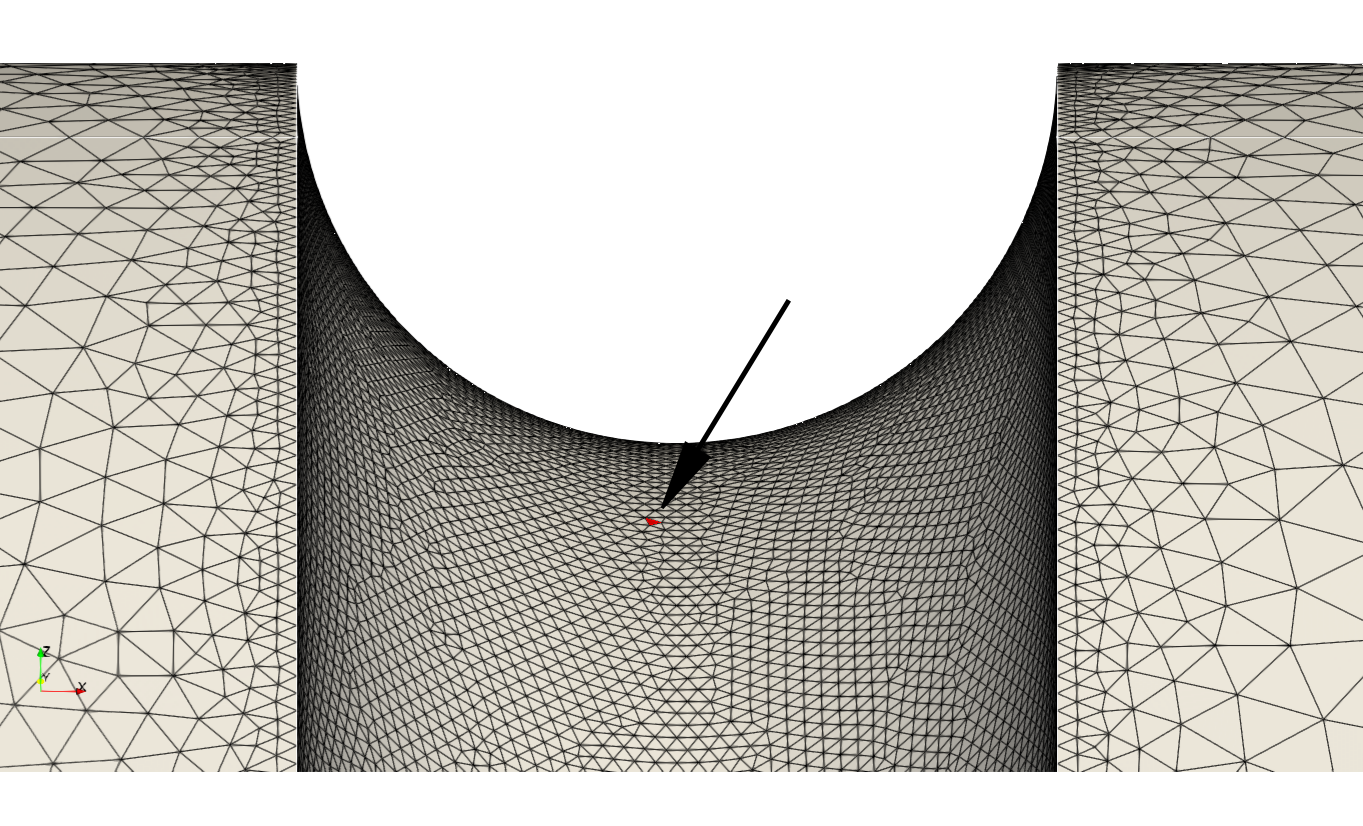}
            \caption{Selected point in the mesh, shown by the arrow}
        \end{subfigure}           
        \begin{subfigure}[b]{0.48\textwidth}
            \includegraphics[width=\textwidth]{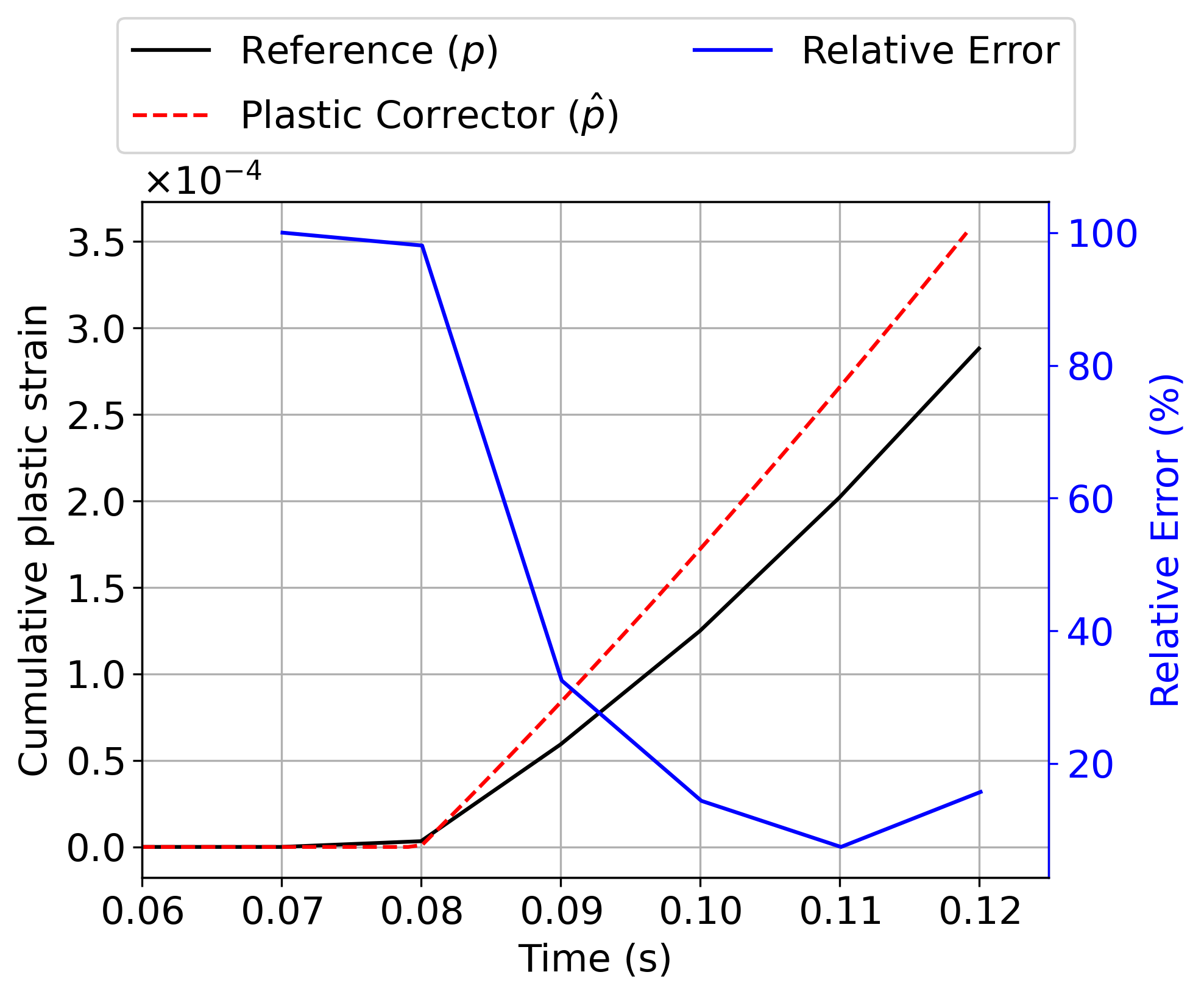}
            \caption{Time-evolution of the cumulative plastic strain at the selected point}
        \end{subfigure}
        \caption{\textcolor{black}{Test case 1 (notched geometry): Comparison of the time-evolution of the cumulative plastic strain at the notch tip obtained by the plastic corrector and the reference computation during monotonic loading.}}
        \label{fig:notch_comparison_p_zset_id_323901}
\end{figure}

The time-evolution of a maximally loaded point on the notch region is shown in Figure \ref{fig:notch_comparison_p_zset_id_323901}. \black{The plastic corrector over-estimates the cumulative plastic strain throughout the loading sequence because stress redistribution was not considered. The relative error is very high at first, due to very low values of plasticity. During loading at the notch between $127\%-152\%\sigma_y$, the relative error lowers to around 15-20\%. These over-estimations match with the literature \cite{Molski1981,jones1998}.}

\begin{figure*}[!htbp]
    \centering
        \begin{subfigure}[b]{0.49\textwidth}
            \adjincludegraphics[width=14.5cm,Clip={.3\width} {.2\height} {0.3\width} {.10\height}]{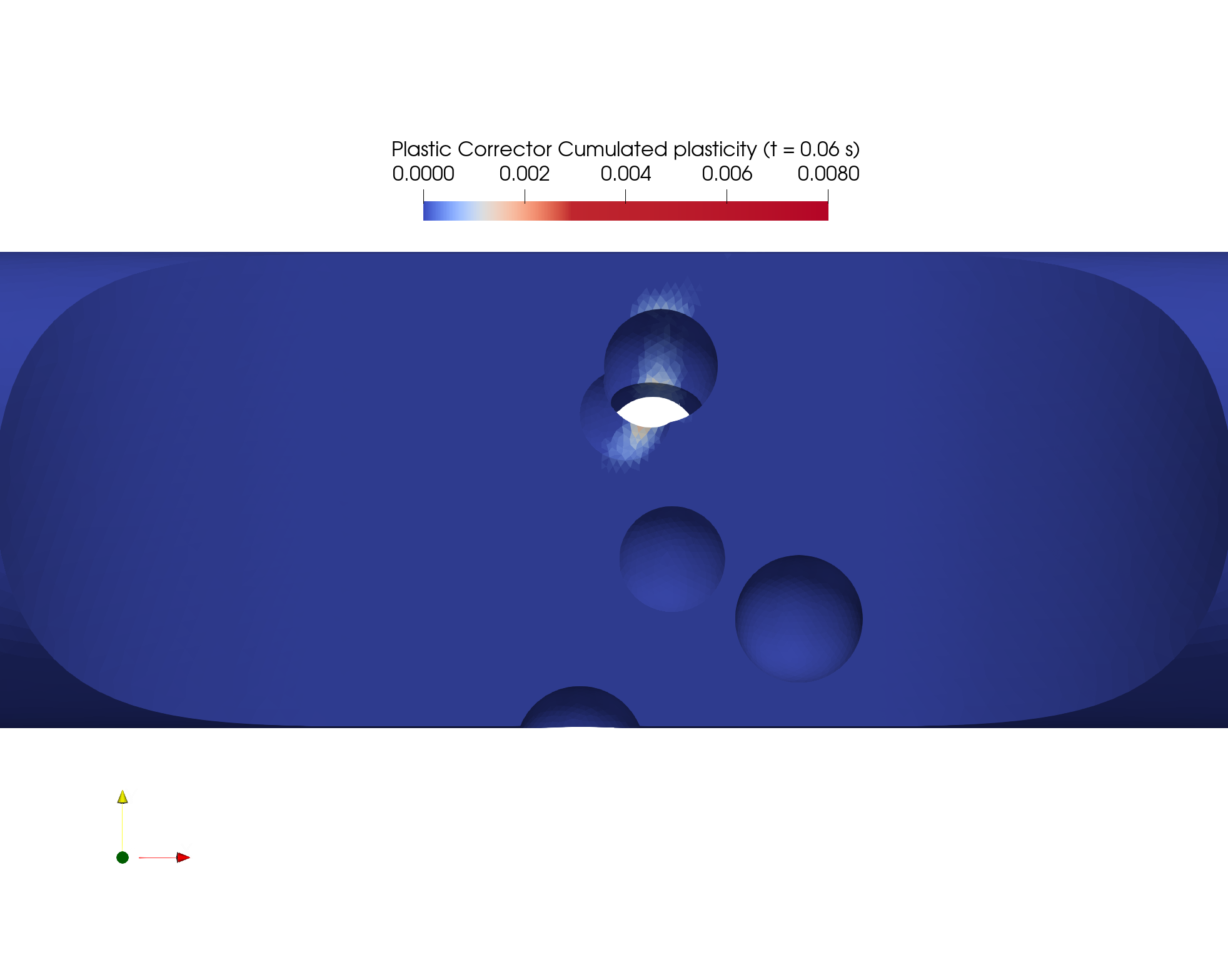}
            \caption{Plastic corrector at t=0.06 s (applied loading 0.4$\sigma_y$)}
        \end{subfigure}           
        \begin{subfigure}[b]{0.49\textwidth}
            \adjincludegraphics[width=14.5cm,Clip={.3\width} {.2\height} {0.3\width} {.10\height}]{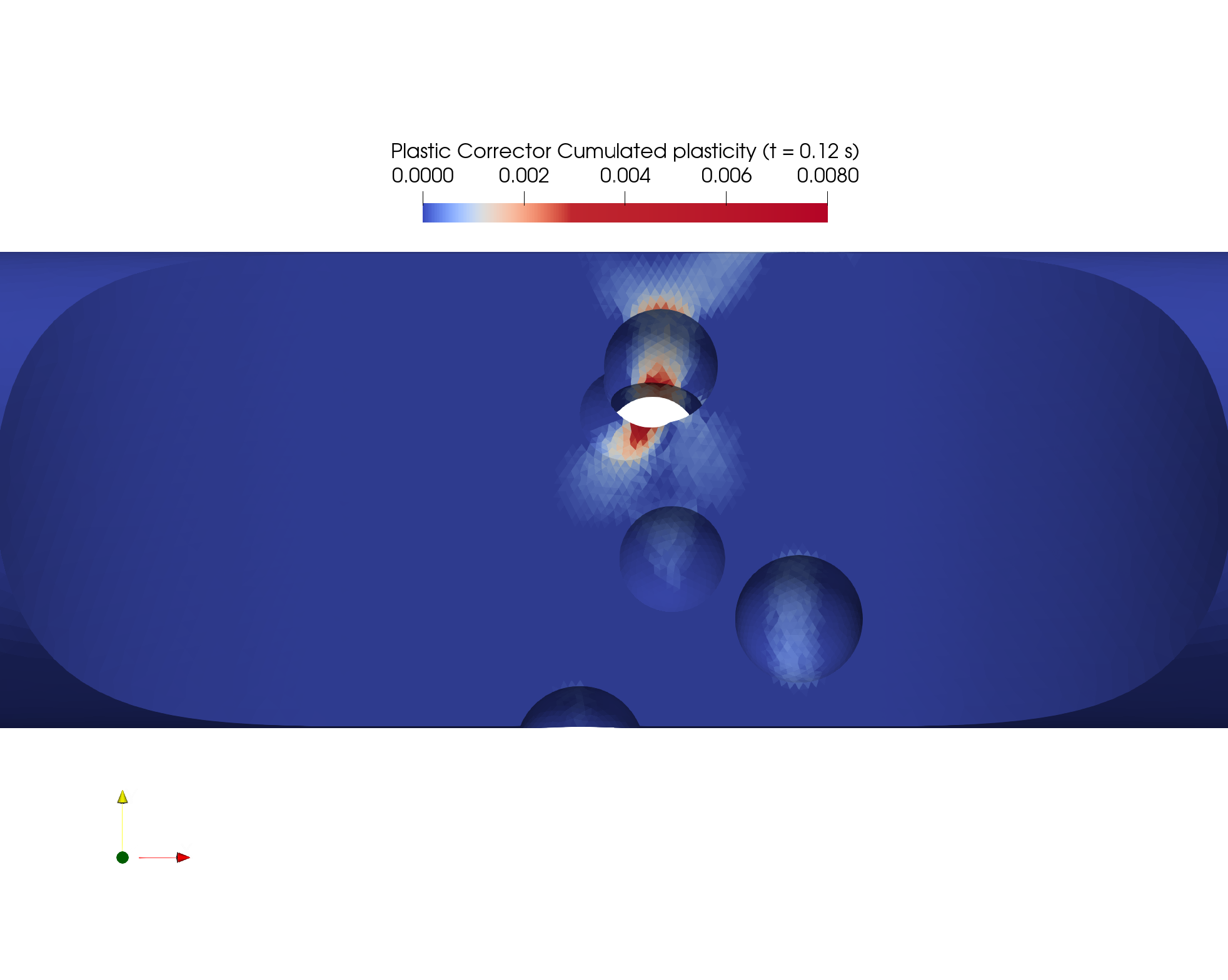}
            \caption{Plastic corrector at t=0.12 s (applied loading 0.8$\sigma_y$)}
        \end{subfigure}
        \begin{subfigure}[b]{0.49\textwidth}
            \adjincludegraphics[width=14.5cm,Clip={.3\width} {.2\height} {0.3\width} {.10\height}]{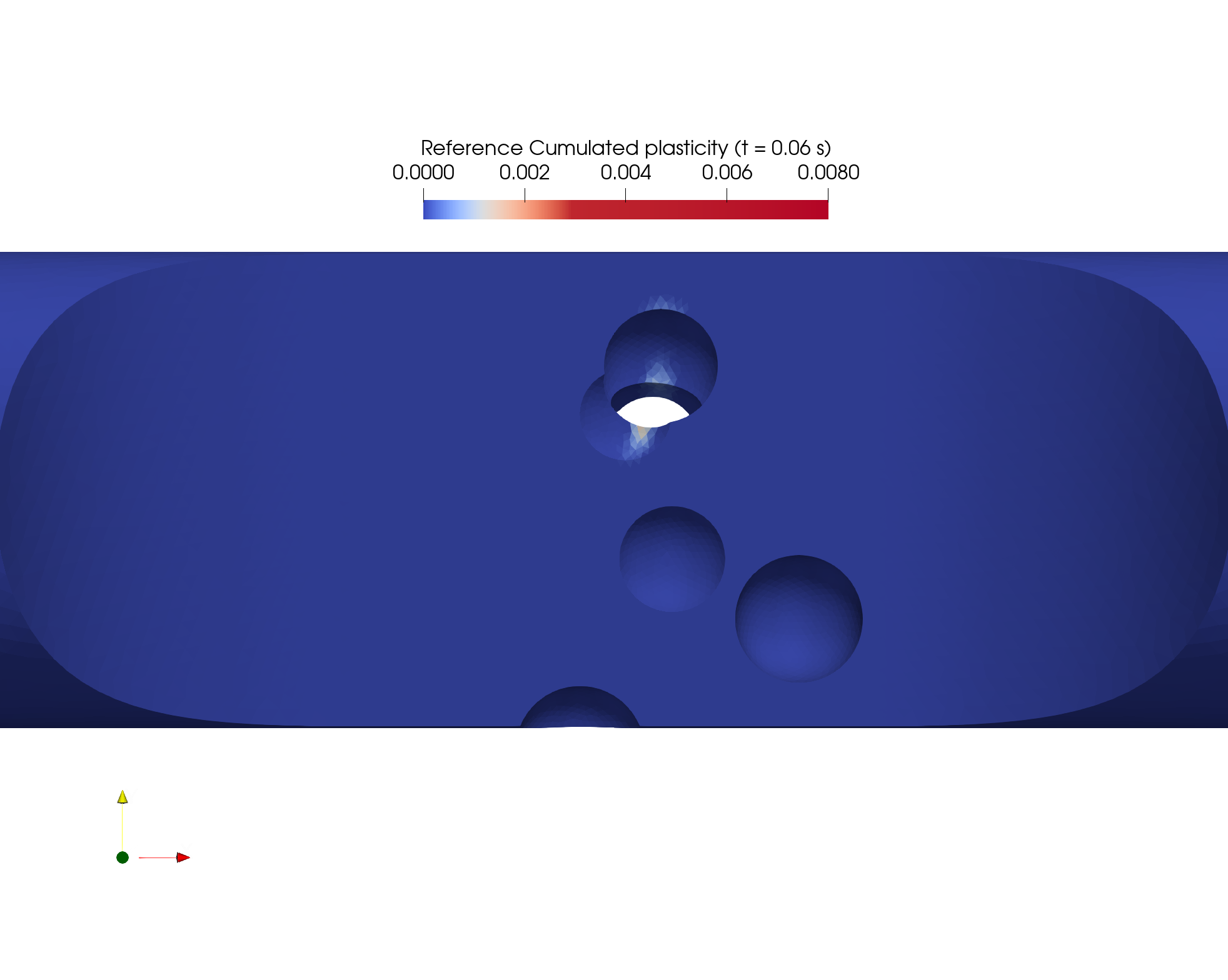}
            \caption{Reference computation at t=0.06 s (applied loading 0.4$\sigma_y$)}
        \end{subfigure}           
        \begin{subfigure}[b]{0.49\textwidth}
            \adjincludegraphics[width=14.5cm,Clip={.3\width} {.2\height} {0.3\width} {.10\height}]{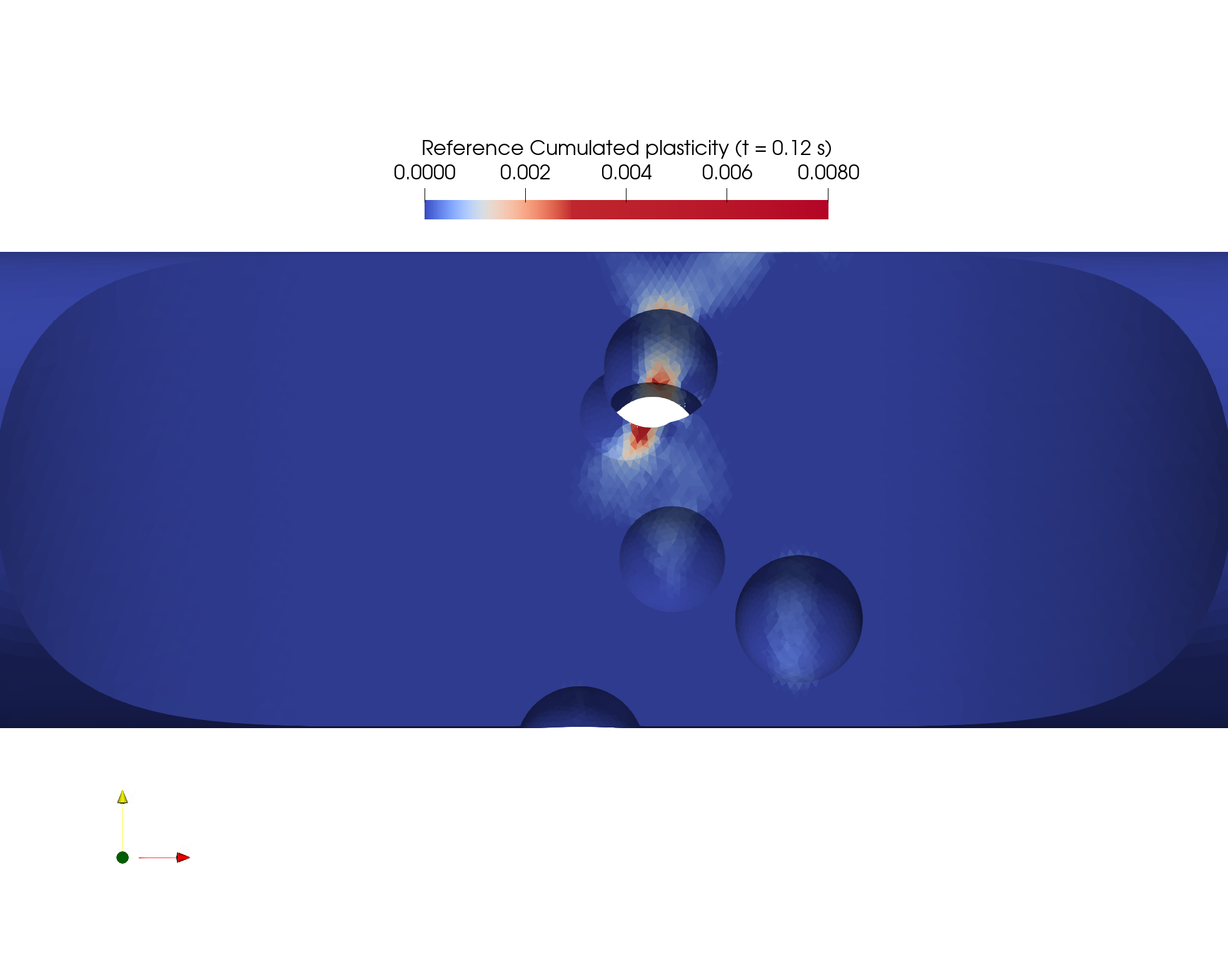}
            \caption{Reference computation at t=0.12 s (applied loading 0.8$\sigma_y$)}
        \end{subfigure}
        \caption{Full-field cumulative plastic strain results computed by the plastic correction algorithm compared to the reference, for two time steps corresponding to successively higher levels of loading.}
        \label{Fig:FullFieldEvolution_p_plasticcorrector}
\end{figure*}

\begin{figure*}[h]
    \centering
        \begin{subfigure}[b]{0.49\textwidth}
            \includegraphics[width=\textwidth]{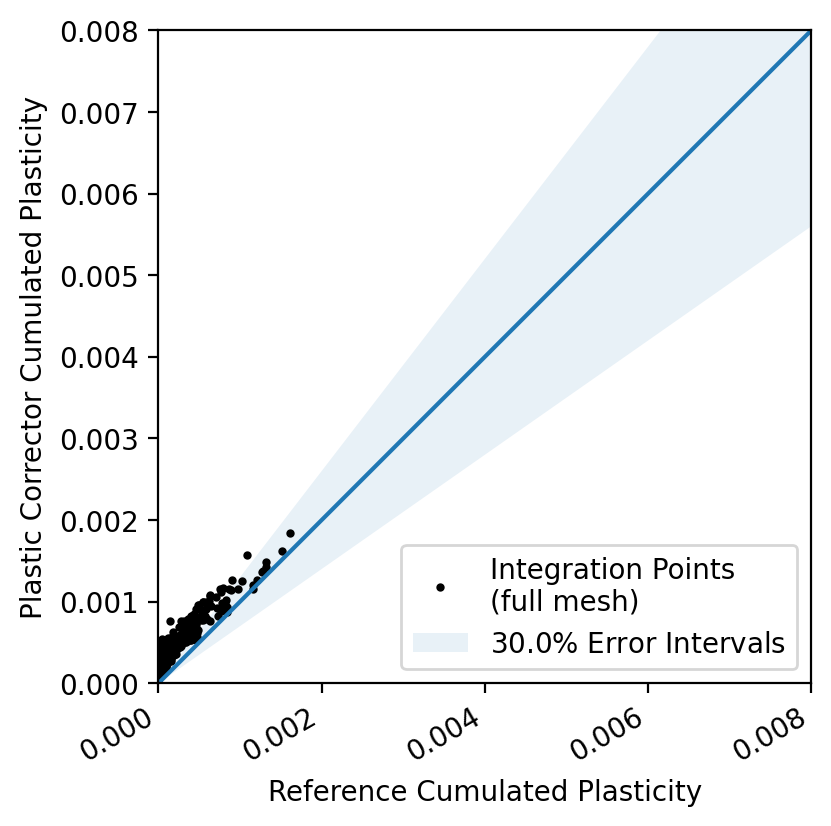}
            \caption{t=0.06 s (applied loading 0.4$\sigma_y$)}
        \end{subfigure}           
        \begin{subfigure}[b]{0.49\textwidth}
            \includegraphics[width=\textwidth]{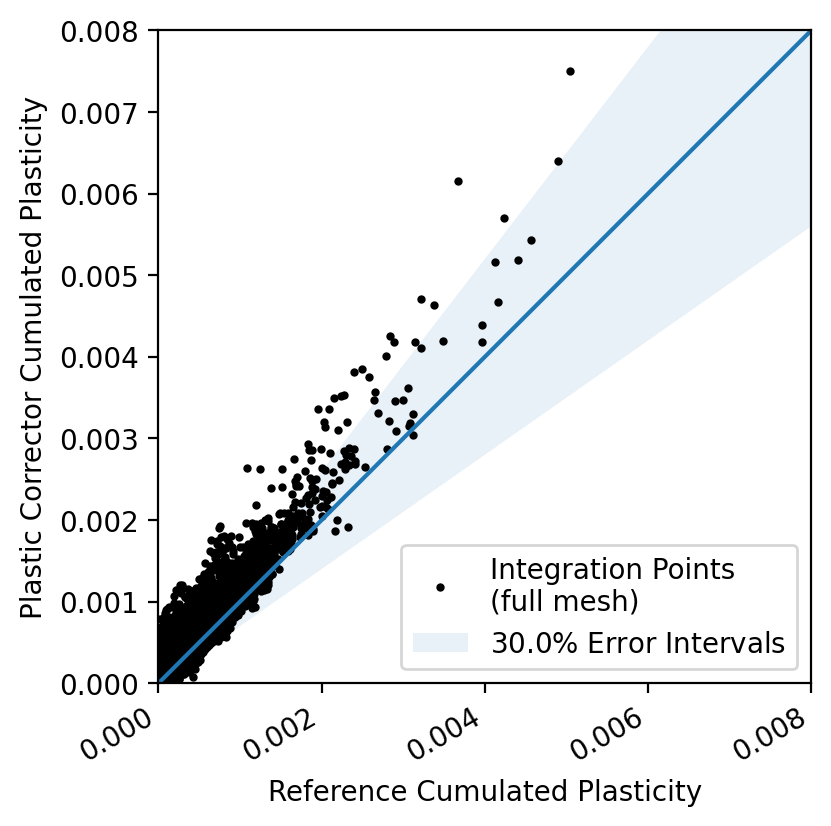}
            \caption{t=0.12 s (applied loading 0.8$\sigma_y$)}
        \end{subfigure}
        \caption{Scatter plots of cumulative plastic strain in all integration points, at time steps corresponding to successively higher levels of loading.}
        \label{Fig:p_scatter_comparison_timeframes}
\end{figure*}

\color{black}
For test case 2, the cumulative plastic strains obtained using the plastic corrector at a couple of time steps of the applied loading (corresponding to 40\% and 80\% of the yield stress of the material) are shown in Figure \ref{Fig:FullFieldEvolution_p_plasticcorrector} and compared with reference computations at the same levels of applied loading. Plasticity develops around the pores, and the plastic corrector \black{once again} predicts overall higher cumulative plastic strain in a slightly larger region, as compared to the reference. Scatter plots of the cumulative plastic strain in all integration points in the mesh are shown in Figure \ref{Fig:p_scatter_comparison_timeframes}. \color{black} The full-field predictions match well with the reference, especially for the time instants corresponding to lower levels of loading. This is because the plastic corrector's accuracy improves when the plasticity is relatively confined. However, a slight over-estimation by the plastic correction algorithm is observed.

\begin{figure*}[!htbp]
    \centering
        \begin{subfigure}[t]{0.40\textwidth}
            \centering
            \adjincludegraphics[width=9.63cm,Clip={.3\width} {.2\height} {0.3\width} {.10\height}]{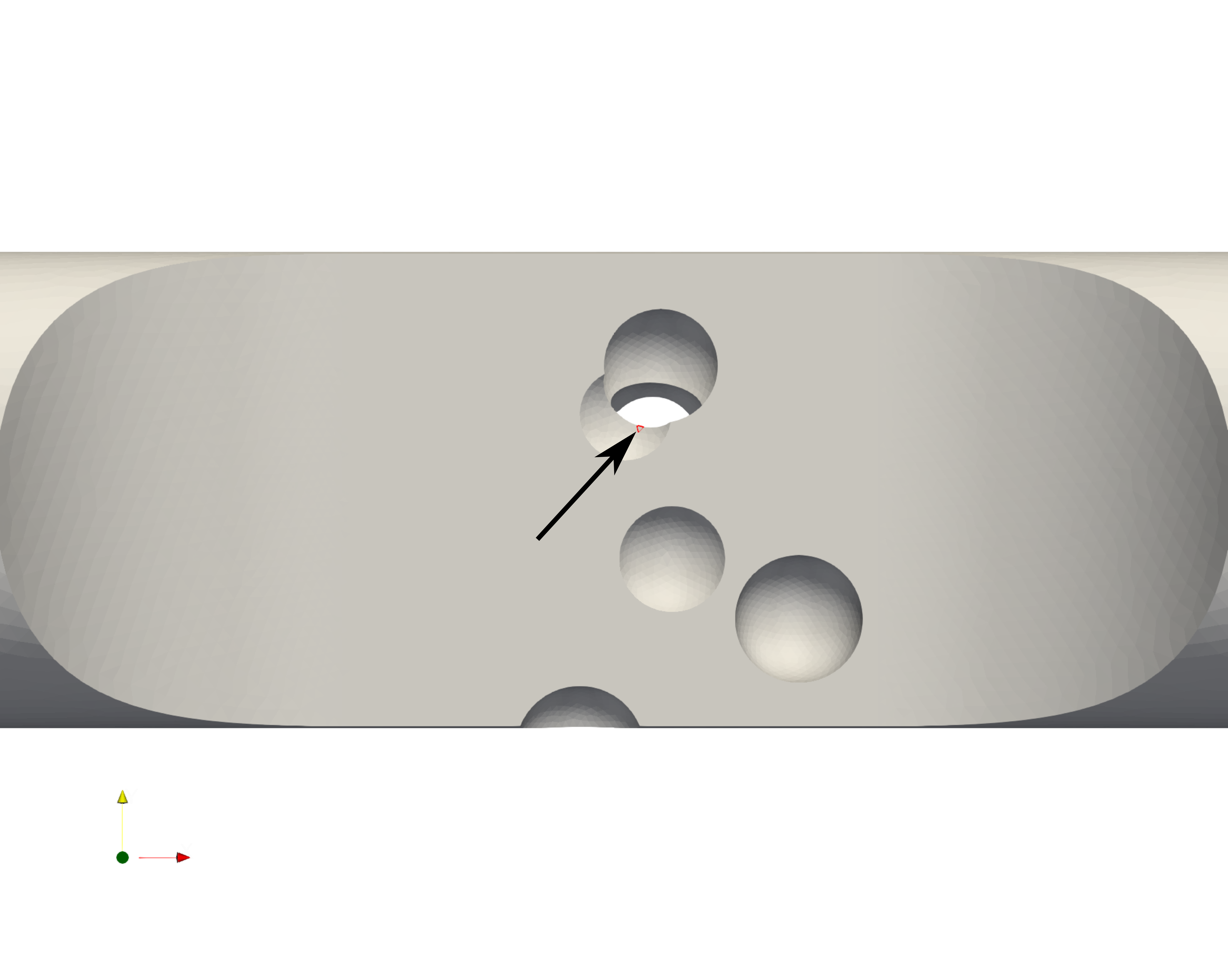}
            \caption{Selected point in the mesh with high loading and high relative error in $p$}
        \end{subfigure}  
        \quad
        \begin{subfigure}[t]{0.5\textwidth}
            \centering
            \adjincludegraphics[width=5.57cm,Clip={.0\width} {.0\height} {0.0\width} {.0\height}]{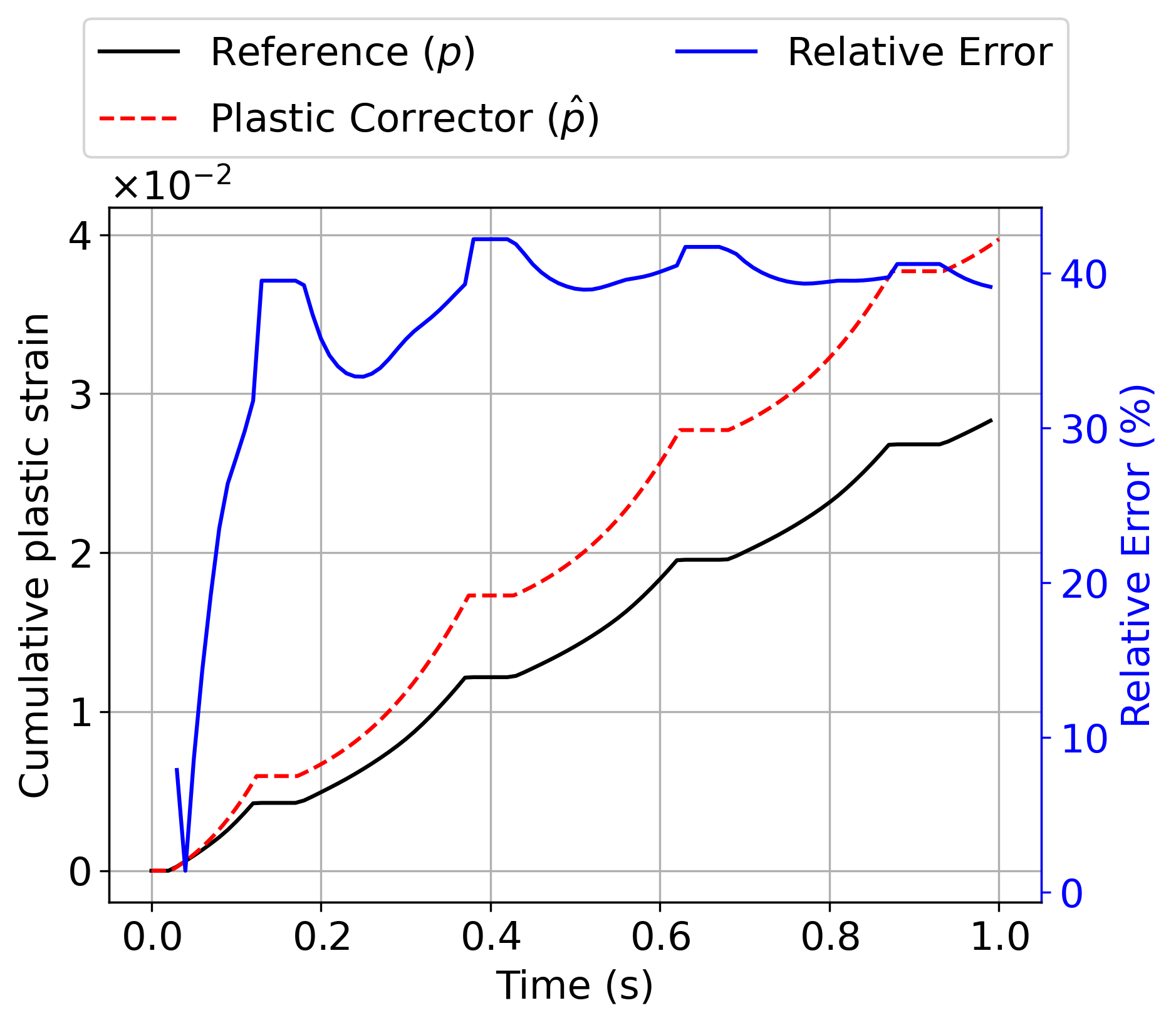}
            \caption{Evolution of cumulative plastic strain at the shown point with high loading and high error}
        \end{subfigure}
        \begin{subfigure}[t]{0.40\textwidth}
            \centering
            \adjincludegraphics[width=9.63cm,Clip={.3\width} {.2\height} {0.3\width} {.10\height}]{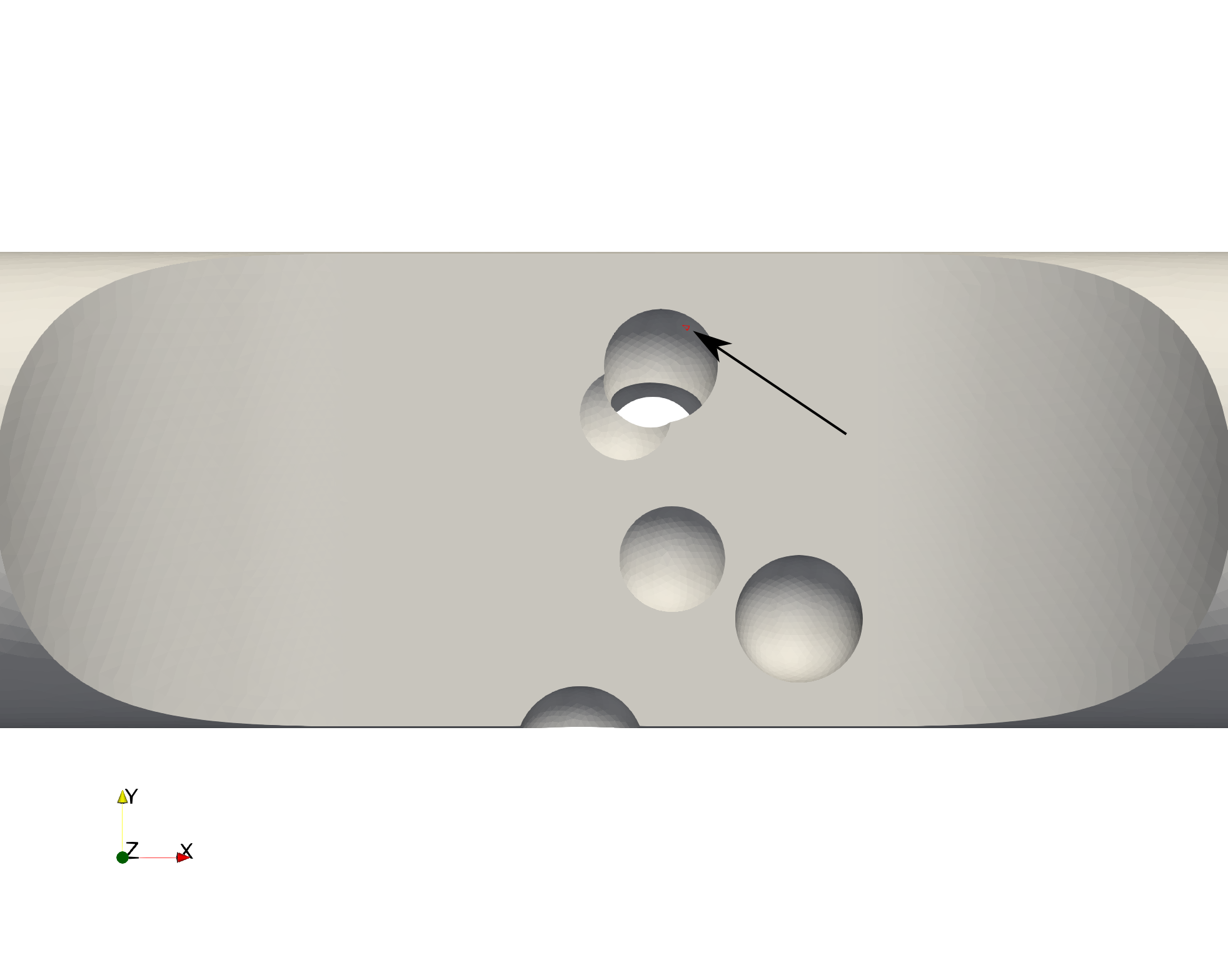}
            \caption{Selected point in the mesh with low loading and low relative error in $p$}
        \end{subfigure}  
        \quad
        \begin{subfigure}[t]{0.5\textwidth}
            \centering
            \adjincludegraphics[width=5.57cm,Clip={.0\width} {.0\height} {0.0\width} {.0\height}]{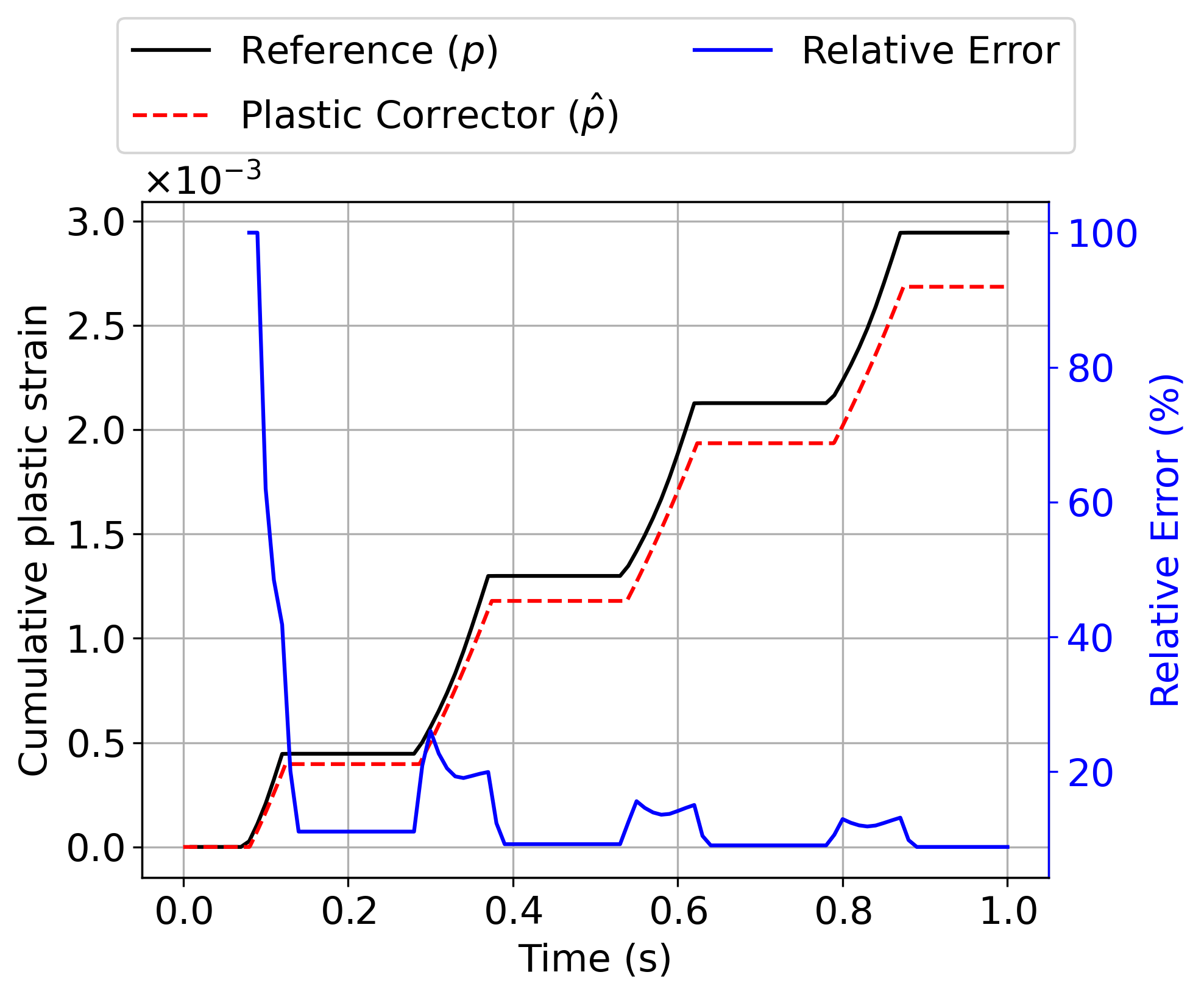}
            \caption{Evolution of cumulative plastic strain at the shown point with low loading and low error}
        \end{subfigure}
        \caption{\black{Comparison of the time-evolution cumulative plastic strain obtained by the plastic corrector and a reference computation at two points in the mesh of test case 2, showing an over-estimation and under-estimation of the plastic corrector}}
        \label{Fig:Comparison_p_paraview_id_438694}
\end{figure*}

The time-evolution of a maximally loaded point near a pore is shown in Figure \ref{Fig:Comparison_p_paraview_id_438694}(a-b). Similar to case 1, the plastic corrector over-estimates the cumulative plastic strain throughout the loading sequence. Interestingly, the relative error is lower at the beginning of loading and rises afterwards, contrary to case 1. The relative error in the prediction at the highest stressed points is around $35-40\%$ during the cyclic loading. The error oscillates with load reversals, with the oscillations reducing with successive cycles. The same analysis is done with a point seeing lesser loading, shown in Figure \ref{Fig:Comparison_p_paraview_id_438694}(c-d). Conversely, the plastic corrector under-estimates the cumulative plastic strain throughout the loading sequence. Under-estimation at points away from highly loaded regions is also a result of stress re-distribution, and has rarely been reported in the literature as Neuber-type methods have primarily been studied for highly loaded points or surfaces.

\color{black}

\clearpage 

\begin{figure*}[h]
    \centering
        \begin{subfigure}[b]{0.49\textwidth}
            \hspace{0.5cm}\adjincludegraphics[width=9.5cm,Clip={.3\width} {.2\height} {0.3\width} {.10\height}]{Figures/elem_paraview_id_438694_arrow.png}
            \caption{Selected point in the mesh with high error}
        \end{subfigure}
        \begin{subfigure}[b]{0.49\textwidth}
            \includegraphics[width=\textwidth]{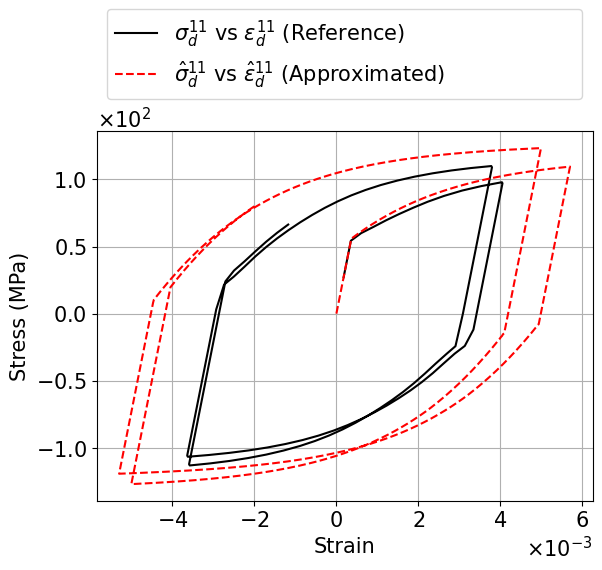}
            \caption{Axial component}
        \end{subfigure}           
        \begin{subfigure}[b]{0.49\textwidth}
            \includegraphics[width=\textwidth]{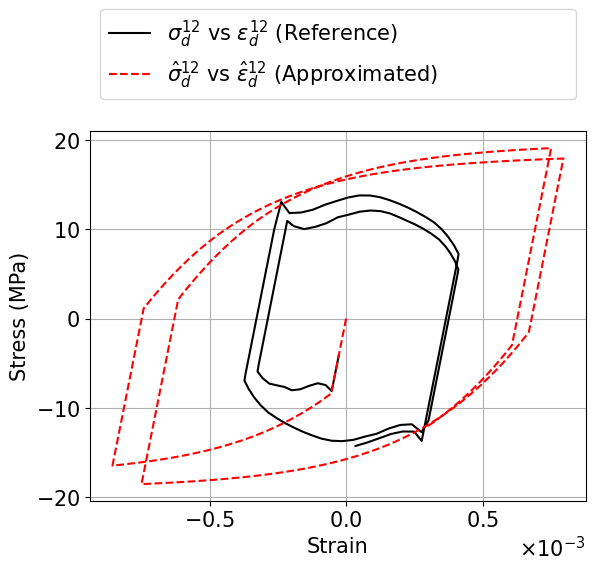}
            \caption{Shear component}
        \end{subfigure}
        \caption{\textcolor{black}{Point in the mesh of test case 2 with a high amount of plasticity and high error in $p$ (30\% relative error between reference $p$ and plastic corrector $\hat{p}$ at t=0.12 s): evolution of the individual components of the approximated deviatoric stress and strain tensors by the plastic corrector (denoted by $\hat{\sigma}_{d}^{ij}$, $\hat{\varepsilon}_{d}^{ij}$) compared to the respective reference curves obtained via a complete elasto-plastic computation (denoted by $\sigma_{d}^{ij}$, $\varepsilon_{d}^{ij}$)}}
        \label{Fig:DeviatoricStressesStrainsPointPore}
\end{figure*}

\begin{figure*}[h]
    \centering
        \begin{subfigure}[b]{0.49\textwidth}
            \hspace{0.5cm}\adjincludegraphics[width=9.5cm,Clip={.3\width} {.2\height} {0.3\width} {.10\height}]{Figures/elem_farfrompore_411685_arrow.png}
            \caption{Selected point in the mesh with low error}
        \end{subfigure}
        \begin{subfigure}[b]{0.49\textwidth}
            \includegraphics[width=\textwidth]{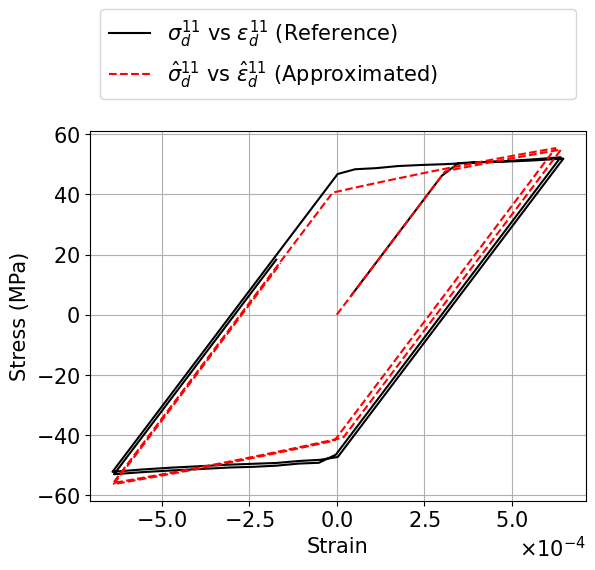}
            \caption{Axial component}
        \end{subfigure}           
        \begin{subfigure}[b]{0.49\textwidth}
            \includegraphics[width=\textwidth]{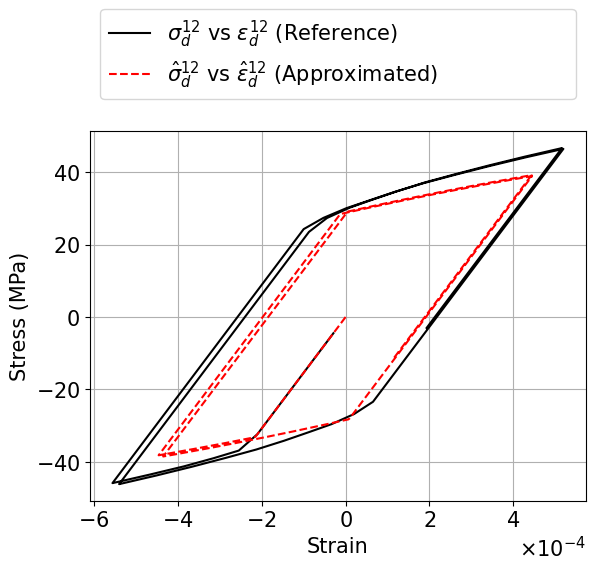}
            \caption{Shear component}
        \end{subfigure}
        \caption{\textcolor{black}{Point in the mesh of test case 2 with a low amount of plasticity and low error in $p$ (9\% relative error between reference $p$ and plastic corrector $\hat{p}$ at t=0.12 s): evolution of the individual components of the approximated deviatoric stress and strain tensors by the plastic corrector (denoted by $\hat{\sigma}_{d}^{ij}$, $\hat{\varepsilon}_{d}^{ij}$) compared to the respective reference curves obtained via a complete elasto-plastic computation (denoted by $\sigma_{d}^{ij}$, $\varepsilon_{d}^{ij}$)}}
        \label{Fig:DeviatoricStressesStrainsPointAwayPore}
\end{figure*}

\paragraph*{Time-evolution of stresses and strains}\mbox{}\\
\black{We limit the study of stresses and strains to the more complex loading case, i.e. the test case 2 with spherical pores. We evaluate the accuracy of the plastic corrector} on components of the deviatoric stress and strain tensors. The same two points are selected as the previous section, i.e. a highly loaded one with high relative error in cumulative plastic strain (Figure \ref{Fig:DeviatoricStressesStrainsPointPore}(a)) and another point with lesser loading and lower relative error (Figure \ref{Fig:DeviatoricStressesStrainsPointAwayPore}(a)).

Figure \ref{Fig:DeviatoricStressesStrainsPointPore}(b-c) shows the evolution of the axial and shear components of the stress and strain tensors for the point under high local loading.

\black{While the stresses are relatively well approximated, the strains have a much higher error -- this is a known result \cite{Demorat2002, Chouman2014,jones1998}. The axial component is overestimated by the plastic corrector. As aforementioned, the Neuber-type method used does not take into account redistribution of stresses. The shear component is not well predicted due to the proportional evolution rule in the plastic corrector -- this will be explained in detail in a later section.}

Figure \ref{Fig:DeviatoricStressesStrainsPointAwayPore}(b-c) shows the evolution of the axial and shear components of the stress and strain tensors for the point under low local loading.

Both the axial and shear components are reasonably well-estimated by the plastic corrector. \black{This is because this point, experiencing lesser loading than the first one, is not as affected by the proportional evolution rule -- again, this is proved later on.}

\begin{figure*}[!htbp]
    \centering
        \begin{subfigure}[b]{0.49\textwidth}
            \adjincludegraphics[width=14.5cm,Clip={.3\width} {.2\height} {0.3\width} {.10\height}]{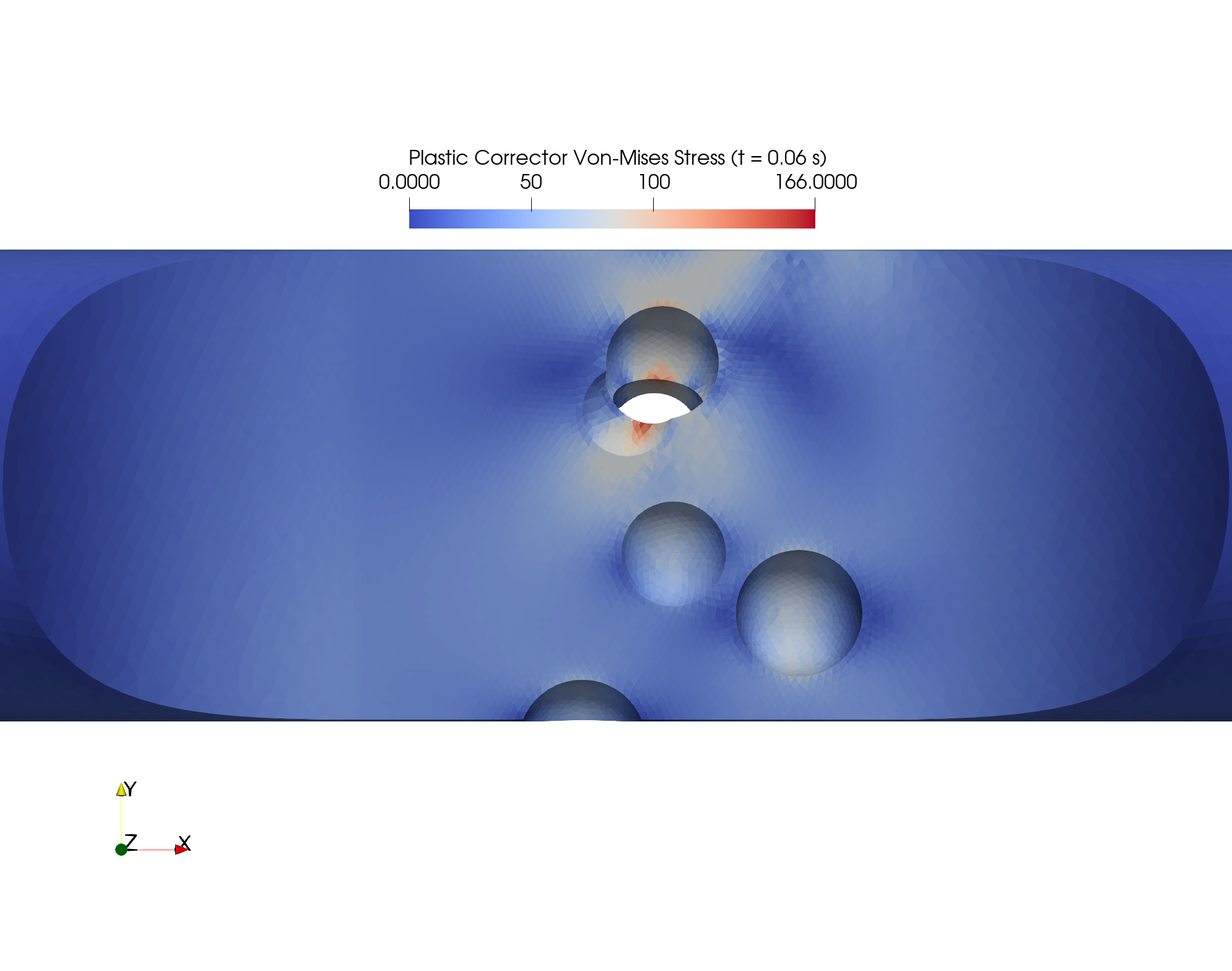}
            \caption{Plastic corrector at t=0.06 s (applied loading 0.4$\sigma_y$)}
        \end{subfigure}           
        \begin{subfigure}[b]{0.49\textwidth}
            \adjincludegraphics[width=14.5cm,Clip={.3\width} {.2\height} {0.3\width} {.10\height}]{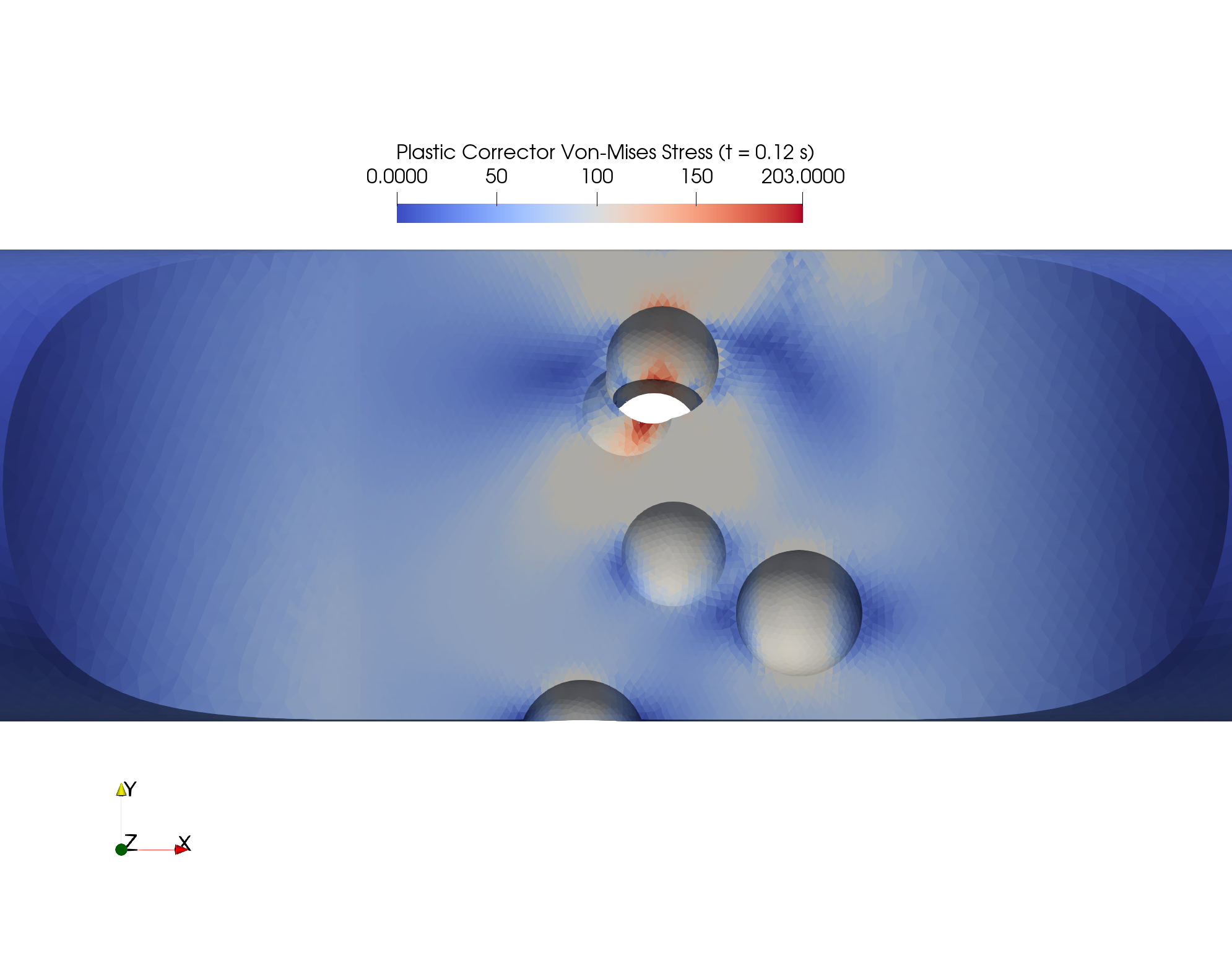}
            \caption{Plastic corrector at t=0.12 s (applied loading 0.8$\sigma_y$)}
        \end{subfigure}
        \begin{subfigure}[b]{0.49\textwidth}
            \adjincludegraphics[width=14.5cm,Clip={.3\width} {.2\height} {0.3\width} {.10\height}]{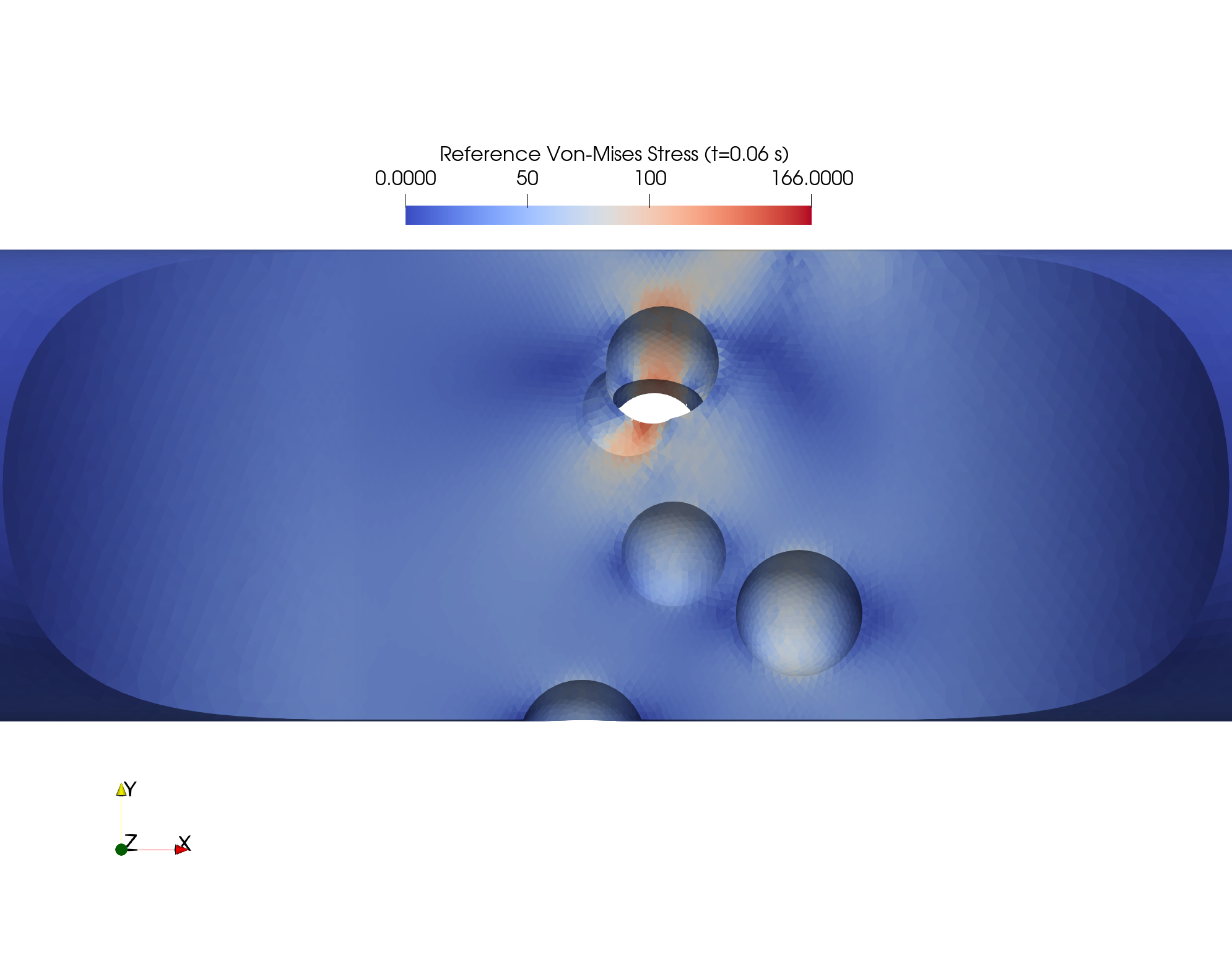}
            \caption{Reference computation at t=0.06 s (applied loading 0.4$\sigma_y$)}
        \end{subfigure}           
        \begin{subfigure}[b]{0.49\textwidth}
            \adjincludegraphics[width=14.5cm,Clip={.3\width} {.2\height} {0.3\width} {.10\height}]{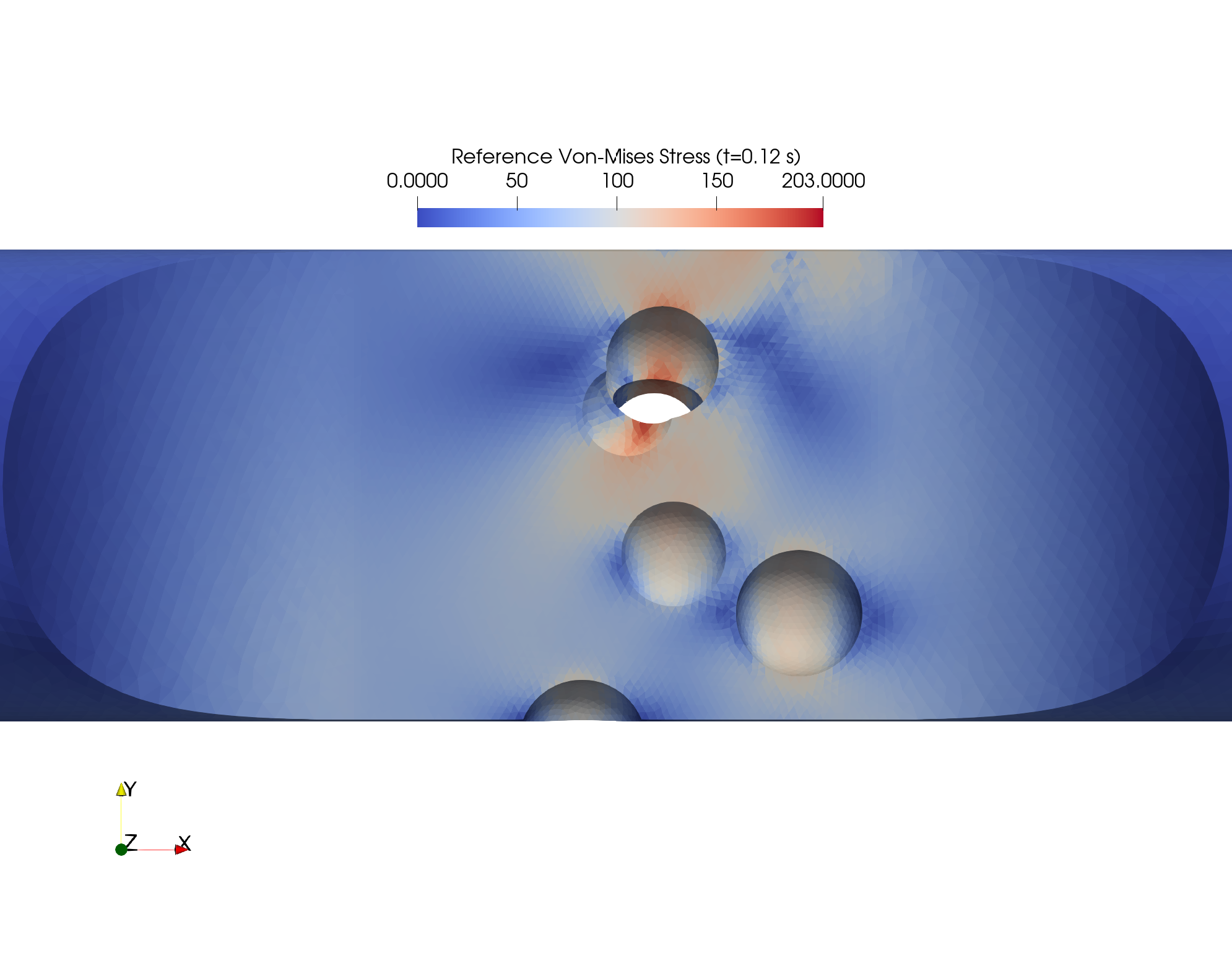}
            \caption{Reference computation at t=0.12 s (applied loading 0.8$\sigma_y$)}
        \end{subfigure}
        \caption{Full-field von Mises stress results computed by the plastic correction algorithm compared to the reference, for two time steps corresponding to successively higher levels of loading.}
        \label{Fig:FullFieldEvolution_sigvm_plasticcorrector}
\end{figure*}

Finally, the von Mises stress approximated by the plastic corrector is compared to the reference computation in Figure \ref{Fig:FullFieldEvolution_sigvm_plasticcorrector}. The results show an excellent match between the plastic corrector and the reference, for progressively higher loading.

\begin{figure*}[!htbp]
    \centering
        \begin{subfigure}[b]{0.49\textwidth}
            \adjincludegraphics[width=16cm,Clip={.3\width} {.2\height} {0.3\width} {.10\height}]{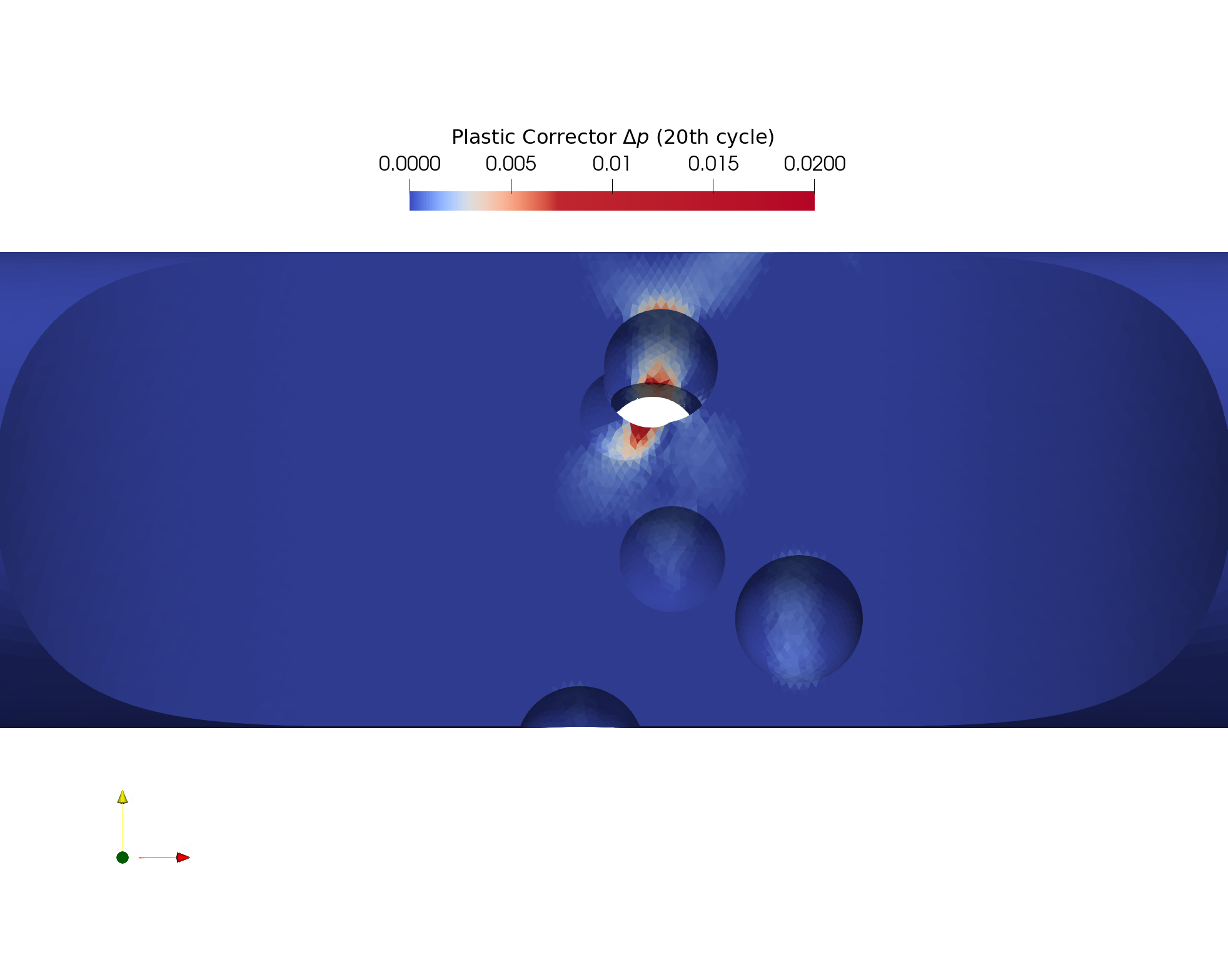}
            \caption{Plastic corrector}
        \end{subfigure}           
        \begin{subfigure}[b]{0.49\textwidth}
            \adjincludegraphics[width=16cm,Clip={.3\width} {.2\height} {0.3\width} {.10\height}]{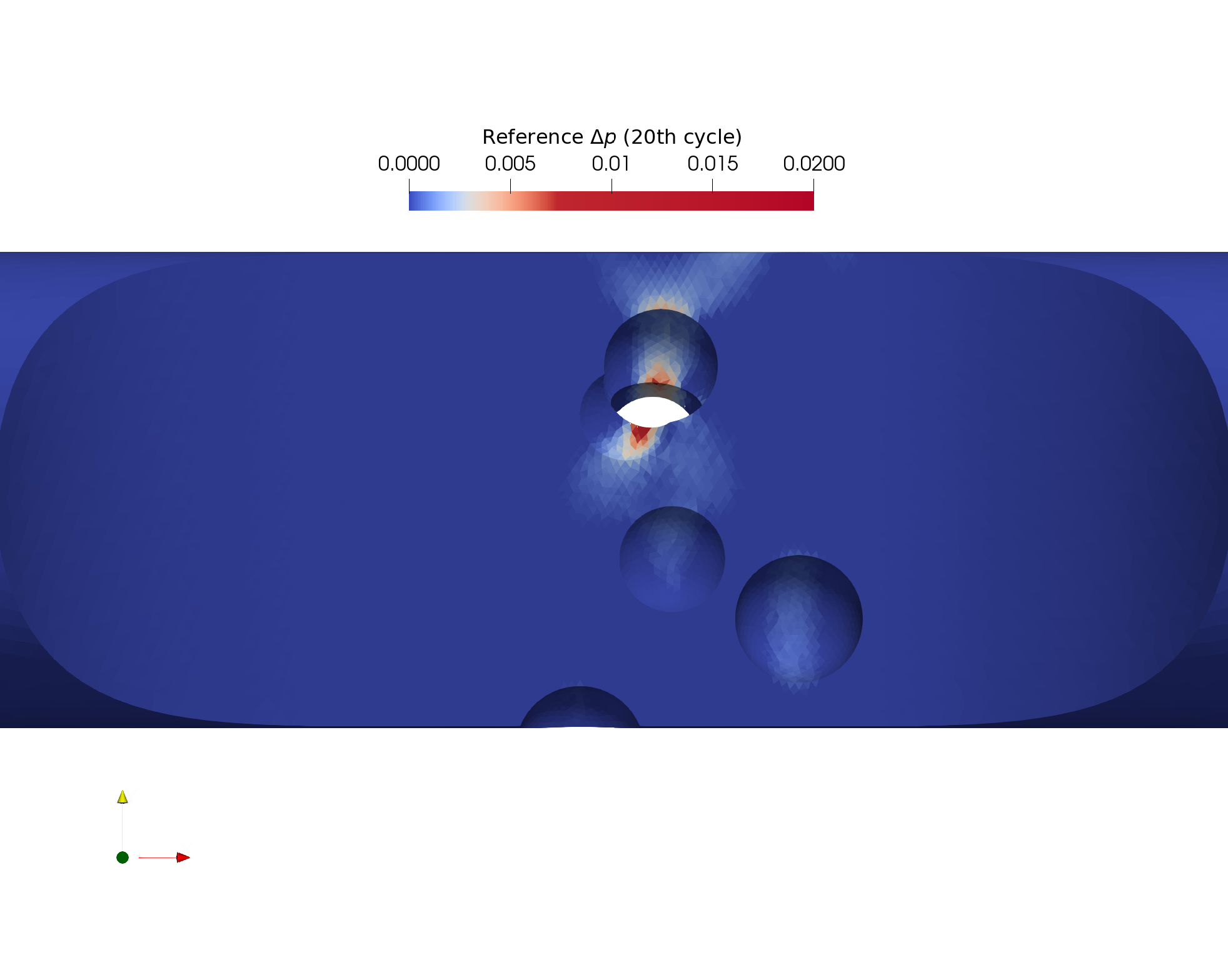}
            \caption{Reference computation}
        \end{subfigure}
        \begin{subfigure}[b]{1\textwidth}
            \centering
            \adjincludegraphics[width=8.42cm,Clip={.0\width} {.0\height} {0.0\width} {.0\height}]{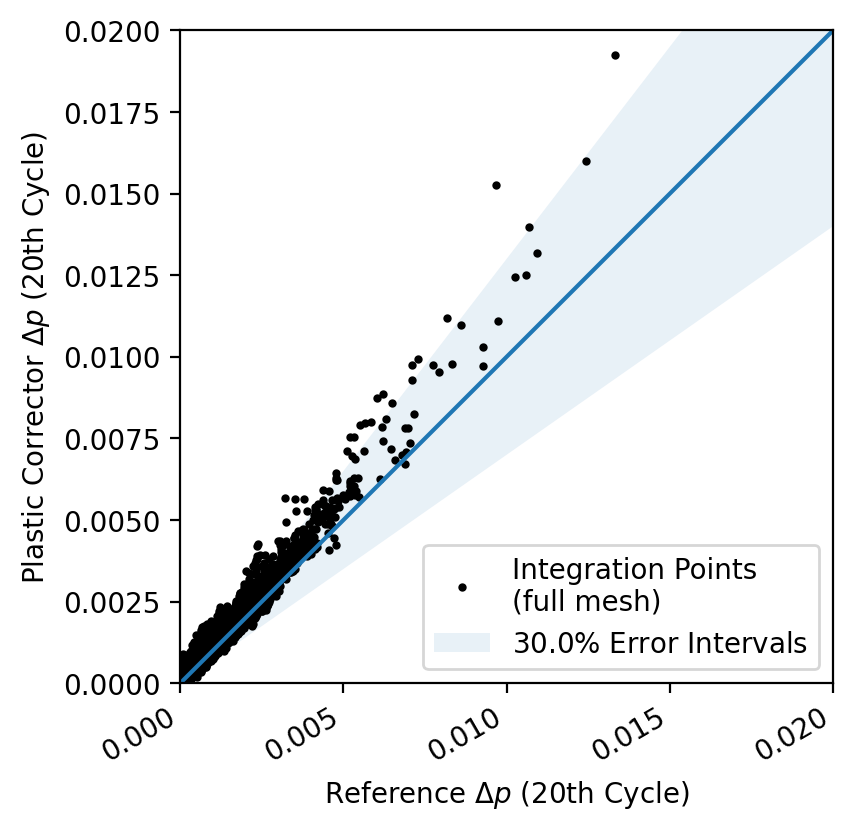}
            \caption{Scatter plot of $\Delta p$ values in all the integration points in mesh}
        \end{subfigure}           
        \caption{Full-field comparison of $\Delta p$ in the 20\textsuperscript{th} cycle in the test case 2 mesh}
        \label{Fig:stabcycledeltap}
\end{figure*}

\clearpage
\paragraph*{Cumulative plastic strain range in the 20\textsuperscript{th} cycle}\mbox{}\\
\black{
The field of cumulative plastic strain range $\Delta p$ over the 20\textsuperscript{th} cycle, obtained by the plastic correction algorithm and the reference elasto-plastic FE computation, is shown respectively in Figures \ref{Fig:stabcycledeltap}(a) and \ref{Fig:stabcycledeltap}(b) for the test case number 2, specimen with spherical pores. A comparison of the values at all the quadrature points in the mesh is shown in Figure \ref{Fig:stabcycledeltap}(c). A good overall match is found in the values of the cumulative plastic strain range obtained by the plastic correction algorithm and the reference computation. Like the previously presented cases, there is over and under-estimation by the plastic corrector -- the number of over-estimated points is significantly higher than the under-estimated ones. The order of relative error in the loaded zones is around $30-40\%$. However, we acknowledge that the errors may be much higher if the material presents ratcheting behaviour, which was not the case here \cite{Chouman2014}.
}

\color{black}
\subsection{Computational time needed for a full-field plastic corrector computation}
A breakdown of the computational costs required for obtaining the elasto-plastic solution by the two methods, i.e. a full reference computation and a plastic corrector computation, is compared in table \ref{tab:CPUtimes_refplas}. The mesh considered is the same porous mesh as test case 2, and has around 661k quadrature points (one per linear element). The loading considered here comprises 1000 time-steps.

\begin{table}[!htbp]
\color{black}
\caption{\color{black}CPU times (in seconds) for full-field $p$ computation for the test case 2: mesh of a specimen with spherical pores (661771 quadrature points, $1000$ time-steps)\color{black}}
\label{tab:CPUtimes_refplas}
\centering
\begin{tabular}[t]{llllllll}
\hline\noalign{\smallskip}
Operation & Reference FEA (s) & Plastic corrector (s) \\
\noalign{\smallskip}\hline\noalign{\smallskip}
Elastic FEA for $\bar{\sigma}_{\textrm{VM}}^{\#}$ & - & 236 \\
 \noalign{\smallskip}\hline\noalign{\smallskip}
Elasto-plastic computation & 52850 & 156 \\
 \noalign{\smallskip}\hline\noalign{\smallskip}
Total CPU time & 52850 & 392 \\
\noalign{\smallskip}\hline
\end{tabular}
\color{black}
\end{table}
\color{black}
\clearpage
\subsection{Errors due to the rule of local proportionality}
\color{black}
We will now analyze the error that arises solely from the local proportionality rule. Our objectives are to evaluate (i) the contribution of this proportionality rule to the error in the plastic corrector method and (ii) the characteristics of this error. To achieve this, we will construct a projected elasto-plastic solution based exclusively on the local proportionality rule, omitting the Neuber-type rule. Specifically, we will project the reference stress history at a given point onto the direction of the stress derived from the elastic computation. This projected stress will be compared to the stress calculated using the plastic corrector, which incorporates both the Neuber-type rule and the local proportionality rule. This comparison will allow us to isolate the part of the error that is due to the local proportionality rule. Additionally, we will compare the projected stress computed using only the proportionality rule to the reference stress to assess the acceptability of the proportionality rule in terms of error. Deviations between the stresses computed with these two methods will indicate significant local non-proportionality, which our algorithm may not adequately capture. We will show the characteristics of this error, i.e. the conditions and reasons for which usage of the proportionality rule could result in high relative error.

The projected stress tensor resulting from the local proportionality rule is computed using the following projection of the reference stress ${\utilde\sigma}_d$:

\begin{equation}\label{Error1}
    \hat{\hat{\utilde\sigma}}_d = \frac{(\utilde{\bar{\sigma}}_d : {\utilde\sigma}_d)}{(\utilde{\bar{\sigma}}_d : \utilde{\bar{\sigma}}_d)}\utilde{\bar{\sigma}}_d
\end{equation}
\color{black}
Quantities denoted by $\hat{\hat{\bullet}}$ represent projected quantities assuming local proportionality; this approximation is thus different from the plastic corrector approximations developed in the previous sections as it directly operates on the reference stress (which is, of course, not available in practice). The absolute error $\xi$ and the relative error $\xi_{rel}$ between the projected deviatoric stress tensor using the hypothesis of proportionality ($\hat{\hat{\utilde\sigma}}_d$), and the reference deviatoric stress tensor via a full non-linear FE computation (${\utilde\sigma}_d$) can be calculated as follows:

\begin{equation}\label{Error2}
    \utilde{\xi} = {\utilde\sigma}_d - \hat{\hat{\utilde\sigma}}_d
\end{equation}

The relative error in stress is computed by taking the Frobenius norm of the tensors:

\begin{equation}
    \xi_{rel} = \frac{\| \utilde{\xi} \|_F}{\| {\utilde\sigma}_d \|_F}
\end{equation}

\begin{figure*}[h]
    \centering
        %\begin{subfigure}[b]{0.49\textwidth}
        %\centering
        %    \hspace{0.5cm}\adjincludegraphics[width=9.5cm,Clip={.3\width} {.2\height} {0.3\width} {.10\height}]{Figures/elem_paraview_id_438694_arrow.png}
        %    \caption{Selected point in the mesh for error analysis (test case 2)}
        %\end{subfigure}
        \begin{subfigure}[b]{0.49\textwidth}
            \centering
            \includegraphics[width=\textwidth]{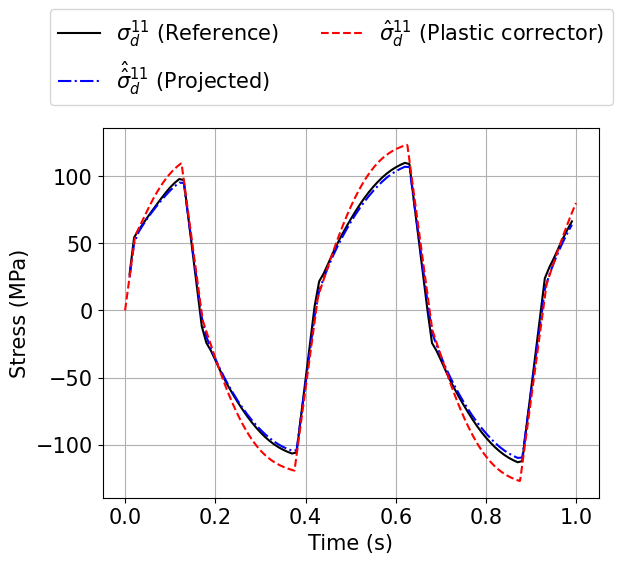}
            \caption{Axial component}
        \end{subfigure}           
        \begin{subfigure}[b]{0.49\textwidth}
            \centering
            \includegraphics[width=\textwidth]{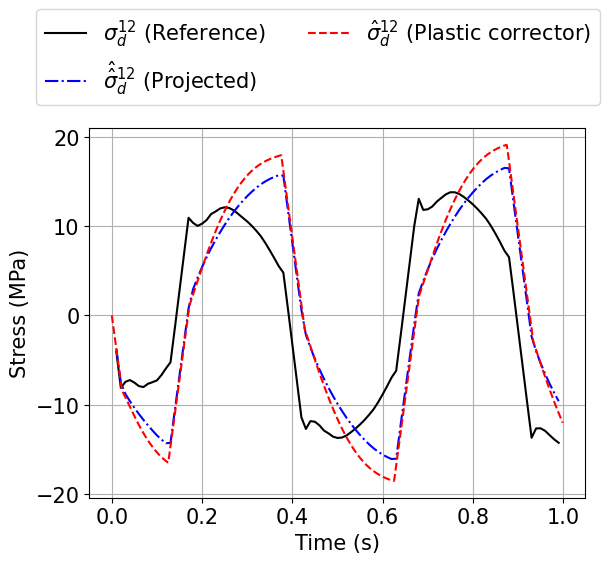}
            \caption{Shear component}
        \end{subfigure}
        \begin{subfigure}[b]{0.49\textwidth}
            \centering
            \includegraphics[width=\textwidth]{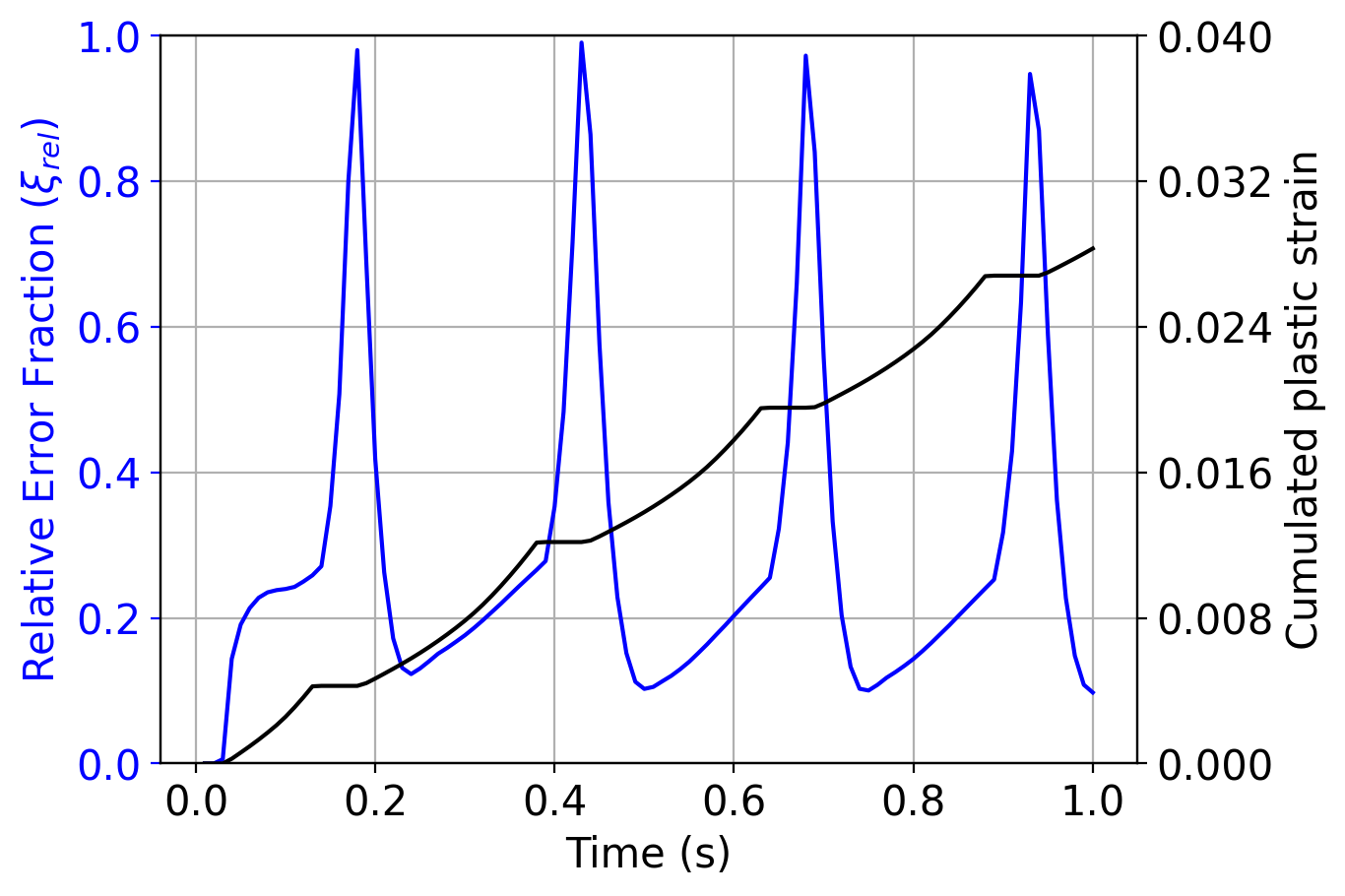}
            \caption{Time evolution of $\xi_{rel}$ and $p$}
        \end{subfigure}
        \caption{\textcolor{black}{For a point with high plasticity (shown in Figure \ref{Fig:Comparison_p_paraview_id_438694}(a)): Evolution of the individual components of the projected deviatoric stress tensor (denoted by $\hat{\hat{\sigma}}_{d}^{ij}$) compared to the respective reference curves obtained via a complete elasto-plastic computation without any reduction (denoted by $\sigma_{d}^{ij}$).} The plastic corrector solution is recalled in red. The relative error fraction for the stresses is shown alongside the cumulative plastic strain.}
        \label{Fig:RelativeError_and_DeviatoricStressesStrainsPointPore}
\end{figure*}

\begin{figure*}[h]
    \centering
        %\begin{subfigure}[b]{0.49\textwidth}
        %    \centering
        %    \hspace{0.5cm}\adjincludegraphics[width=9.5cm,Clip={.3\width} {.2\height} {0.3\width} {.10\height}]{Figures/elem_farfrompore_411685_arrow.png}
        %    \caption{Selected point in the mesh for error analysis (test case 2)}
        %\end{subfigure}
        \begin{subfigure}[b]{0.49\textwidth}
            \centering
            \includegraphics[width=\textwidth]{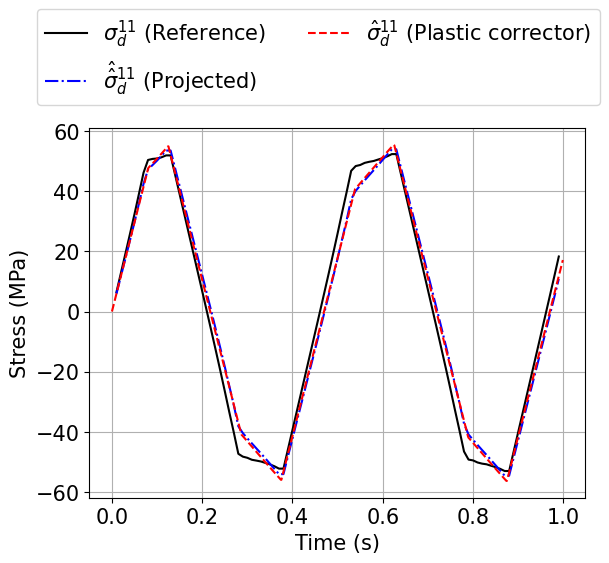}
            \caption{Axial component}
        \end{subfigure}           
        \begin{subfigure}[b]{0.49\textwidth}
            \centering
            \includegraphics[width=\textwidth]{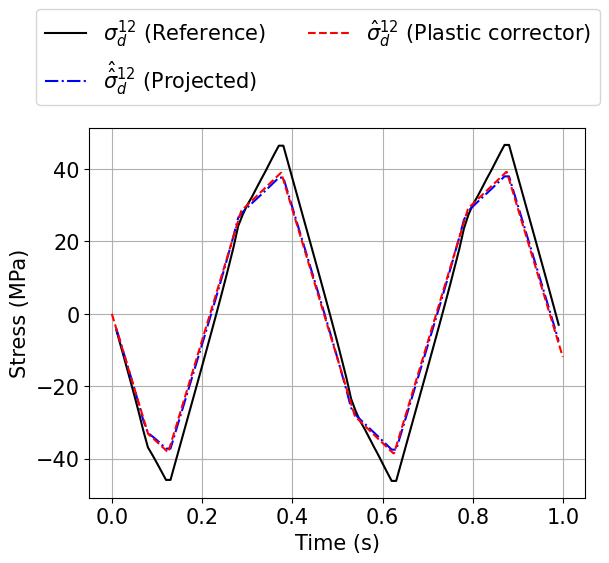}
            \caption{Shear component}
        \end{subfigure}
        \begin{subfigure}[b]{0.49\textwidth}
            \centering
            \includegraphics[width=\textwidth]{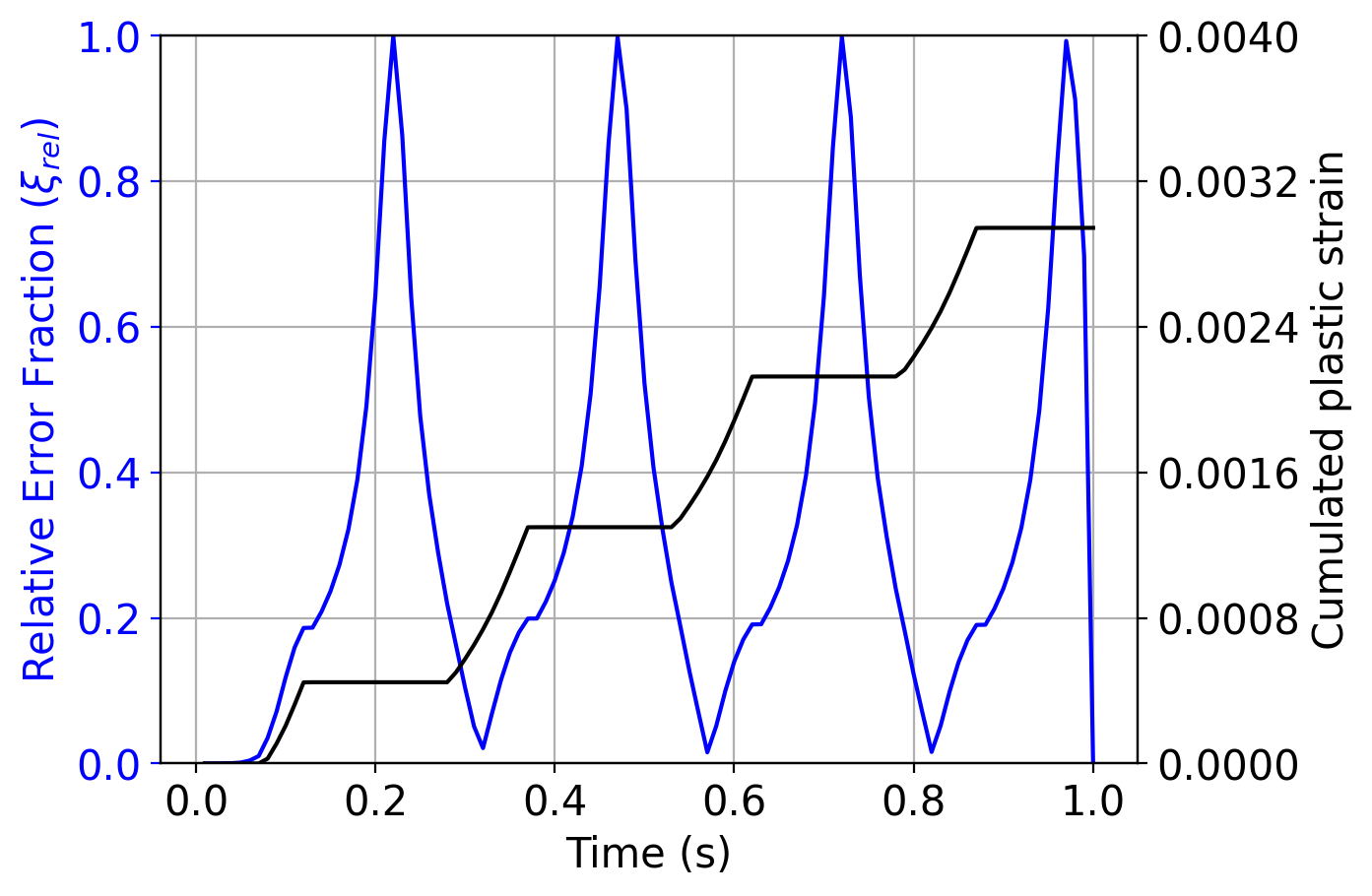}
            \caption{Time evolution of $\xi_{rel}$ and $p$}
        \end{subfigure}
        \caption{\textcolor{black}{For a point with low plasticity (shown in Figure \ref{Fig:Comparison_p_paraview_id_438694}(c)): Evolution of the individual components of the projected deviatoric stress tensor calculated by assuming a hypothesis of proportionality (denoted by $\hat{\hat{\sigma}}_{d}^{ij}$) compared to the respective reference curves obtained via a complete elasto-plastic computation without any reduction (denoted by $\sigma_{d}^{ij}$).} The plastic corrector solution is recalled in red. The relative error fraction for the stresses is shown alongside the cumulative plastic strain.}
        \label{Fig:RelativeError_and_DeviatoricStressesStrainsPointAwayPore}
\end{figure*}

\paragraph{Contribution of the local proportionality rule to the error in the plastic corrector solution}\mbox{}\\
We will now analyze the contribution of the proportional evolution rule in terms of error in the plastic corrector solution, at a few points of the computational domain of test case 2. The same two points previously analyzed in Figure \ref{Fig:Comparison_p_paraview_id_438694}(a,c) are considered. Figure \ref{Fig:RelativeError_and_DeviatoricStressesStrainsPointPore}(a-b) shows a comparison of the evolution of the axial and shear components of the deviatoric stress tensors for the first point near a pore, with high plasticity and high error, computed using (i) purely the proportional evolution rule (denoted by
$\hat{\hat{\sigma}}_{d}^{ij}$
%,$\hat{\hat{\varepsilon}}_{d}^{ij}$
) (ii) the plastic corrector (denoted by
$\hat{\sigma}_{d}^{ij}$
%, $\hat{\varepsilon}_{d}^{ij}$
) (iii) reference (denoted by $\sigma_{d}^{ij}$
%, $\varepsilon_{d}^{ij}$
). For the components computed using purely the local proportional evolution rule, we observe that the axial component is very well predicted. The shear component is badly predicted, but the maximum stress value is acceptable. This shows that in the plastic corrector solution, the proportional evolution rule does not contribute significantly to the overall solution, and it is the Neuber-type rule that causes most of the error. A similar comparison is done for the second point with low plasticity and low error (Figure \ref{Fig:RelativeError_and_DeviatoricStressesStrainsPointAwayPore}(a-b)). For this point, both the axial and shear components computed with the local proportional evolution rule are reasonably well predicted, and also do not differ from the plastic corrector solution.

The relative error $\xi_{rel}$ due to the local proportionality rule is shown for the two points in Figures \ref{Fig:RelativeError_and_DeviatoricStressesStrainsPointPore}(c) and \ref{Fig:RelativeError_and_DeviatoricStressesStrainsPointAwayPore}(c), alongside their cumulative plastic strain. \color{black} The load reversal naturally causes $\xi_{rel}$ to increase sharply. \textcolor{black}{$\xi_{rel}$ becomes close to 1, which indicates a 100\% relative error.
The explosion of relative errors at these load inversion points is because the stress values are very close to zero (see Figures \ref{Fig:RelativeError_and_DeviatoricStressesStrainsPointPore}(a,b) and \ref{Fig:RelativeError_and_DeviatoricStressesStrainsPointAwayPore}(a,b)). The absolute error at these inversion points is very low. This shows that even if the relative error is significant, in reality, its effect is negligible.}

\black{Furthermore, the cumulative plastic strain does not increase during these load reversals. In other words, the error during load reversal does not contribute significantly to the error in plasticity computations. For the first point with a higher $p$ error, cumulative plastic strain starts increasing again despite a high $\xi_{rel}$. For the second point, with lower $p$ error, we observe that the $\xi_{rel}$ always stays low during increase in plasticity. Thus, some points of loading, which are closer to a proportional loading sequence, are better predicted than others.}

\begin{figure*}[h]
    \centering
        \begin{subfigure}[b]{0.6\textwidth}
            \includegraphics[width=\textwidth]{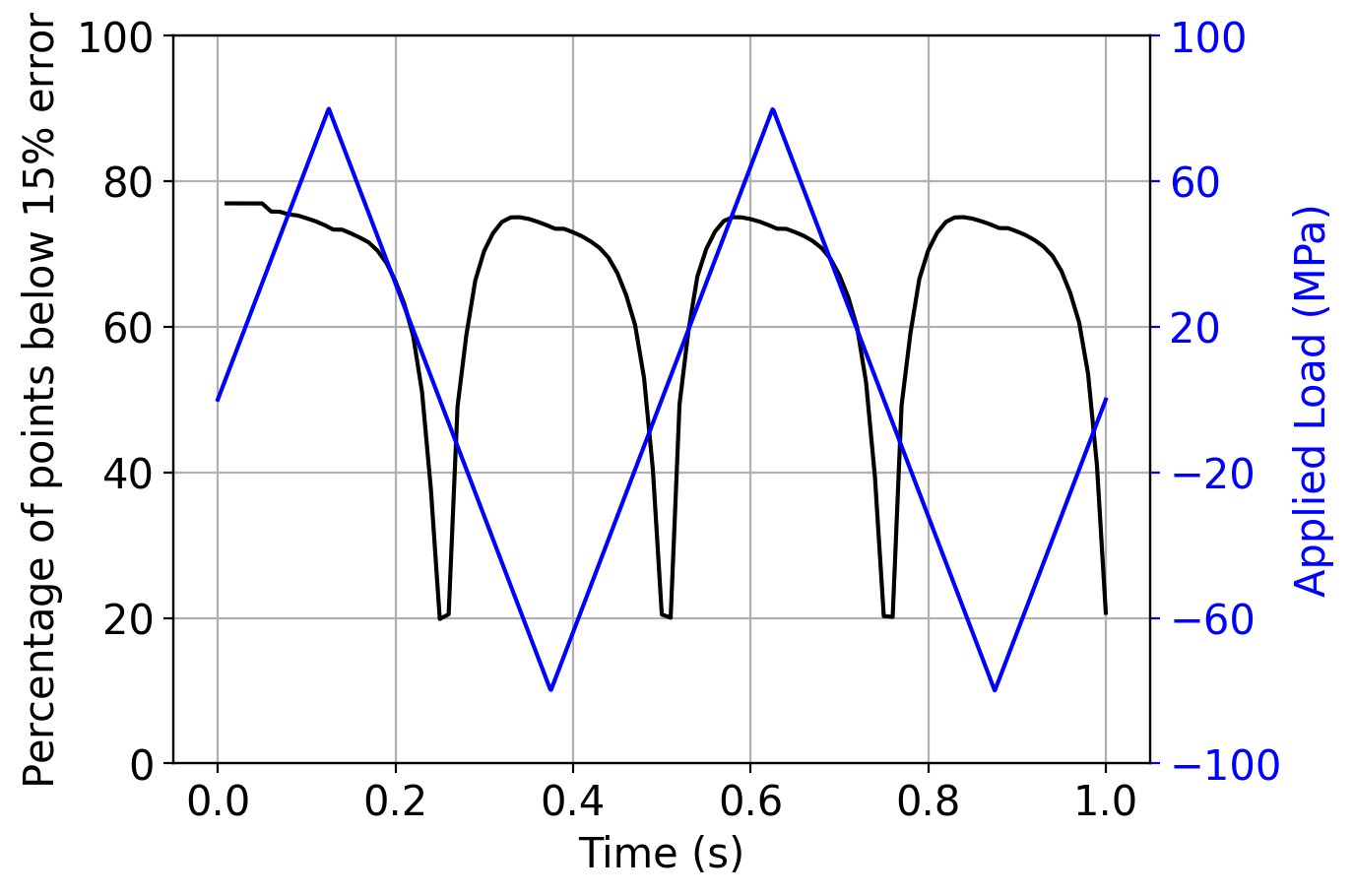}
        \end{subfigure}
        \caption{Time evolution of the percentage of elements in the FE computation (test case 2: spherical pores) below 15\% relative error, alongside the applied loading history in the gauge section away from pores}
        \label{Fig:RelativeError_vs_time}
\end{figure*}

\clearpage
\color{black}
\paragraph{Relative error due to the local proportionality rule evaluated in the full mesh}\mbox{}\\
\textcolor{black}{Only test case number 2, i.e. the specimen with spherical pores, is considered.} 
The relative error $\xi_{rel}$ due to the local proportionality rule is calculated for all the loading history. The percentage of points in the mesh that stay below 15\% relative error during the loading sequence is shown in Figure \ref{Fig:RelativeError_vs_time}. During the first branch of monotonic loading, around 70-75\% of all the points in the mesh remain below 15\% relative error, and during the load reversal, this number goes down to 20\%. \black{As previously discussed, this occurs because the near-zero stress values at these load inversion points lead to a sharp increase in relative errors.} The percentage of points below 15\% relative error goes back up to 70\% after the load reversal, when plasticity starts developing again.

\clearpage
\section{Machine learning-based acceleration of plastic corrector computations}
\medskip

\color{black}

We wish to compute a scalar quantity of interest (QoI) from the elasto-plastic solution at arbitrary time $t \in [ 0 , T]$. More precisely, we wish to extract a scalar value $Q \in \mathbb{R}$ from $(\hat{\utilde{\sigma}}(t),\hat{\utilde{\varepsilon}}(t),\hat{\utilde{\varepsilon}}^p(t),\hat{p}(t),\hat{\utilde{X}}(t))$ at an arbitrary number of quadrature points of the finite element mesh.

During the process of acquiring the QoI, the computation of $(s(t),e(t),e_p(t),\hat{p}(t),x(t))$ requires implicit time integration, as described in section \ref{sec:plasreduced} and the appendix \hyperref[appendixA]{A}, and needs to be done for every quadrature point of the finite element mesh, which may lead to significant computational expense, especially for long time analysis. Remarkably, variables  $(s(t),e(t),e_p(t),\hat{p}(t),x(t))$ are dependent on the solution of the elastic finite element prediction through the von Mises stress only, which can be seen by inspection of the system of equations in Table \ref{tab:tensortoscalarplascorr}, right column. 
\black{Therefore, any elasto-plastic quantity of interest, which depends solely on these proportionality ratios also depends on the sole von Mises stress $\bar{\sigma}_{\text{VM}}^{\#}$ stemming from the elastic finite element computation, for a given $f(t)$.}

\black{We propose to acquire the scalar QoI with a one-dimensional Gaussian process regression (GP) algorithm \cite{rasmussen2006} trained on plastic corrector computations. One may use several independent GP regression models for several quantities of interest.}

\color{black}
\paragraph{Gaussian process regression algorithm}\mbox{}\\
The GP needs to capture the dependence of the scalar QoI on $\bar{\sigma}_{\text{VM}}^{\#}$. To this end, we need to define a training interval for $\bar{\sigma}_\textrm{VM}^{\#}$.
A uniformly distributed set of $n_s$ von Mises stress values, between 0 and $s^+$ times the $\sigma_y$, is created in logarithmic space. As stress values can span several orders of magnitude, usage of logarithmic space ensures that the GP has good interpolation ability. These values are input to the plastic corrector to calculate the QoI. $\bar{\sigma}_{\text{VM}}^{\#}$ and QoI constitute the training data for the GP. For numerical stability, we applied a logarithmic transformation to both the input and target data. Any zero values in the target data were replaced with a small positive value to avoid undefined logarithmic values. The model was then fitted to the log-transformed data.

Once trained, the GP can be applied to all integration points within a given mesh of a structure without requiring any time integration.

\begin{figure*}[h!tb!p]
    \centering
        \begin{subfigure}[b]{0.49\textwidth}
            \includegraphics[width=\textwidth]{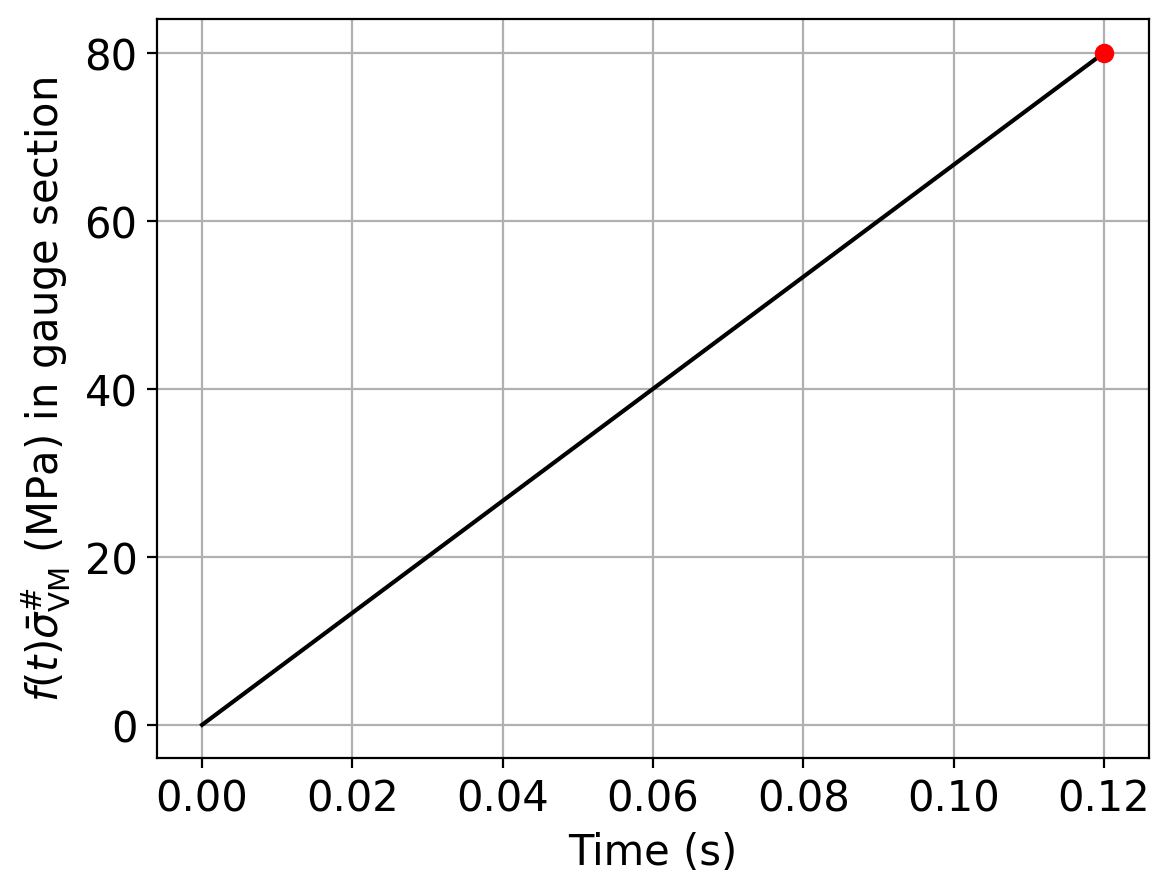}
            \caption{}
        \end{subfigure}  
        \begin{subfigure}[b]{0.49\textwidth}
            \includegraphics[width=\textwidth]{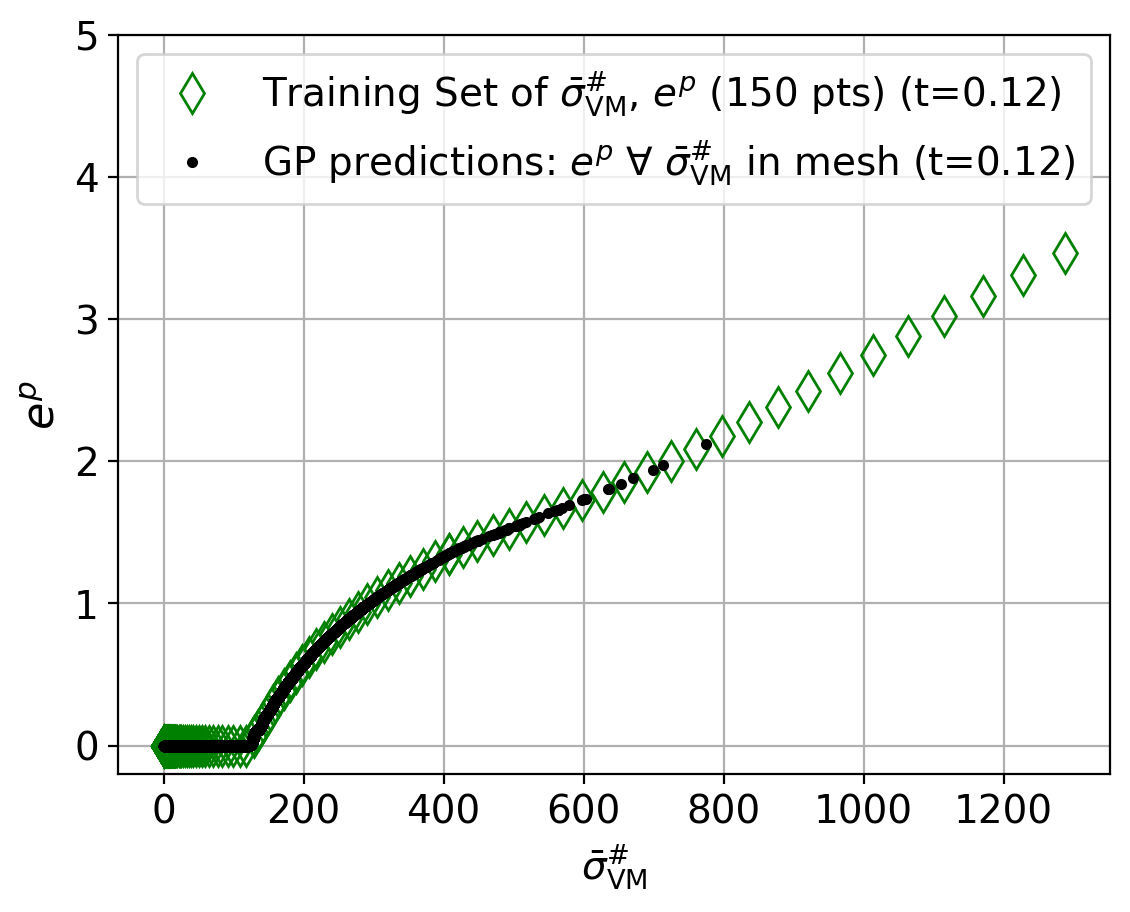}
            \caption{}
        \end{subfigure}           
        \caption{\textcolor{black}{(a) The load function chosen, with 1000 time-steps, with $e^p$ being extracted at the last time-step at the peak of loading (highlighted in red) (b) Training data created for the Gaussian Process (GP) using the plastic corrector, along with the GP's predicted $e^p$ for the last time-step
        }}
        \label{Fig:FiniteElementSimulatorGP}
\end{figure*}

\paragraph{Numerical example on a QoI}\mbox{}\\
As an example we train a GP to predict $e^p$ at the 1000\textsuperscript{th} time-step of a given $f(t)$ (shown in Figure \ref{Fig:FiniteElementSimulatorGP}(a)) for the mesh of a specimen with spherical pores (test case 2). \black{The training data set for learning the scalar relation between $\bar{\sigma}_{\text{VM}}^{\#}$ and $e^p$ is created using $n_s=150$ points and $s^+=12$. This is shown in Figure \ref{Fig:FiniteElementSimulatorGP}(b). The choice of 150 points ensures sufficient resolution for accurately capturing the relationship between the two quantities, which is either zero or a simple monotonically increasing function. The upper limit, set at 12 times the yield stress, is chosen to be slightly higher than the maximum stress concentration factor induced by the pores in the mesh, which is approximately 8.}

\begin{figure*}[h!t!bp]
  \begin{center}
  \includegraphics[trim=0cm 0 0 0, clip, width=0.6\textwidth]{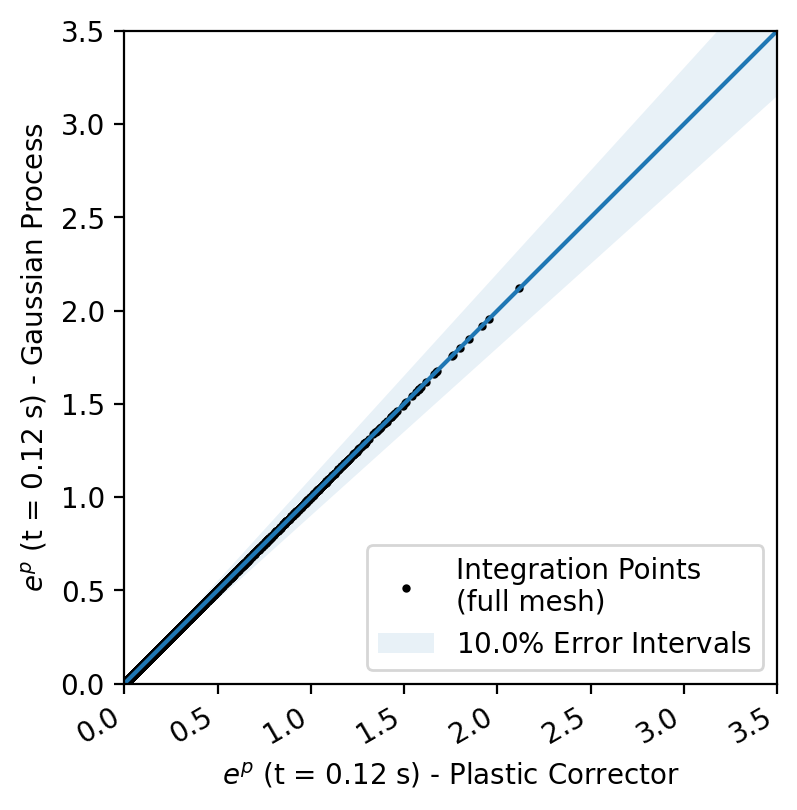}
  \end{center}
\caption{\textcolor{black}{Validation of the Gaussian process regression: Comparison between the results obtained via the plastic correction algorithm (pure Neuber-type algorithm) and the predicted results via the Gaussian process regression.}}
\label{Fig:GP_vs_pureNeuber}
\end{figure*}

A comparison between the results obtained via the plastic correction algorithm and the predicted results via the Gaussian process regression is shown in Figure \ref{Fig:GP_vs_pureNeuber} for all integration points. The results show virtually no difference, indicating that the 1D meta-model can be used to further accelerate computation of scalar QoIs of the plastic corrector with no added error.

The Gaussian process regression only needs a few seconds in total to use for the prediction of a scalar QoI, including the cost of computing the training set via the plastic corrector, training time (computation and factorisation of data covariance matrix), and prediction time (shown in table \ref{tab:CPUtimes_GP}).

\begin{table}[!htbp]
\color{black}
\caption{\color{black}CPU times (in seconds) for QoI computation for the test case 2: mesh of a specimen with spherical pores (661771 quadrature points, $1000^{th}$ time-step). The training set comprises 150 points integrated over all 1000 time-steps.}
\label{tab:CPUtimes_GP}
\centering
\begin{tabular}{lccccc}
\hline\noalign{\smallskip}
Operation & Computing training set & Training & QoI computation & Total CPU time \\
\noalign{\smallskip}\hline\noalign{\smallskip}
Plastic corrector & - & - & 156 & 156 \\
\noalign{\smallskip}\hline\noalign{\smallskip}
Gaussian process regression & 0.6 & 0.3 & 1.7 & 2.6 \\
\noalign{\smallskip}\hline
\end{tabular}
\color{black}
\end{table}

\color{black}

\clearpage
\section{Neural plastic corrector: learning from examples}

\color{black}

The aim of this section is to investigate the relative merits of the previously described Neuber-type \black{plastic corrector that corrects elasto-static simulations} and that of an AI-based alternative that learns plastic corrections from full elasto-plastic simulations, before attempting to merge the two approaches.

The application of machine learning as a surrogate or correction technique for finite element analysis (FEA) has been successful across various fields, including the acceleration of nonlinear computations in porous specimens \cite{Krokos2024}, structural analysis \cite{Nie2019,deshpande2024}  
and other nonlinear mechanics problems \cite{raissi2019, rabczuk2024}. However, these methods tend to perform poorly when applied to cases that are geometrically very different from the cases in the training set \cite{Krokos2022}. Taking this knowledge on-board, we restrict our AI approach using Convolutional Neural Networks (CNNs) to the correction of elasto-plastic fields in 
specimens with given nominal geometries but with random geometrical defects, the morphology and positioning statistics of which are known in advance. We will concentrate here on a tensile specimen including uniformly distributed spherical pores of fixed radius.

The AI strategy is the following: we voxelise the neighbourhood of points for which plastic correction is to be performed. Next, we provide either (i) the von Mises stress from the elastic computation or (ii) the result of the proposed Neuber-type plastic corrector as input to a CNN designed to predict a pointwise output of interest stemming from an elasto-plastic simulation. Training is done by acquiring reference quantities of interests (QoI) from a series of full elasto-plastic simulations using the Z-Set suite \cite{Besson1998}, for a sufficiently large number of specimens with random pore placement.

\color{black}
\subsection{Dataset generation}
The objective here is to generate a robust dataset that contains sufficient mechanical information about pore-surface and pore-pore interactions, so that the QoI in specimens containing a new distribution of spherical random pores is well predicted. 

\begin{figure*}[h]
    \centering
        \begin{subfigure}[t]{0.49\textwidth}
            \includegraphics[width=\textwidth]{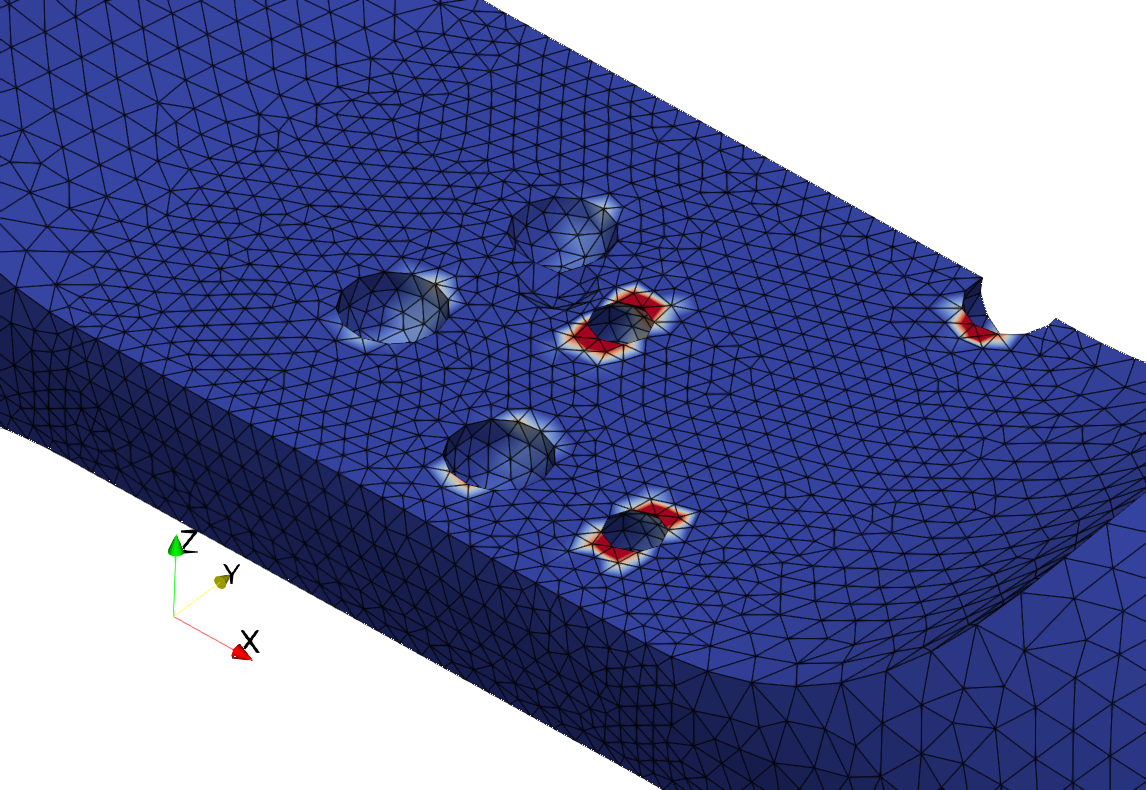}
            \caption{Finite Element results on a continuous space}
        \end{subfigure}           
        \begin{subfigure}[t]{0.49\textwidth}
            \includegraphics[width=\textwidth]{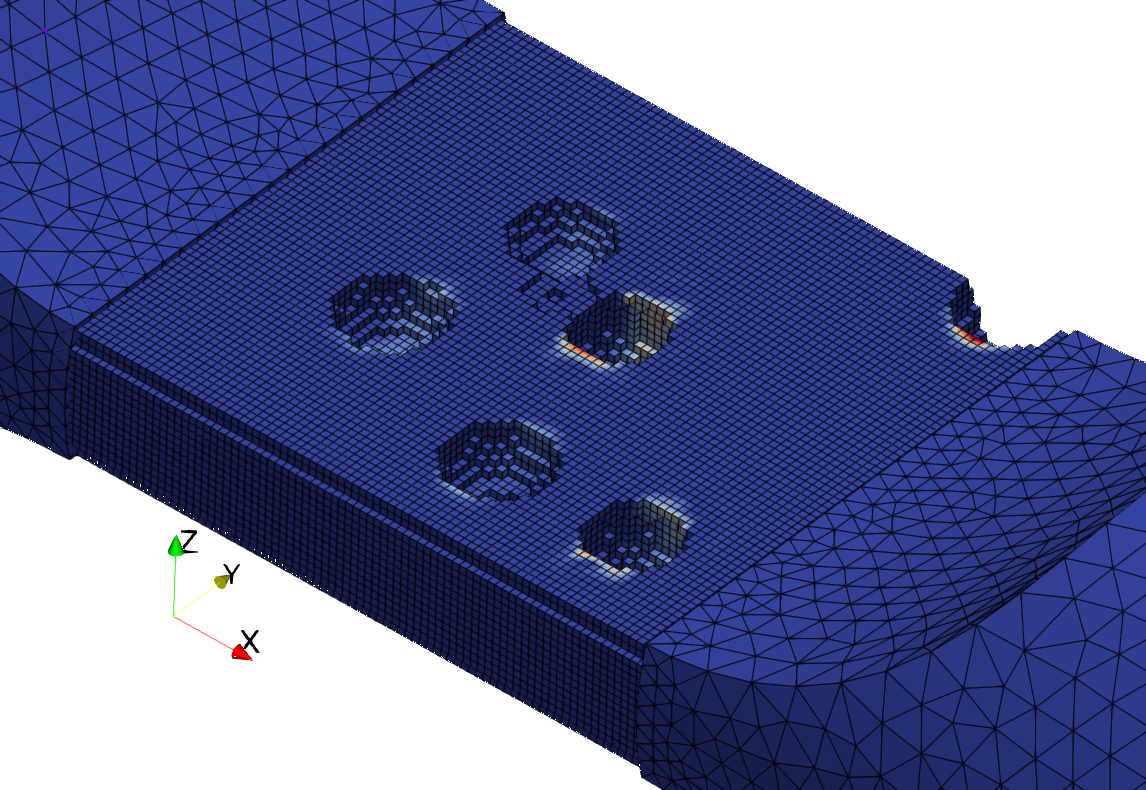}
            \caption{Transformation to a uniform grid of voxels}
        \end{subfigure}
        \begin{subfigure}[t]{0.49\textwidth}
            \includegraphics[width=\textwidth]{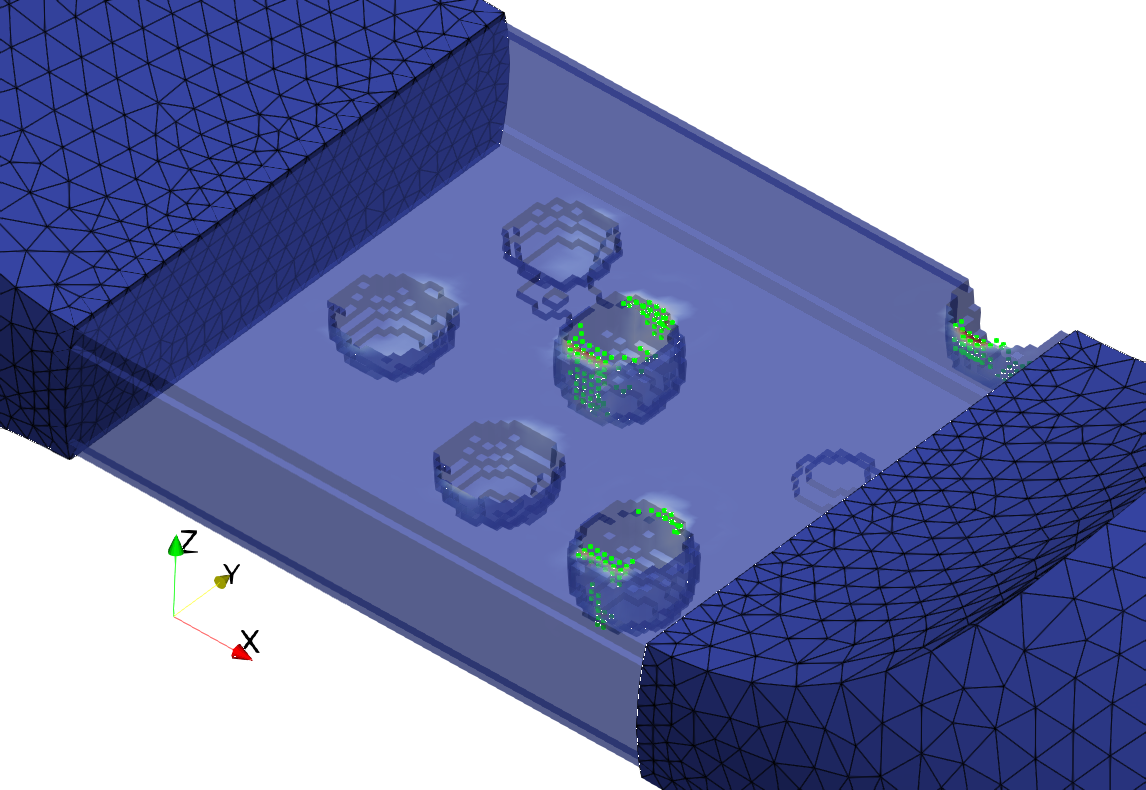}
            \caption{Clusters using the K-medoids algorithm, at which the QoI is to be predicted}
        \end{subfigure}
        \begin{subfigure}[t]{0.49\textwidth}
            \includegraphics[width=\textwidth]{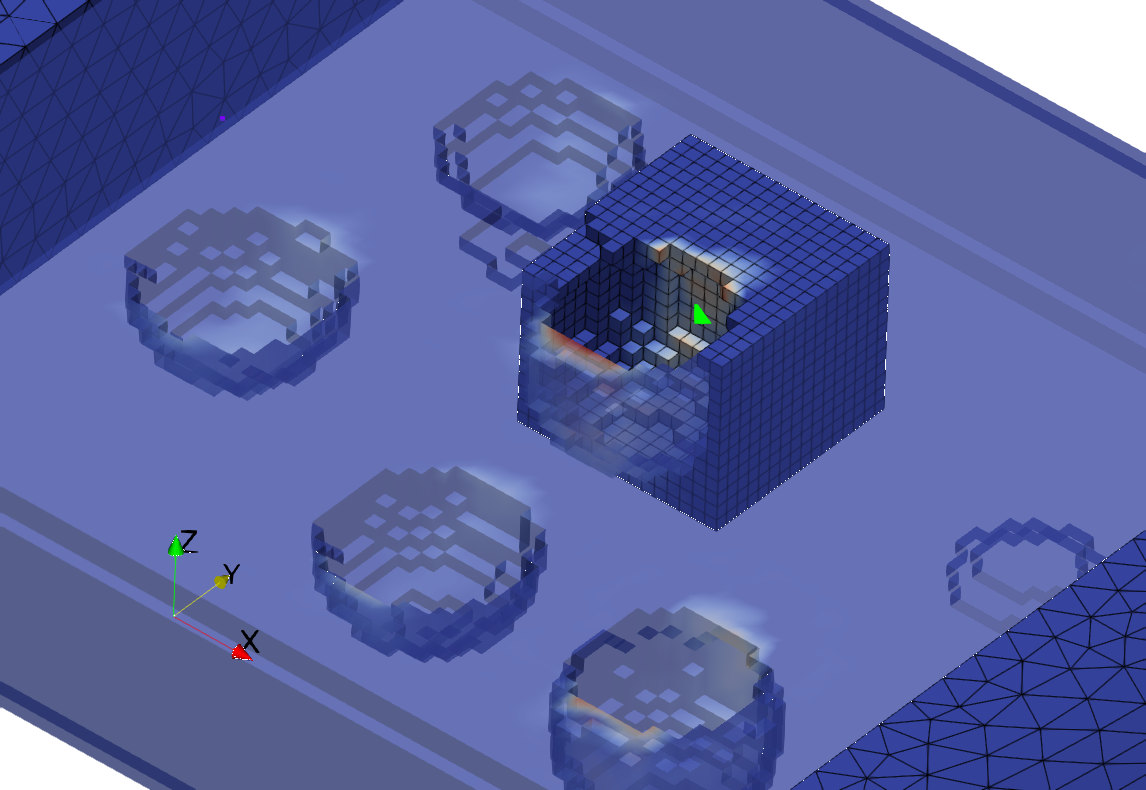}
            \caption{A subvolume of voxels around a cluster voxel, serving as low-fidelity input to CNN}
        \end{subfigure}
        \caption{\black{Obtaining training data from a mesh for the Convolutional Neural Network}}
        \label{Fig:DataTreatment}
\end{figure*}

The Convolutional Neural Network (CNN) used here is designed to work with 3D images. A procedure was developped in order to convert finite element results into uniform grids of voxels (see Figure \ref{Fig:DataTreatment}(a-b)). CNNs are resource-heavy when working with 3D images. Therefore, a size restriction is necessary for the images to fit in memory. We choose the CNN prediction \textcolor{black}{of the QoI} at a given voxel to be conditioned on the surrounding sub-volume of 16x16x16 voxels \textcolor{black}{of the low-fidelity simulation. This sub-volume} is assumed to contain enough information about the mechanical behaviour of the voxel's surroundings. In other words, the influence of the mechanical state beyond this sub-volume on the center point is assumed to be sufficiently well-represented by the mechanical fields that will be provided as input channel for the 16x16x16 input volume. 
The QoI in the subvolumes is normalised by the maximum value of the QoI across the training set. \black{Empty space due to the pores or due to regions beyond the specimen's surfaces are encoded with a negative value, as input to our CNN is positive. If a negative input is to be given, an additional binary channel can be used to encode the geometric information of the specimens \cite{Krokos2022}.}

\black{Next, the QoI computed by the plastic corrector is extracted specifically for the points that experience plastic deformation (i.e. all points with the von Mises stress exceeding $\sigma_y$). A K-medoids clustering algorithm (scikit-learn) is used to extract clusters from these points. K-medoids is preferred because it chooses actual data points as cluster centers, rather than creating new, synthetic points like K-means. This approach ensures that each cluster center is a real sample from the dataset, and is necessary as the mechanical field around the cluster center is needed.} The number of clusters per realisation of a random porous specimen was chosen to be 200, to extract a reasonable amount of data per specimen. An example of the cluster points in a mesh is shown in Figure \ref{Fig:DataTreatment}(c).

Finally, an input-output pair consists of a 16x16x16 volume of real values (the von Mises stress from the elastic prediction, or the actual QoI computed using the Neuber-type plastic corrector), associated with a reference value for the QoI corresponding to the center of this volume. 

\subsection{Architecture}
\begin{figure*}
  \includegraphics[trim=0cm 0 0 0, clip, width=1\textwidth]{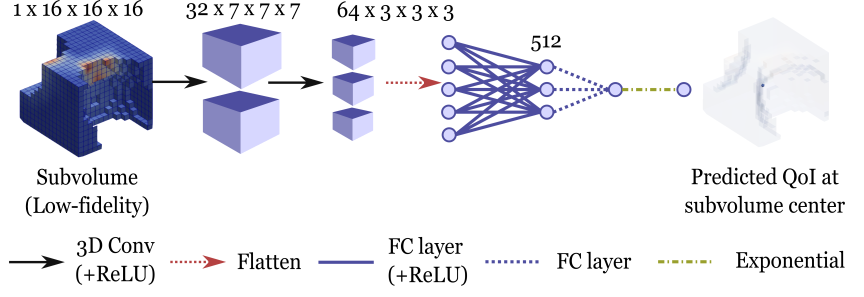}
\caption{\black{Architecture of the CNN developed for multi-fidelity corrections of a quantity of interest (QoI)}}
\label{Fig:ArchitectureCNN}
\end{figure*}

The input of the CNN (Figure \ref{Fig:ArchitectureCNN}) consists of 16x16x16 subvolumes \textcolor{black}{of low-fidelity information} around a voxel of the \textcolor{black}{QoI} to be corrected.
The convolutional blocks used consist of 3D convolutional operations with zero padding and a stride of two voxels, which reduces the size of the volumes after each block, followed by ReLU activation functions. After two blocks, the volumes are flattened and reduced through a fully connected layer followed by a ReLU activation function, and a last fully connected layer reduces the dimensions of the pseudo-output to unity. \textcolor{black}{An exponential function can be used on the pseudo-output, to ensure that the CNN predictions remain positive, which is activated if the QoI is required to be strictly positive}. The model parameters consist of the weights of the kernel and the fully connected layers. The output of the network is the value of the predicted QoI at the centre of the given subvolume. A Mean Squared Error between the \textcolor{black}{CNN predictions and the reference data summed over the batch (size of 32) is used as the loss function for training.}

\subsection{Numerical results}
\textcolor{black}{For the numerical study, the intrinsic dissipated energy $\phi$ integrated over a loading cycle from time $t_i$ to $t_f$ \cite{BCCF2010} is considered as a QoI to be corrected via a CNN:}
\begin{equation}\color{black}
    \phi = \int_{t_i}^{t_f} \left( f_y\textrm(\utilde{\sigma};\utilde{X},p) + \sigma_{y} + \frac{(R(p))^{2}}{2Q} + \frac{D}{2C}(\mathcal{J}(\utilde{X}))^{2} \right) \dot{p} \, dt
\end{equation}
\color{black}with $\mathcal{J}(\utilde{X}) = \sqrt{\frac{3}{2}\utilde{X}:\utilde{X}}$.

\begin{figure*}[h]
    \centering
        \begin{subfigure}[b]{0.29\textwidth}
            \includegraphics[width=\textwidth]{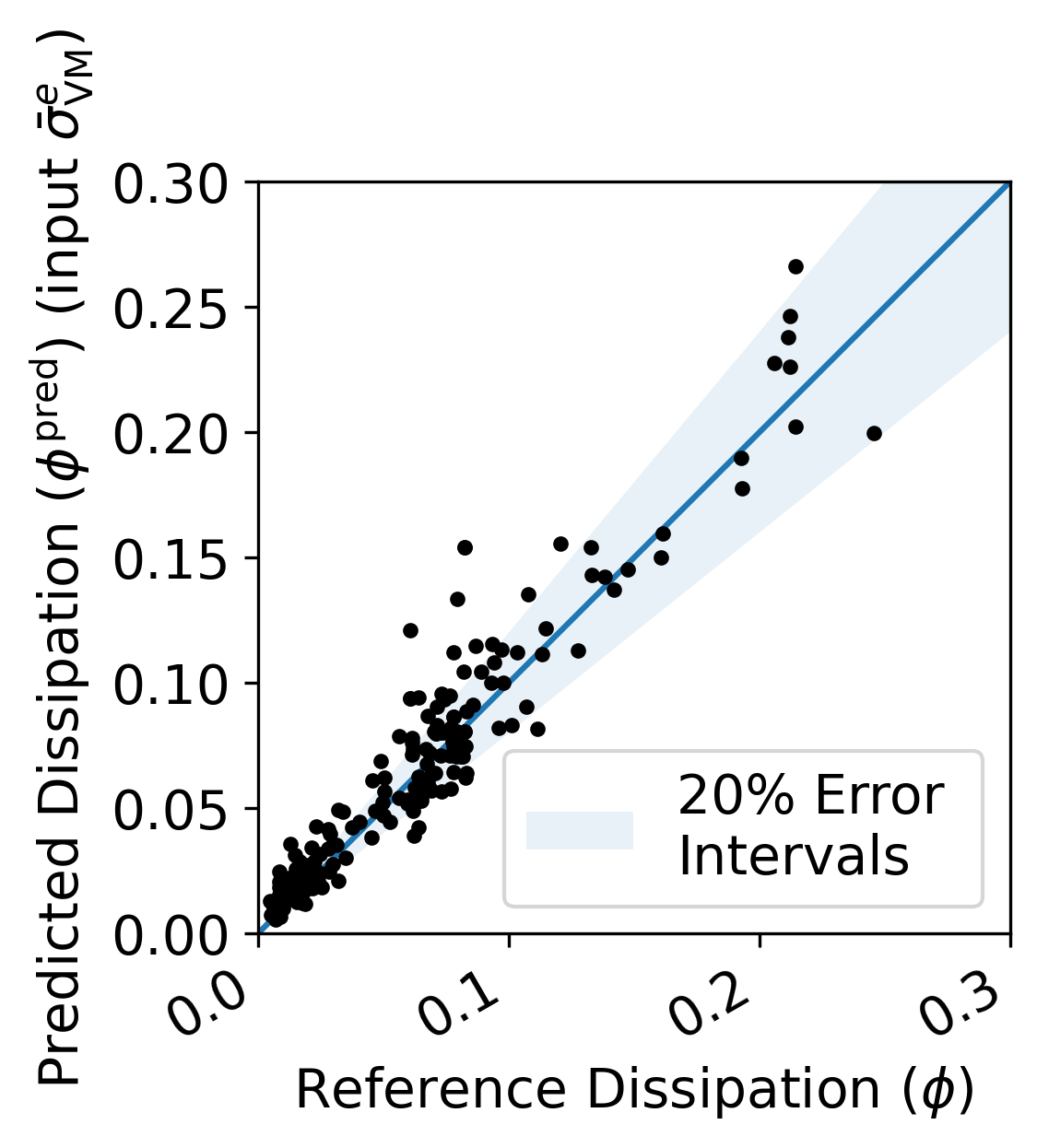}
            \caption{20 meshes}
        \end{subfigure}
        \begin{subfigure}[b]{0.29\textwidth}
            \includegraphics[width=\textwidth]{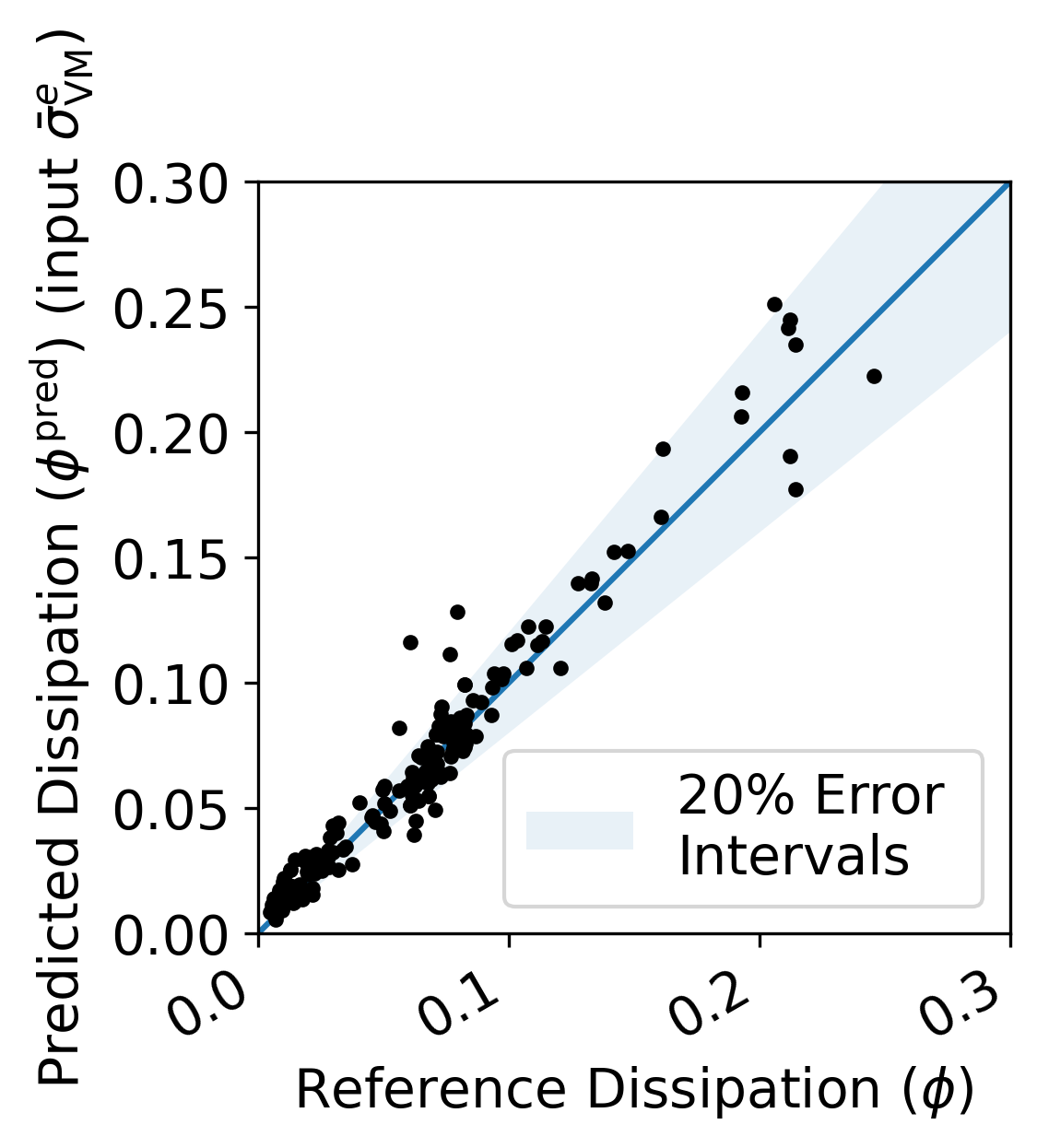}
            \caption{40 meshes}
        \end{subfigure}
        \begin{subfigure}[b]{0.29\textwidth}
            \includegraphics[width=\textwidth]{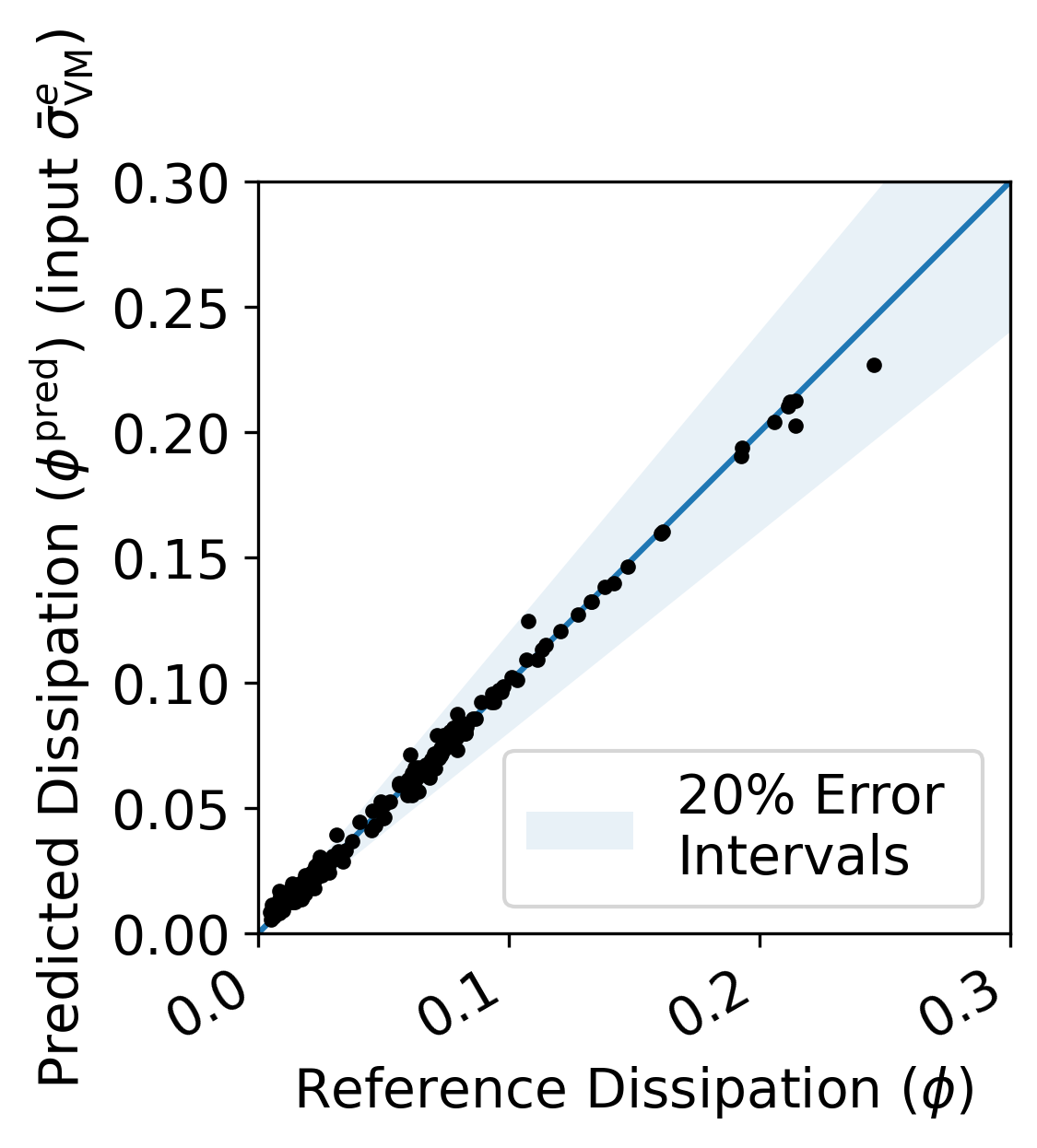}
            \caption{100 meshes}
        \end{subfigure}
        \caption{\black{Using von Mises stress coming from elasto-static simulations ($\bar{\sigma}_{\textrm{VM}}^{\#}$) as input: CNN predictions of dissipation for the clusters of a new, unseen mesh, after training for 2000 epochs on a varying number of meshes.}}
        \label{Fig:ResultsCNN_elasinput}
\end{figure*}

\color{black}
\paragraph{Impact of mechanical inputs on the performance of the CNN}\mbox{}\\
The CNN is trained using the von Mises stress from the elasto-static simulation ($\bar{\sigma}_{\textrm{VM}}^{\#}$) as input. The CNN learns to predict the reference dissipation $\phi$. \black{We evaluate the accuracy of the CNN on only the clusters of the K-medoids algorithm, as we expect that this reflects well the CNN accuracy on full fields due to the good representative nature of the clusters.} Figure \ref{Fig:ResultsCNN_elasinput} shows the CNN predictions on the clusters of a mesh that is not in the training set after being trained on a progressively increasing number (20 to 100) of meshes.  The results show that the accuracy of the prediction increases as a function of the amount of training data used. These results highlight that the CNN is able to predict the QoI using only $\bar{\sigma}_{\textrm{VM}}^{\#}$ as input.

\begin{figure*}[h]
    \centering
        \begin{subfigure}[b]{0.29\textwidth}
            \includegraphics[width=\textwidth]{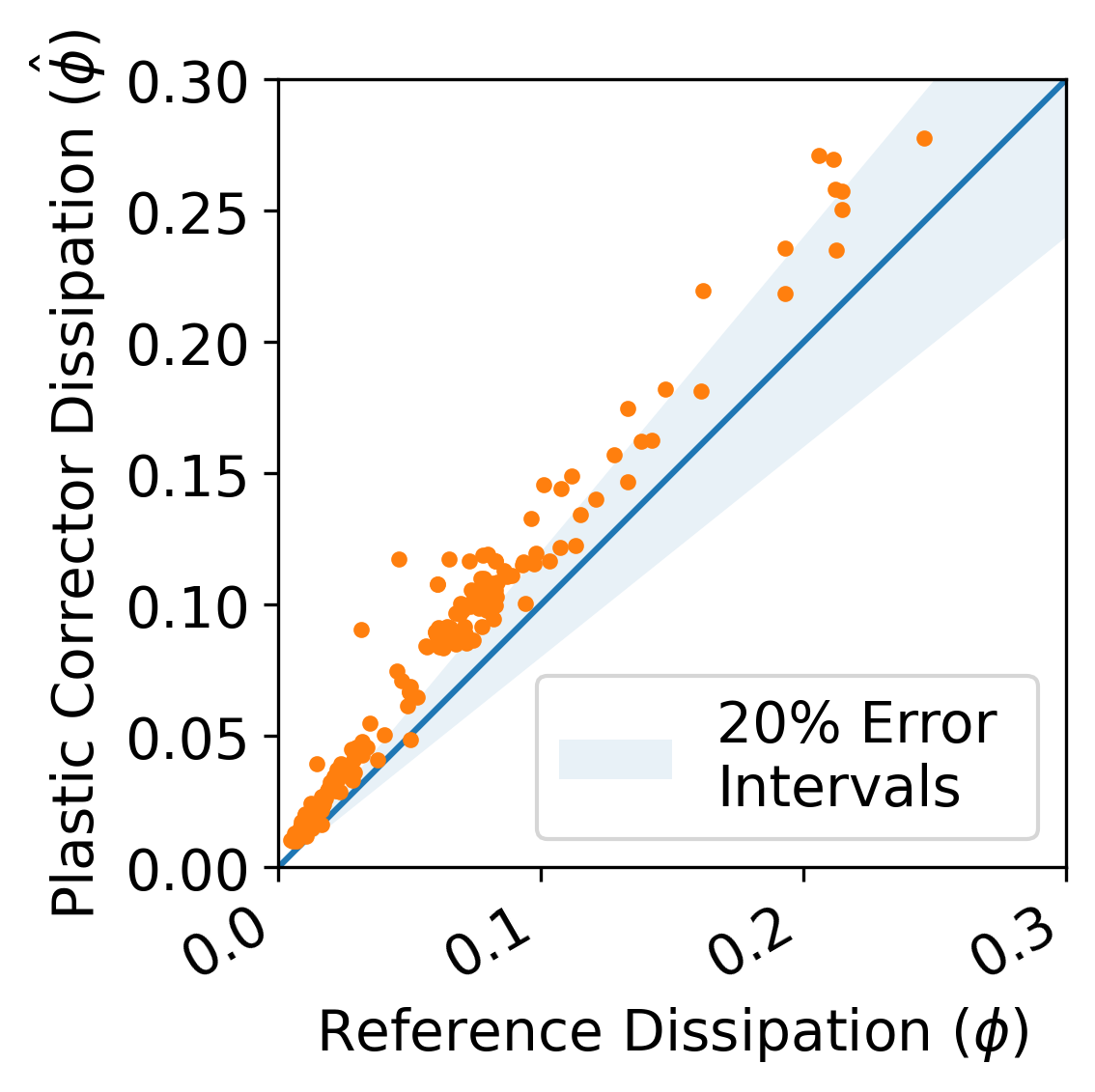}
            \caption{Without CNN}
        \end{subfigure}           
        \begin{subfigure}[b]{0.29\textwidth}
            \includegraphics[width=\textwidth]{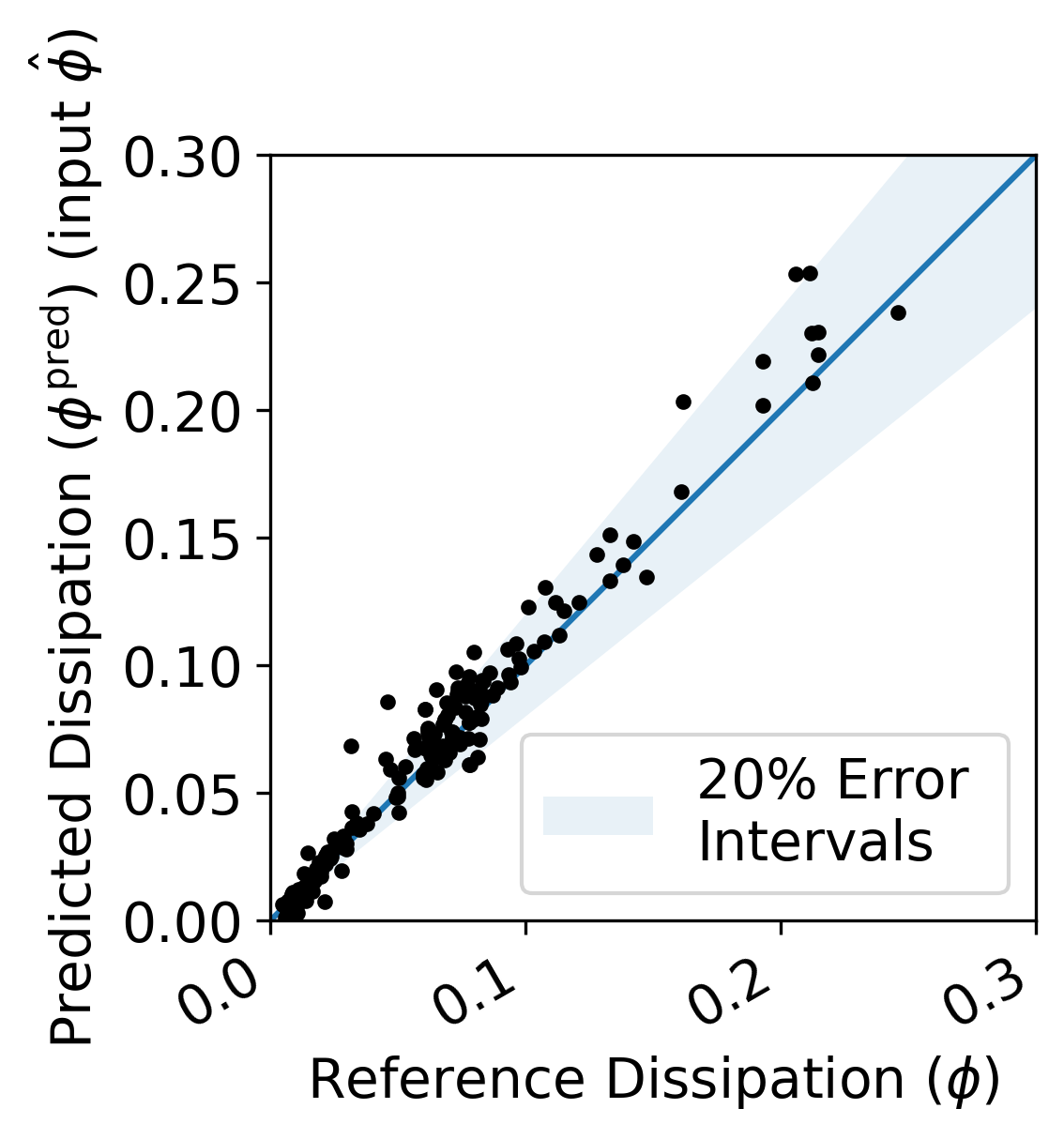}
            \caption{20 meshes}
        \end{subfigure}
        \begin{subfigure}[b]{0.29\textwidth}
            \includegraphics[width=\textwidth]{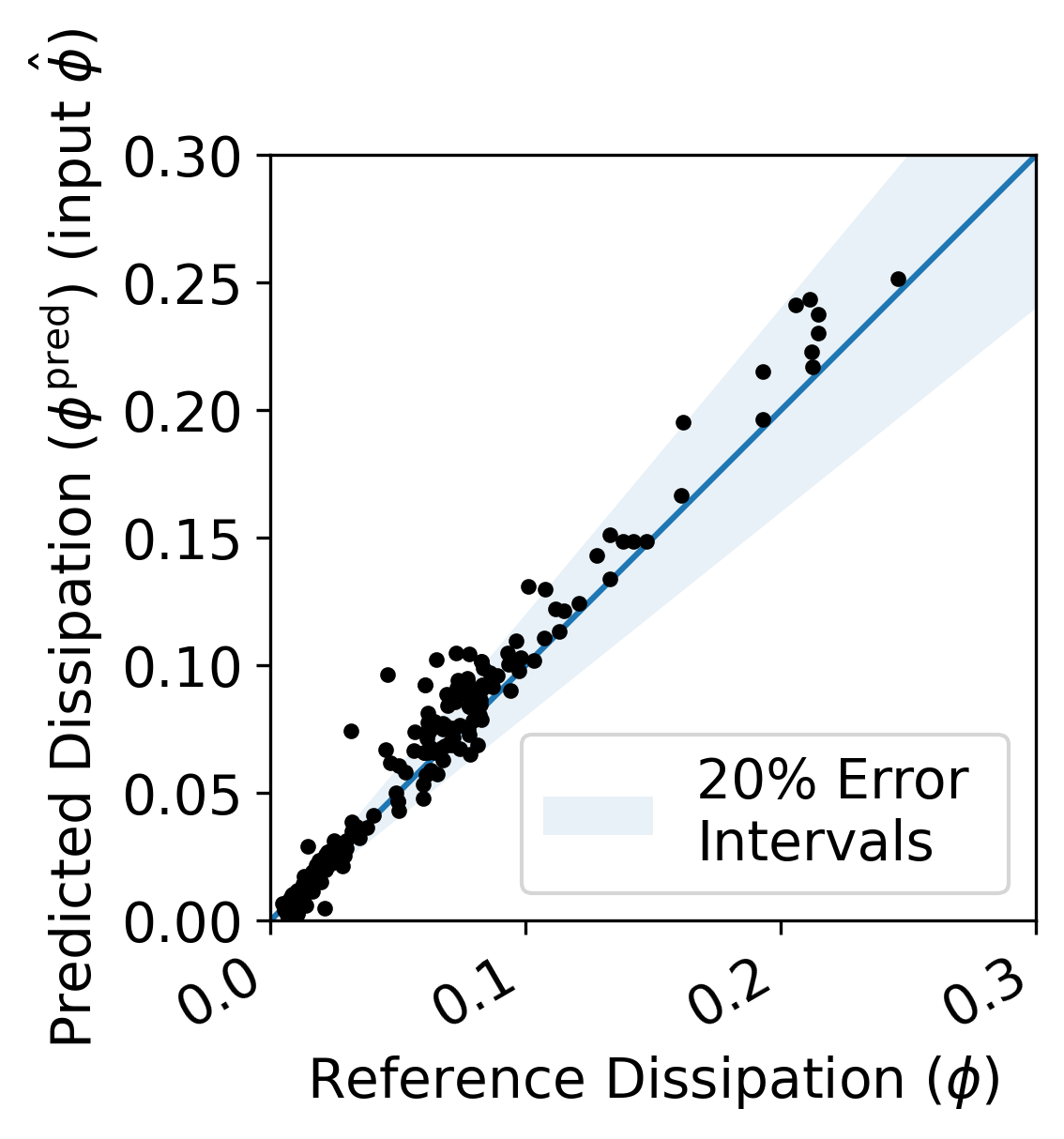}
            \caption{40 meshes}
        \end{subfigure}
        \begin{subfigure}[b]{0.29\textwidth}
            \includegraphics[width=\textwidth]{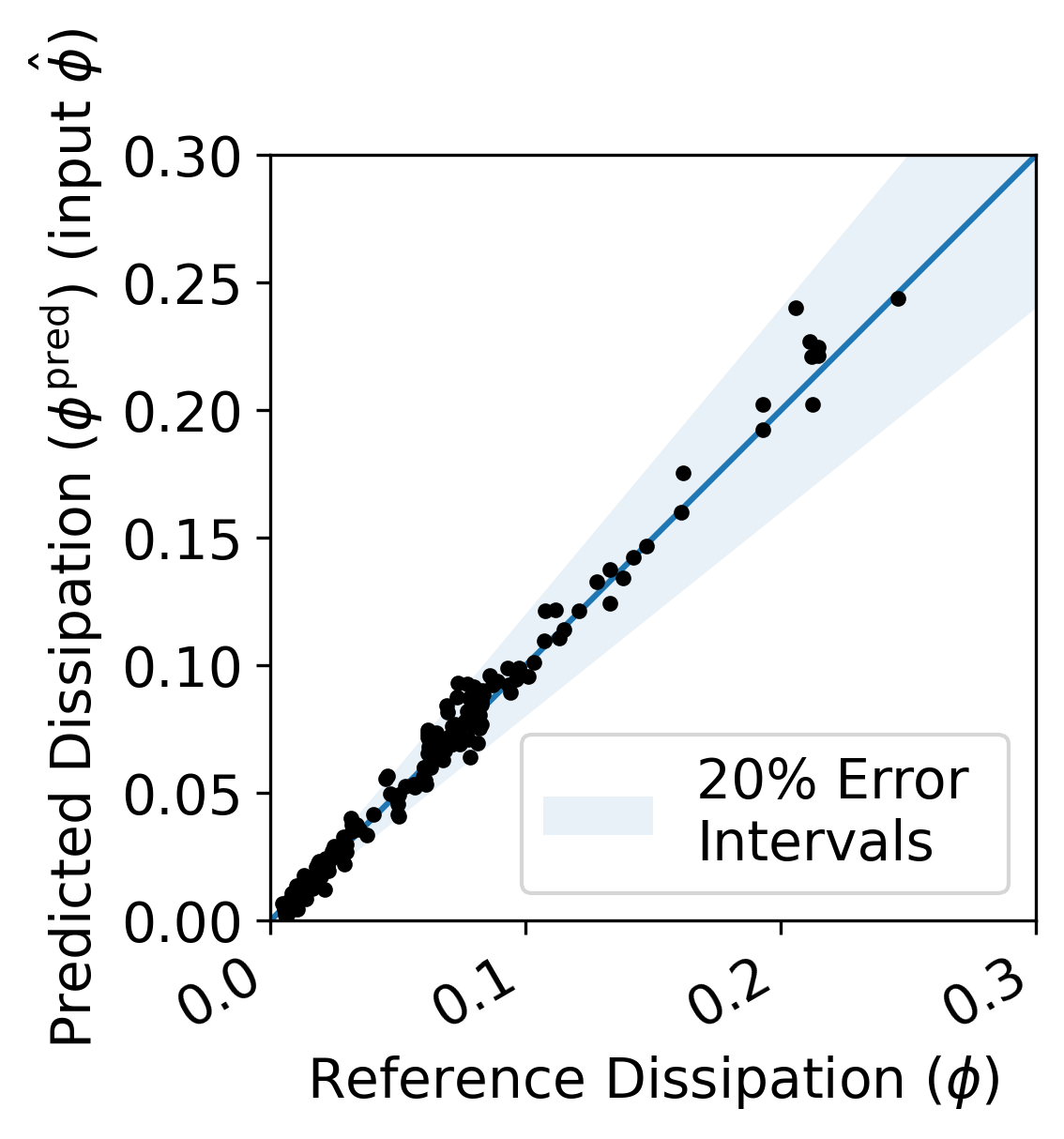}
            \caption{100 meshes}
        \end{subfigure}
        \begin{subfigure}[b]{0.29\textwidth}
            \includegraphics[width=\textwidth]{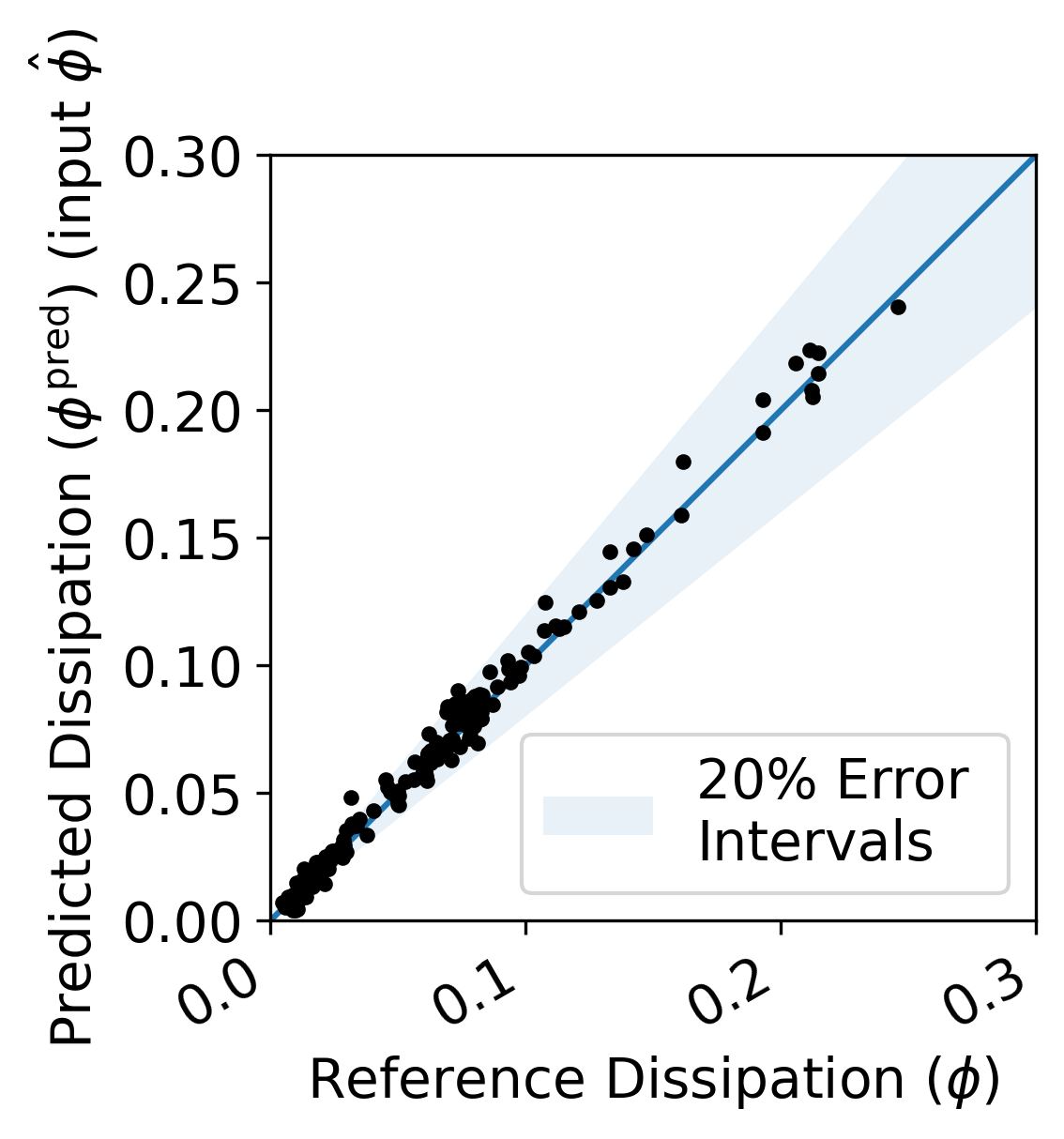}
            \caption{380 meshes}
        \end{subfigure}
        \begin{subfigure}[b]{0.30\textwidth}
            \includegraphics[width=\textwidth]{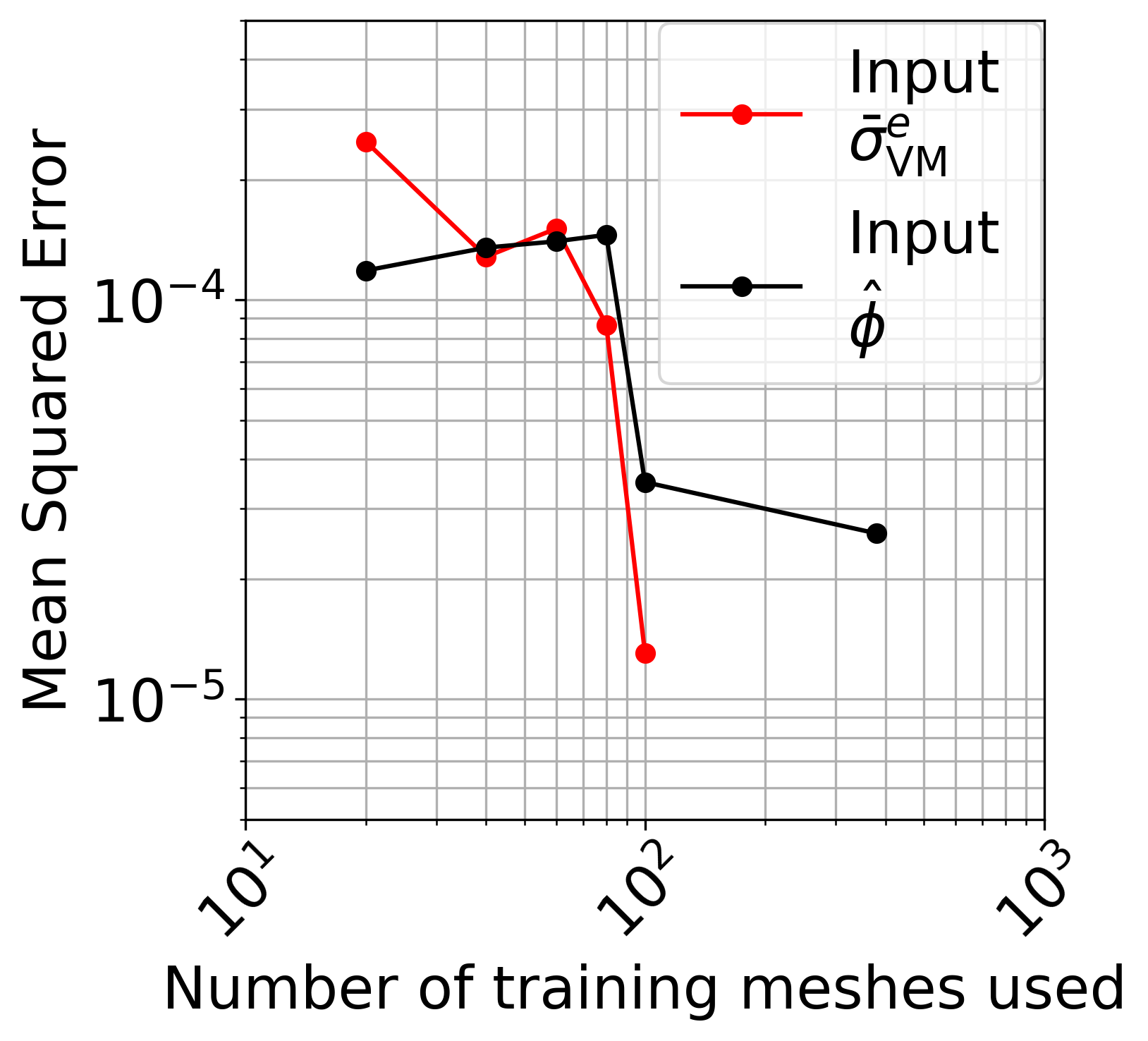}
            \caption{Sensitivity analysis}
        \end{subfigure}
        \caption{\black{Using plastic corrector dissipation ($\hat{\phi}$) as input: CNN predictions of dissipation in the clusters of a new, unseen mesh: (a) Plastic corrector, before passing into CNN, (b-e) CNN predictions after training for 2000 epochs on a varying number of meshes, and (f) Sensitivity analysis showing decrease of the MSE between the CNN predictions and reference data with increase in training data.}}
        \label{Fig:ResultsCNN}
\end{figure*}

Next, the CNN is trained using the plastic corrector dissipation ($\hat{\phi}$) as input to the CNN. The CNN learns to predict the difference between the plastic corrector dissipation $\hat{\phi}_i$ and the reference dissipation $\phi$. After the prediction, this difference is then subtracted from the plastic corrector dissipation $\hat{\phi}_i$ to get the final CNN-predicted dissipation. \black{We focus on learning the difference because it typically yields better results than directly learning the QoI \cite{Chinesta2020}.}
\color{black}
Figure \ref{Fig:ResultsCNN} (a) shows the plastic corrector dissipation as a function of the reference dissipation, and Figures \ref{Fig:ResultsCNN} (b)-(e) show the CNN predictions on the K-medoid clusters of a mesh that is not in the training set after being trained on a progressively increasing number (20 to 380) of meshes. The results show the accuracy of the predictions improved as a function of the amount of training data used. The CNN trained with 380 meshes has 89\% of cluster points of a mesh, not included in the training set, falling in a $\pm20\%$ error cone, and the highest values of dissipation, which are generally of the most interest, have less than a percent of error.

Figure \ref{Fig:ResultsCNN} (f) shows the mean square error (MSE) calculated on all the clusters of the mesh that is not in the training set, as a function of the number of meshes used for training. \textcolor{black}{The downward trend of the CNNs trained with the two inputs is a clear indicator that the CNN is able to perform better if given more data during training.}

\color{black}
We observe that training the CNN with the plastic corrector is beneficial when there is a low amount of available training data. The percentage of cluster points of a mesh not included in the training set, falling in a $\pm20\%$ error cone, for the CNN trained on different amounts of training data, is shown in the second column of Table \ref{tab:CNNpercentpoints_elas_plascorr}. The CNN trained with $\bar{\sigma}_{\textrm{VM}}^{\#}$ as input, on 20 meshes, performs significantly worse than the CNN trained with $\hat{\phi}$ as input. The difference of the accuracy of the predictions between the CNNs, however, becomes smaller and smaller as the amount of training data increases. In our opinion, this is because the corrections provided by the proposed Neuber-type approach are purely local. The plastic corrections do not contain any additional information as compared to elastic results about the source of the discrepancy between the input and output of the CNN model from the topology of the neighbourhood of a point.
\color{black}

\begin{table}[!htbp]
\color{black}
\caption{\color{black}Percentage of cluster points in a mesh not included in the training set falling in a $\pm20\%$ error cone for the networks trained with different inputs ($\bar{\sigma}_{\textrm{VM}}^{\#}$ and $\hat{\phi}$) at different amounts of training data used}
\label{tab:CNNpercentpoints_elas_plascorr}
\centering
\begin{tabular}[t]{lll}
\hline\noalign{\smallskip}
Training data used \newline (number of meshes) & Input $\bar{\sigma}_{\textrm{VM}}^{\#}$ \newline (\% points) & Input $\hat{\phi}$ \newline (\% points) \\
\noalign{\smallskip}\hline\noalign{\smallskip}
20 & 62\% & 80.5\% \\
\noalign{\smallskip}\hline\noalign{\smallskip}
40 & 73\% & 77.5\% \\
\noalign{\smallskip}\hline\noalign{\smallskip}
60 & 69\% & 76.0\% \\
\noalign{\smallskip}\hline\noalign{\smallskip}
80 & 76.5\% & 79.0\% \\
\noalign{\smallskip}\hline\noalign{\smallskip}
100 & 89\% & 87.5\% \\
\noalign{\smallskip}\hline\noalign{\smallskip}
380 & - & 89.0\% \\
\noalign{\smallskip}\hline
\end{tabular}
\color{black}
\end{table}

\begin{figure*}[h!]
    \centering
        \begin{subfigure}[b]{0.6\textwidth}
            \includegraphics[width=\textwidth]{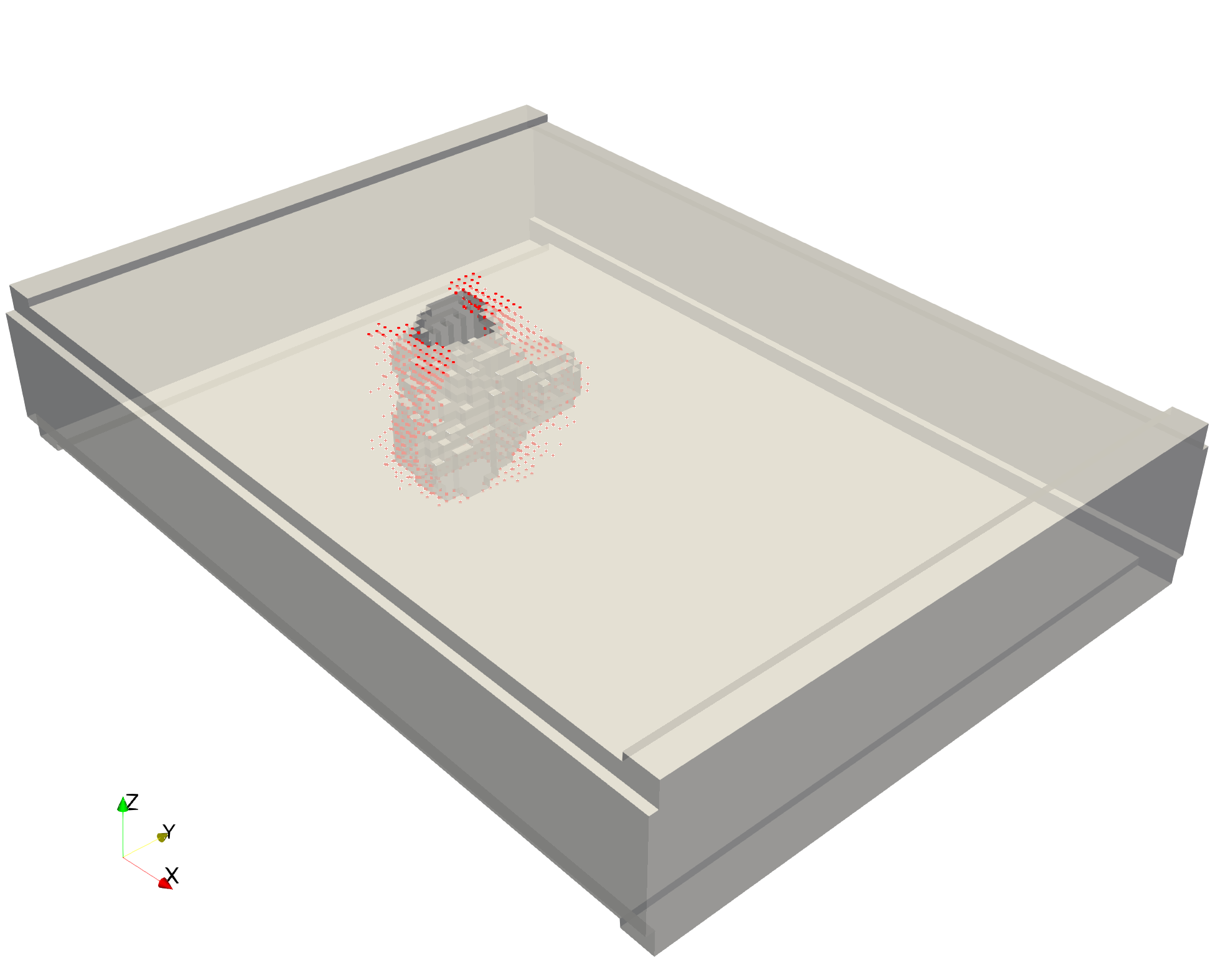}
            \caption{Points in mesh selected for correction}
        \end{subfigure}           
        \begin{subfigure}[b]{0.4\textwidth}
            \includegraphics[width=\textwidth]{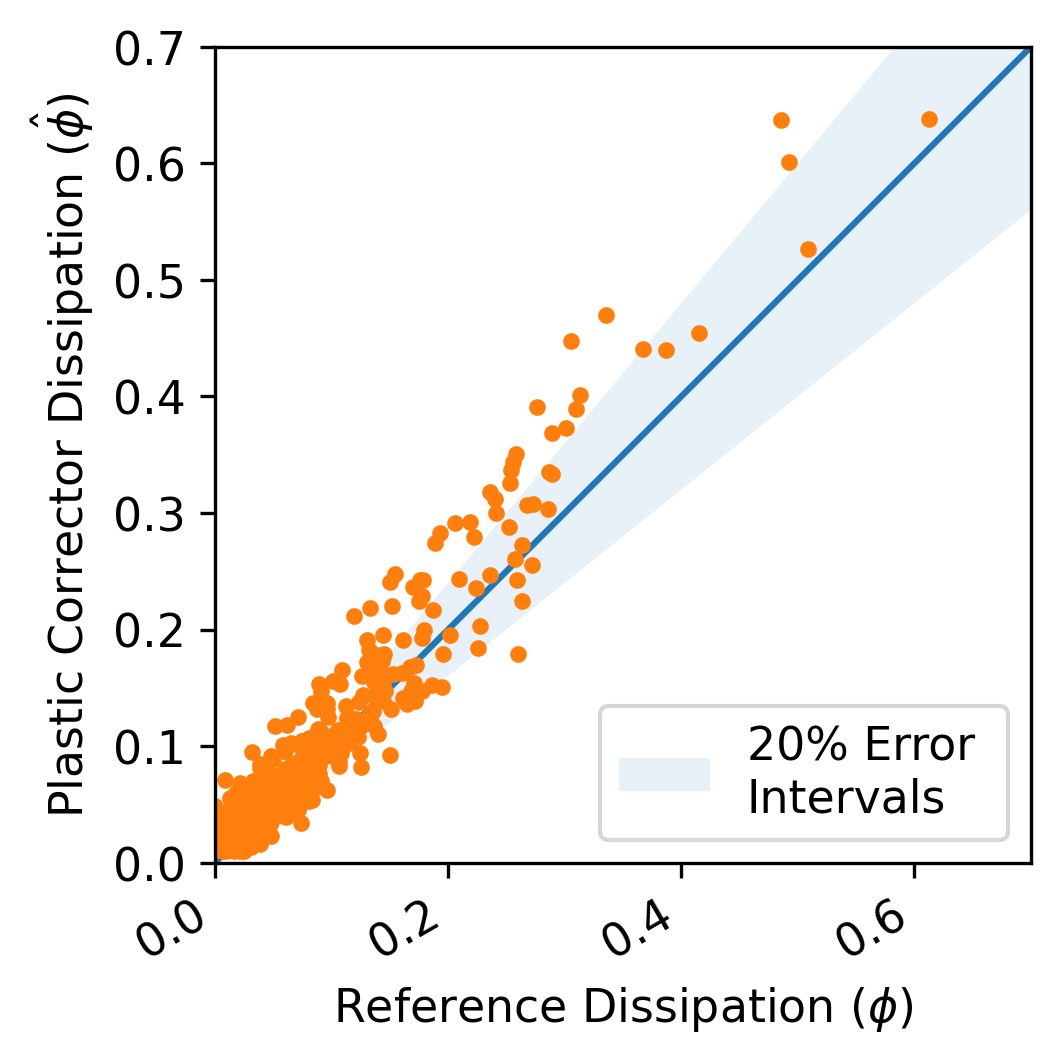}
            \caption{Before passing into CNN}
        \end{subfigure}
        \begin{subfigure}[b]{0.4\textwidth}
            \includegraphics[width=\textwidth]{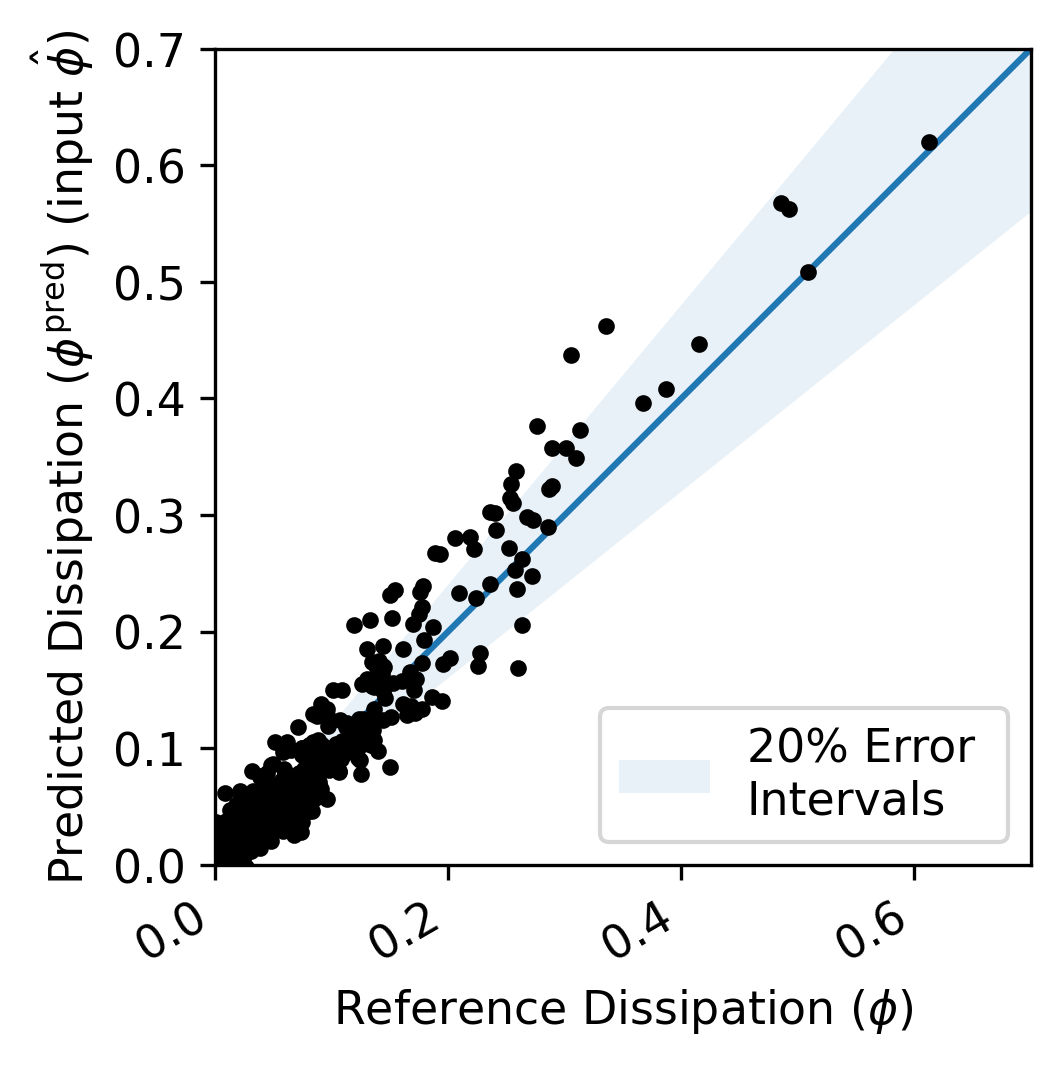}
            \caption{CNN predictions}
        \end{subfigure}
        \caption{Accuracy of the CNN trained on 380 specimens with spherical pores on a specimen containing a defect of non-spherical morphology. The highlighted red points indicate points that undergo plastic deformation, that are selected for correction, in a mesh not included in the training set with a defect of complex morphology}
        \label{Fig:ScaledDefectCNN}
\end{figure*}

\paragraph{\black{Accuracy of the CNN predictions for a non-spherical defect}}\mbox{}\\
\black{We now illustrate the behaviour of the CNN, trained successfully on a data-set consisting of 380 specimens with spherical pores, when making predictions on a specimen containing a defect of non-spherical morphology. To this hand, we will generate a new specimen with a single non spherical defect whose geometry is illustrated in Figure \ref{Fig:ScaledDefectCNN}(a). The geometric morphology of this defect comes from a computed tomography scan of a cast aluminium alloy. We ask the CNN to perform the plastic correction on all the points of the mesh that plastify (shown by the red points in Figure \ref{Fig:ScaledDefectCNN}(a)).} After being corrected by CNN, the number of predictions that are within the 20\% error cone reaches 40.4\%, starting from $40.8\%$ for correction-less Neuber-type predictions, as shown in Figure \ref{Fig:ScaledDefectCNN}. This behavior is in agreement with previous scientific findings \cite{Krokos2022} that report a poor performance of neural networks when used to make predictions for cases outside the distribution of the training set.

%\begin{figure*}[h!]
%    \centering
%        \begin{subfigure}[b]{0.6\textwidth}
%            \includegraphics[width=\textwidth]{Figures/wholemesh100_thresholdpoints.png}
%            \caption{}
%        \end{subfigure}           
%        \begin{subfigure}[b]{0.4\textwidth}
%            \includegraphics[width=\textwidth]{Figures/0_whole_mesh100.png}
%            \caption{}
%        \end{subfigure}
%        \begin{subfigure}[b]{0.4\textwidth}
%            \includegraphics[width=\textwidth]{Figures/0_pred_whole_mesh100.png}
%            \caption{}
%        \end{subfigure}
%        \caption{ (a) Highlighted red points indicate positive dissipation, greater than a threshold of 0.01 that are selected for correction, in a mesh not included in the training set with spherical defects (b) Before passing into CNN (c) CNN predictions.}
%        \label{Fig:ResultsCNN_FullMeshThreshold}
%\end{figure*}

\clearpage
\section{Conclusion}

\black{A plastic correction algorithm has been proposed that rapidly post-processes elasto-statics simulations to approximate the full-field elasto-plastic response of structures subjected to proportional loading. The classical Neuber's rule has been modified to operate on the deviatoric part of stress and strain tensors. Cyclic loading is handled via a change of origin at every loading peak, using the method in \cite{Chaudonneret1985}. A pointwise rule of proportional evolution of the deviatoric stress and strain tensors is used to reduce J2 elasto-plastic constitutive laws to a set of equations operating on scalar representations of stresses and strains. A fully implicit time integration algorithm for this system of equations has been developed, which leads to approximations of elasto-plastic solutions for arbitrary (proportional) loading sequences.}

\textcolor{black}{Numerical investigations on two test cases, one being a notched structure and the other one being a specimen with spherical pores, have shown good overall approximations of cumulative plastic strain fields, for both monotonic and cyclic loading, with errors in the notched structure matching previous studies \cite{Molski1981,jones1998}.} \textcolor{black}{The overly stiff behavior observed in these plastic corrector predictions arises because plastic accommodation and stress redistribution effects have not been taken into account. \cite{Molski1981,Chouman2014}.
For cyclic loading sequences, quantities computed over later cycles, like the cumulative plastic strain range, are reasonably well approximated. The plastic corrector's accuracy improves when the plasticity is relatively confined.}

\black{Furthermore, we have attempted to isolate the error due to the local proportional evolution rule in the plastic corrector algorithm. Investigations reveal that the proportionality rule contributes significantly less error compared to the Neuber-type rule. We have also shown that the major part of the error owed to the rule of local proportional evolution occurs during load reversal, when there is no accumulating plasticity.}

\black{Despite the plastic corrector's low computational cost compared to a full nonlinear finite element simulation, time-integration over long time histories may remain expensive. However, as a consequence of the rules underlying the proposed plastic corrector (J2 plasticity, pointwise proportionality of deviatoric strain and stress tensors, scalar Neuber rule), any output of the plastic correction algorithm depends on the elastic finite element simulation via the von Mises stress only. We have shown that this property can be exploited to build 1D meta-models that accelerate the plastic correction method even further, for any quantity of interest, by sampling the output of the plastic corrector for a relatively small number of von Mises stresses, training the meta-model and using it at almost no cost in place of the time integration algorithm.}

\black{Lastly, we have investigated the possibility of performing plastic correction using deep learning. These neural networks need reference elasto-plastic computations on a set of geometries for training, and do not generalise well to other types of geometries. We have shown that the output of the Neuber-type plastic corrector can be improved by using a CNN that leverages information about the local topology of material point neighbourhoods. We have also shown that the output of our Neuber-type methodology, as full-field input to the CNN, is indeed beneficial compared to using inputs of the NPC with a lesser mechanical content, i.e. the von Mises stress from elasto-static simulations.
We have shown that the benefit rapidly becomes negligible with an increase in the training data. Thus, the plastic corrector offers an advantage for deep learning-based corrections, particularly when the amount of training data is limited.}

\color{black}
To further advance the research presented in this paper, several directions can be pursued. First, the accuracy of the plastic corrector could be evaluated on cyclic loading sequences with non-zero mean values and irregular shapes. Second, the plastic corrector algorithm could be extended to accommodate non-proportional applied loading conditions, with an assessment of the algorithm's robustness and limitations under these scenarios. Finally, the potential of the plastic corrector in predicting elasto-viscoplastic material behavior and creep could be explored by incorporating time-dependent plasticity and viscous flow effects into the algorithm.

\color{black}
\section{Acknowledgments}
The authors would like to thank the French National Research Agency  (ANR-20-THIA-0022) for the partial funding of this project. The direct financial support of Mines Paris - PSL is also acknowledged.

\section{Competing Interests}
The authors declare that they have no known competing financial interests or personal relationships that could have appeared to influence the work reported in this paper.

\bibliography{sn-bibliography}

\clearpage
\section*{\black{Appendix A: Algorithm for solving the plastic corrector equations}}\label{appendixA}
\color{black}
This section details the algorithm used to solve numerically the equations of the local plastic corrector, for arbitrary (proportional) loading functions $f(t)$.

\paragraph{Monotonic loading}\mbox{}\\
In that case, the proposed Neuber-type rule (from section \ref{sec:ModNeuber}) reads as:
\begin{equation}\label{eq:app:neuber}
    s e  =  f^2
\end{equation}
The plastic corrector equations derived in section \ref{sec:plasreduced} are reminded here for the sake of readability.
\begin{equation}\label{eq:app:stressstrain}
\text{Stress strain relation} \quad
s = (e - e^p)
\end{equation}

\begin{equation}\label{appeq1}
 \text{Yield surface} \quad {\hat{f_y}}(s;x,\hat{p}) = \mathcal{\hat{J}}(s,x) - {\sigma}_{y} - R(\hat{p})
\end{equation}

\begin{equation}
   \text{Yield surface evolution} \quad  \hat{f}_y \dot{\hat{p}} = 0 \quad \textrm{and} \quad {\hat{f_y}}(s;x,\hat{p}){\leq 0}
\end{equation}

\begin{equation}\label{appeq2}
  \text{von Mises stress} \quad  \hat{\mathcal{J}}(s,x) = \left| s - \frac{x}{2 \mu}\right|{\bar{\sigma}_{\text{VM}}^{\#}}
\end{equation}

\begin{equation}\label{appeq3}
 \text{Kinematic hardening} \quad \dot{x} = \frac{2}{3}C\dot{e}^{p} - Dx\dot{\hat{p}}
\end{equation}

\begin{equation}\label{appeq4}
 \text{Isotropic hardening} \quad R(\hat{p}) = Q(1 - e^{-b\hat{p}})
\end{equation}

\begin{equation}\label{appeq5}
 \text{Cumulative plastic strain} \quad \dot{\hat{p}} = \frac{1}{3\mu} |\dot{e}^{p}| {\bar{\sigma}_{\text{VM}}^{\#}}
\end{equation}

\color{black}
Implicit time integration of this set of equations is done by introducing the following set of discretised equations for  $s_{i+1}$, $e_{i+1}$, $e^p_{i+1}$, $\hat{p}_{i+1}$ and $x_{i+1}$ at time $t_{i+1} \in [0 \ T]$, scalar quantities $e^p_i$, $\hat{p}_i$ and $x_i$ at time $0 \leq t_i < t_{i+1}$ being known.

\begin{equation}\label{eq:app:discretizedneuber}
    s_{i+1} e_{i+1}  =  f_{i+1}^2
\end{equation}

\begin{equation}\label{eq:app:discretizedstressstrain}
s_{i+1} = e_{i+1} - e^p_{i+1}
\end{equation}

\begin{equation}\label{eq:app:pnew}
    \hat{p}_{i+1} = \hat{p}_i + \left| e^p_{i+1} - e^p_{i} \right| \frac{1}{3\mu}{\bar{\sigma}_{\text{VM}}^{\#}}
\end{equation}

\begin{equation}\label{eq:app:xnew}
    x_{i+1} = \frac{x_i + \frac{2}{3} C (e^p_{i+1} - e^p_{i})}{1 + D (e^p_{i+1} - e^p_{i}) }
\end{equation}

\begin{equation}
    \hat{f_y}(s_{i+1},x_{i+1},\hat{p}_{i+1}) \leq 0 \qquad  \hat{f_y}(s_{i+1},x_{i+1},\hat{p}_{i+1}) ( \hat{p}_{i+1} - \hat{p}_i ) = 0
\end{equation}
To solve this system of equations, we first need to express $s_{i+1}$ as a function of $e^p_{i+1}$ by making use of equations \eqref{eq:app:discretizedneuber}, and \eqref{eq:app:discretizedstressstrain}:
\begin{equation}\label{eq:app:snew}
    s_{i+1} = \begin{cases}
\frac{- e^p_{i+1} + \sqrt{\left( e^p_{i+1} \right)^2 + {4 f_{i+1}^2}}}{2} & \text{if} \quad f_{i+1}>f_{i} \\
\frac{- e^p_{i+1} - \sqrt{\left( e^p_{i+1} \right)^2 + {4 f_{i+1}^2}}}{2} & \text{if} \quad f_{i+1}<f_{i} 
\end{cases}
\end{equation}
There are two roots for $s$. 
Substituting the expression for $e$ from the scalar stress-strain relation of the material law in the scalar Neuber-type equation, a second order polynomial equation is obtained:
The positive sign in the root is taken for increasing $f$ (tensile loading), the negative sign is taken in the case of decreasing $f$ (compression loading). \\

Now, the following solution algorithm is proposed
\begin{itemize}
    \item compute
\begin{equation}
    s_{i+1}^\star = \begin{cases}
\frac{- e^p_{i} + \sqrt{\left( e^p_{i} \right)^2 + {4 f_{i+1}^2}}}{2} & \text{if} \quad f_{i+1}>f_{i} \\
\frac{- e^p_{i} - \sqrt{\left( e^p_{i} \right)^2 + {4 f_{i+1}^2}}}{2} & \text{if} \quad f_{i+1}<f_{i} 
\end{cases}
\end{equation}
    \item compute $f_{y,i+1}^* = \hat{f_y}(s_{i+1}^\star,x_{i},\hat{p}_{i}) $, i.e. the value of the yield function assuming that no plastic flow takes place between $t_i$ and $t_{i+1}$.
    \item if $f_{y,i+1}^* \leq 0$, set $e^p_{i+1} = e^p_{i}$ 
    \item if $f_{y,i+1}^* > 0$, find $e^p_{i+1}$ such that 
    \begin{equation}
        \hat{f_y}(s_{i+1},x_{i+1},\hat{p}_{i+1}) = 0
    \end{equation}
    with $s_{i+1}$ given as a function of $e^p_{i+1}$ in equation \eqref{eq:app:snew}, $x_{i+1}$  given as a function of $e^p_{i+1}$ in equation \eqref{eq:app:xnew} and $\hat{p}_{i+1}$  given as a function of $e^p_{i+1}$ in equation \eqref{eq:app:pnew}. The root of this equation is found by a Newton algorithm, which is initialised by setting $e^p_{i+1} = e^p_{i}$, $x_{i+1}=x_{i}$ and $\hat{p}_{i+1} = {p}_{i}$. In our implementation, the derivative of $\hat{f_y}$ with respect to  $e^p_{i+1}$ is computed by finite differences.
    \item set $x_{i+1} = \frac{x_i + \frac{2}{3} C (e^p_{i+1} - e^p_{i})}{1 + D (e^p_{i+1} - e^p_{i}) }$ and $\hat{p}_{i+1} = \hat{p}_i + \left| e^p_{i+1} - e^p_{i} \right| \frac{1}{3\mu}{\bar{\sigma}_{\text{VM}}^{\#}}$ 
\end{itemize}

\color{black}

\paragraph{Cyclic loading}\mbox{}\\

In that case, the proposed Neuber-type rule (from section \ref{sec:ModNeuber}) reads as:
\begin{equation}\label{eq:app:cyclic_neuber}
    ( s - s_o )( e - e_o ) = ( f - f_o)^2
\end{equation}
From the plastic corrector equations derived in section \ref{sec:plasreduced},  the scalar stress-strain relation extended for cyclic loading is:
\begin{equation}
\label{eq:app:cyclic_stressstrain}
    (s - s_o) = (e - e_o) - (e^p - e^p_o)
\end{equation}
The rest of the plastic corrector equations are the same as equations \eqref{appeq1}-\eqref{appeq5}. The equations \eqref{eq:app:cyclic_neuber} and \eqref{eq:app:cyclic_stressstrain} involve the quantities $s_o$, $e_o$, $e^p_o$ and $f_o$ which are updated with $s_i$, $e_i$, $e^p_i$ and $f_i$ respectively, each time a load reversal occurs. The time discretisation of these equations is given here:
\begin{equation}\label{eq:app:discretizedcyclicneuber}
    (s_{i+1}-s_o) (e_{i+1}-e_o)  =  (f_{i+1}-f_o)^2
\end{equation}
\begin{equation}\label{eq:app:discretizedcyclicstressstrain}
s_{i+1}-s_o = (e_{i+1}-e_o) - (e^p_{i+1}-e^p_o)
\end{equation}
And the expression for $s_{i+1}$ as a function of $e^p_{i+1}$ by making use of equations \eqref{eq:app:discretizedcyclicneuber}, and \eqref{eq:app:discretizedcyclicstressstrain}:
\begin{equation}\label{eq:app:snewc}
    s_{i+1}-s_o = \begin{cases}
\frac{- (e^p_{i+1}-e^p_o) + \sqrt{\left( (e^p_{i+1}-e^p_o) \right)^2 + {4 (f_{i+1}-f_o)^2}}}{2} & \text{if} \quad f_{i+1}>f_{i} \\
\frac{- (e^p_{i+1}-e^p_o) - \sqrt{\left((e^p-e^p_o)_{i+1} \right)^2 + {4 (f_{i+1}-f_o)^2}}}{2} & \text{if} \quad f_{i+1}<f_{i} 
\end{cases}
\end{equation}
The same implicit time integrator used for the monotonic case presented in the previous paragraph is used to find the value of $e^p$ point-wise at each time-step.

\paragraph{Python implementation}

A Python implementation of this algorithm is made available under LGPL licence. The code and detailed usage instructions can be found in this  \href{https://github.com/AbhishekPalchoudhary/PlasticCorrector/tree/main}{GitHub repository} (https://github.com/AbhishekPalchoudhary/PlasticCorrector/tree/main). The repository includes scripts and instructions for point-wise plastic correction and correction of full elastic FEA computations. Examples are also included. This will enable reproduction of our results and enable further exploration of the algorithm's capabilities.

\color{black}

\clearpage
\color{black}
\section*{\black{Appendix B: Accuracy of the plastic corrector on a specimen with pores from X-ray tomography loaded in the high-cycle fatigue regime}}\label{appendixB}

\color{black}
As a final test case, we present here a specimen containing pores (arising due to the manufacturing process) that were meshed using information from tomography of a porous AlSi7Mg0.3 alloy. This case was chosen to show the particular suitability of application of the plastic corrector for high-cycle fatigue models that require elasto-plastic fields around pores of non-spherical geometry \cite{Bercelli2021,Lacourt2019,Palchoudhary2024}. The parameters of the Chaboche law are chosen according to reference \cite{LePhD2016}, and are summarised in table \ref{tab:PlasticityModelParameters_realpores}.

For the plastic corrector, an elasto-static FEA computation is computed with prescribed displacements $\bar{\underbar{u}}_a = [u_x,0,0]$ for $x\geq L_c$ and $\bar{\underbar{u}}_a = [-u_x,0,0]$ for $x\leq -L_c$ such that $\bar{\sigma}_{\text{VM}}^{\#}$ in the gauge section away from the pores is at the yield stress of the material (this computation corresponds to $f=1$). The function $f(t)$ is chosen to have 20 cycles. At the peak of the cyclic load, $f(t) \bar{\sigma}_{\text{VM}}^{\#}$ is chosen to reach 47\% of the yield stress of the material in the gauge section away from pores. The boundary conditions are shown in Figure \ref{fig:BCs_SpecimenRealPores}.

For the reference elasto-plastic computation, prescribed displacements are applied to both ends of the porous specimen in the same way. The von Mises stress in the gauge section away from pores reaches 47\% of the material's yield stress at peak loads in the cycles, assuming the body behaves elastically.

\color{black}
The plastic corrector and reference computations for the cumulative plastic strain range $\Delta p$ in the $20\textsuperscript{th}$ cycle for this mesh are respectively shown in Figure \ref{Fig:RealPoresApproximationQuality}(a) and \ref{Fig:RealPoresApproximationQuality}(b). A scatter plot comparing $\Delta p$ in all the integration points in the mesh is shown in Figure \ref{Fig:RealPoresApproximationQuality}(c). The accuracy of the full-field approximation indicates that the plastic corrector performs well for varying material parameters and on different sizes and geometries of pores. Despite the nominal loading being 0.47$\sigma_y$, there are regions with higher stress concentrations arising due to a maximum stress concentration factor of $k_t \sim 4.3$. The plastic corrector approximates the solution well in these regions despite the high $k_t$ as the plasticity remains confined due to the small size of the pores.
 
\begin{table}[!htbp]
\caption{Parameters of the elasto-plastic model detailed in equations \eqref{eq:plas:stressfull}-\eqref{eq:plas:yieldfull2} \cite{Chaboche1989,VietDucLe2018}}
\label{tab:PlasticityModelParameters_realpores}
\centering
\begin{tabular}[t]{llllllll}
\hline\noalign{\smallskip}
E & $\sigma_y$ & b & Q & C & D  \\
MPa & MPa &   & MPa & MPa &  \\
\noalign{\smallskip}\hline\noalign{\smallskip}
75500 & 170 & 19 & 20 & 127499 & 1334\\
\noalign{\smallskip}\hline
\end{tabular}
\end{table}

\color{black}
\begin{figure}[h!tbp]
    \centering
    \begin{tikzpicture}
        \node[anchor=south west,inner sep=0] (image) at (0,0) {\includegraphics[width=0.9\textwidth]{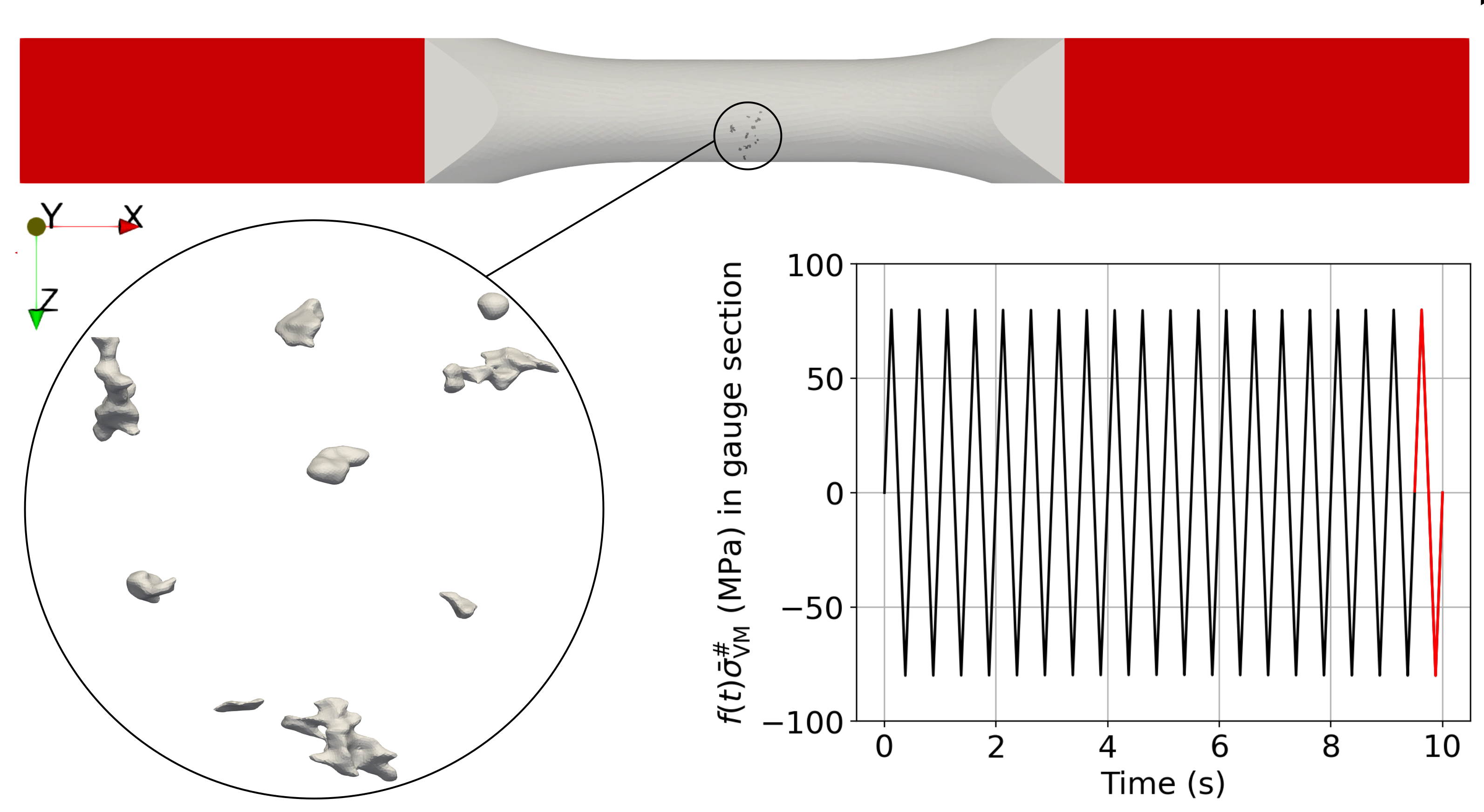}};
        \begin{scope}[x={(image.south east)},y={(image.north west)}]
            % Draw arrows
            \draw[black, ultra thick, <-] (-0.026, 0.86) -- (0.014, 0.86); % Left arrow
            \node at (0.014, 1.01) {$f(t)\bar{\underbar{u}}_a$}; % Left arrow text
            \draw[black, ultra thick, ->] (0.99, 0.86) -- (1.03, 0.86); % Right arrow
            \node at (0.99, 1.01) {$f(t) \bar{\underbar{u}}_a$}; % Right arrow text
            \node at (0.27, 1.01) {$-L_c$}; % BC text
            \node at (0.73, 1.01) {$L_c$}; % BC text
        \end{scope}
    \end{tikzpicture}
    \caption{\label{fig:BCs_SpecimenRealPores} \textcolor{black}{Boundary conditions (shown in red) for a specimen containing a sub-volume of tomography-informed pores, showing where displacement is applied to get a cyclic loading in the gauge section of the specimen (away from pores) with peak equal to 47\% of the yield stress of the material. The 20\textsuperscript{th} cycle is chosen for the computation of $\Delta p$.}}
\end{figure}
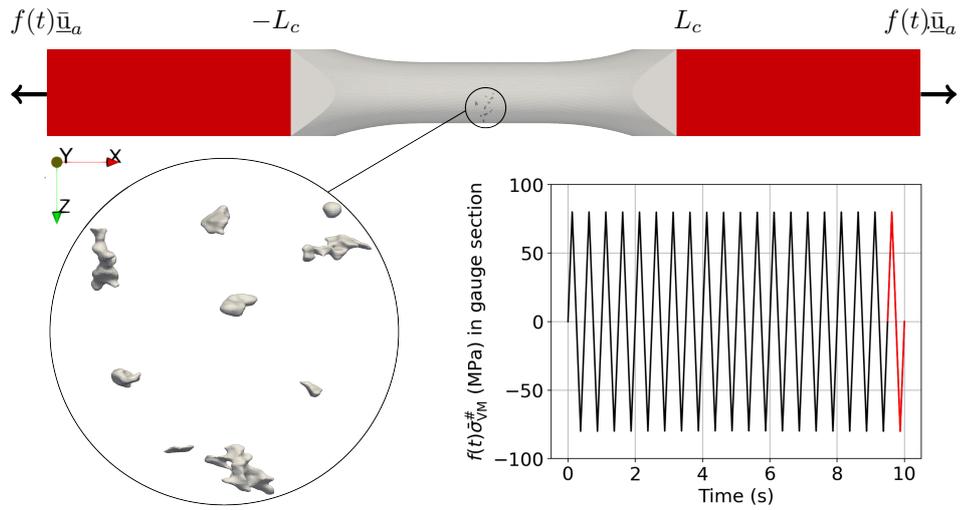
\begin{figure}[htbp]

    \centering
         \begin{subfigure}[b]{0.49\textwidth}
            \includegraphics[width=\textwidth]{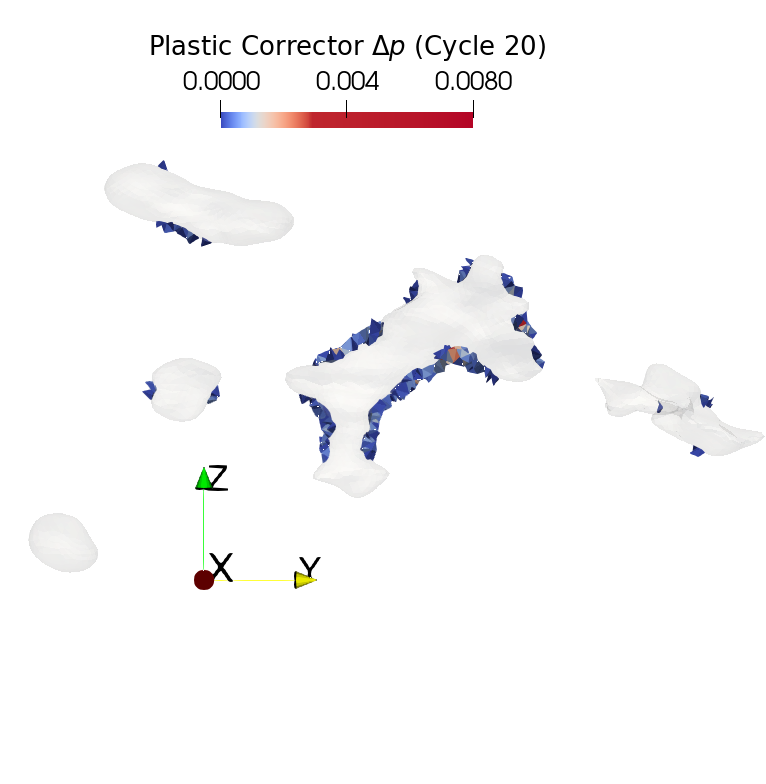}
            \caption{}
        \end{subfigure}
        \hfill
        \centering
        \begin{subfigure}[b]{0.49\textwidth}
            \includegraphics[width=\textwidth]{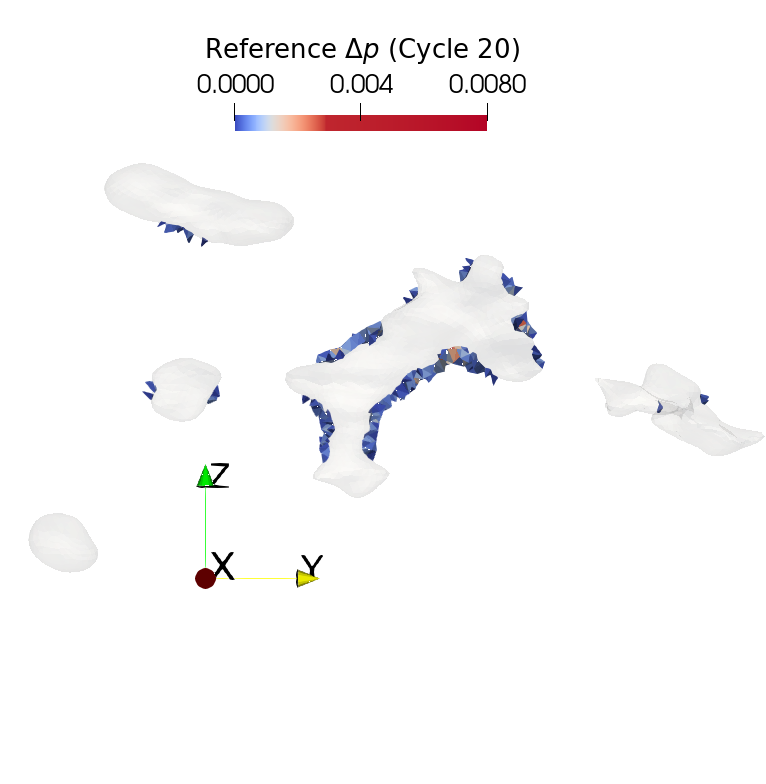}
            \caption{}
        \end{subfigure}
        
        \begin{subfigure}[b]{1\textwidth}
            \centering
            \adjincludegraphics[width=8.42cm,Clip={.0\width} {.0\height} {0.0\width} {.0\height}]{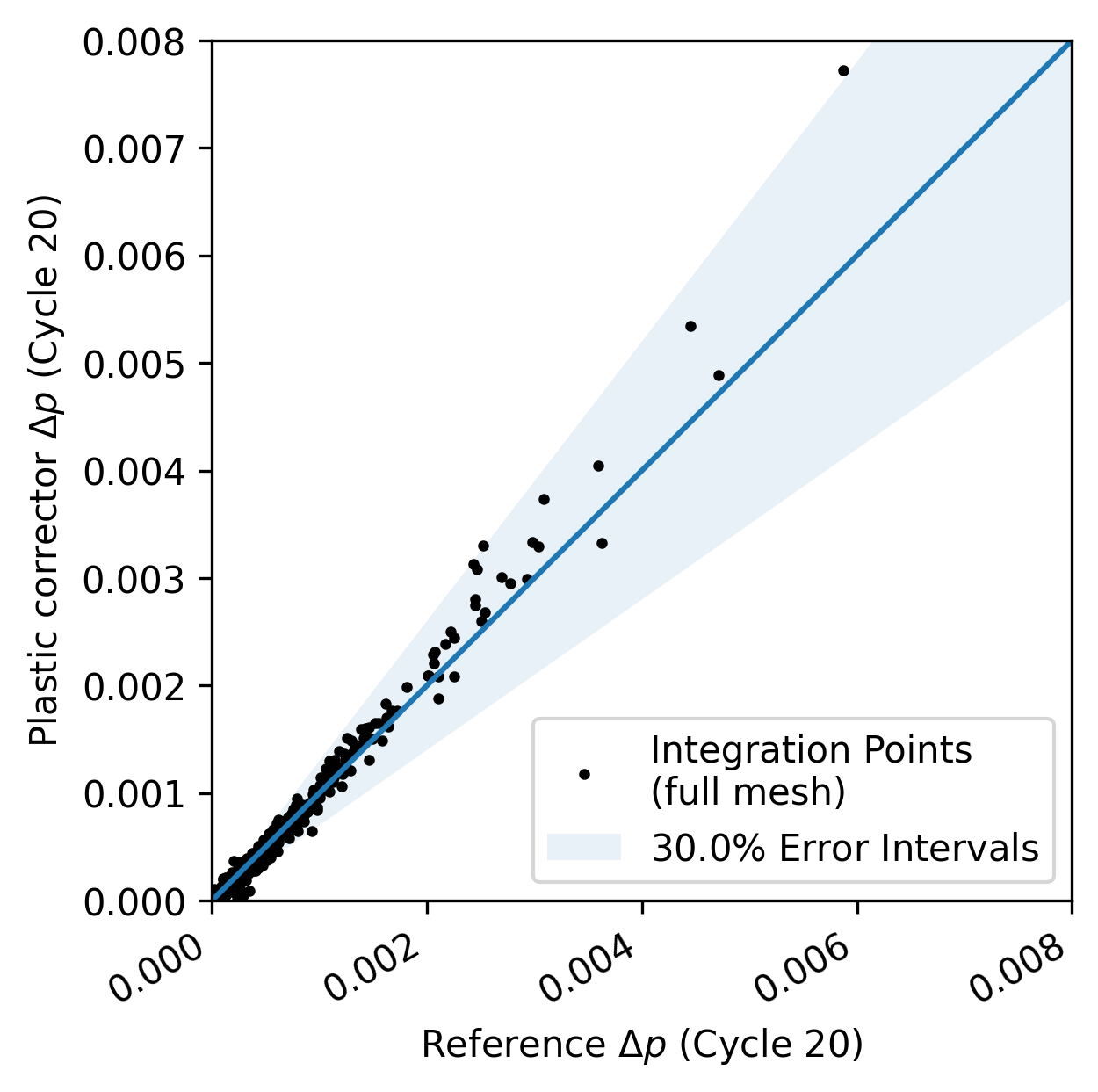}
            \caption{}
        \end{subfigure}   
    
    \caption{\label{Fig:RealPoresApproximationQuality} \textcolor{black}{(a-b) A comparison between $\Delta p$ in the $20\textsuperscript{th}$ cycle calculated via the plastic corrector and a reference computation via Z-Set \cite{Besson1998} in a few pores of a specimen containing a subvolume of pores (with the maximum stress concentration factor being $k_t \sim 4.3$). The loading corresponds to $80$ MPa in the gauge section at the peak of cyclic loading, away from pores (around 47\% of $\sigma_y$) (c) A scatter plot comparing $\Delta p$ calculated via the plastic corrector and a reference computation via Z-Set \cite{Besson1998} in all the integration points of the specimen containing the subvolume of pores.}} 
\end{figure}

\clearpage
\downloadlink{https://cloud.minesparis.psl.eu/index.php/s/1FU3noY0oJRUh76}
\end{document}